\pdfoutput=1
\documentclass[11pt,a4paper]{article}
\usepackage[utf8]{inputenc}
\usepackage[T1]{fontenc}
\usepackage[a4paper,textwidth=15.4cm,textheight=23.5cm,centering,headsep=0.8cm]{geometry}
\usepackage[final]{microtype}
\DisableLigatures[i]{encoding = T1, family = rm*}
\usepackage{amsmath}
\usepackage{amsthm}
\usepackage{newpxtext,newpxmath}

\usepackage{booktabs,graphicx}
\usepackage{adjustbox}
\usepackage{xcolor}
\usepackage{tikz}
\usepackage[font=small,labelfont={bf},labelsep=period]{caption}
\usepackage{titlesec}
\titleformat{\section}{\normalfont\large\bfseries}{\thesection}{0.8em}{}
\titleformat{\subsection}{\normalfont\normalsize\bfseries}{\thesubsection}{0.7em}{}
\titlespacing*{\section}{0pt}{3.2ex plus 1ex minus .2ex}{1.6ex plus .2ex}
\titlespacing*{\subsection}{0pt}{2.4ex plus 1ex minus .2ex}{1.0ex plus .2ex}
\usepackage{fancyhdr}
\usepackage{cite}
\usepackage{url}
\definecolor{linkblue}{RGB}{30,60,140}
\usepackage[colorlinks=true,linkcolor=linkblue,citecolor=linkblue,urlcolor=linkblue,
  pdftitle={The Exhaustive Peres Transform: Extraction and Universal Coding},pdfauthor={Aurelien Garivier},
  pdfkeywords={randomness extraction, universal coding, individual sequences, Peres algorithm, Burrows-Wheeler transform}]{hyperref}
\newtheorem{theorem}{Theorem}
\newtheorem{proposition}[theorem]{Proposition}
\newtheorem{lemma}[theorem]{Lemma}
\newtheorem{corollary}[theorem]{Corollary}
\theoremstyle{definition}
\newtheorem{definition}[theorem]{Definition}
\theoremstyle{remark}
\newtheorem{remark}[theorem]{Remark}

\newcommand{\bits}{\{0,1\}}
\newcommand{\E}{\mathbb E}
\renewcommand{\Pr}{\mathbb P}
\DeclareMathOperator{\wt}{w}
\DeclareMathOperator{\sat}{sat}
\newcommand{\KT}{\mathrm{KT}}
\newcommand{\U}{\mathsf U}
\newcommand{\V}{\mathsf V}
\newcommand{\T}{\mathsf T}

\newcommand{\Rc}{\mathcal R}
\newcommand{\A}{\mathcal A}
\newcommand{\Dec}{\mathrm{Dec}}
\newcommand{\ustep}{\mathrm u}
\newcommand{\vstep}{\mathrm v}
\newcommand{\ee}{\mathrm e}
\newcommand{\KL}{\mathrm D}
\newcommand{\gap}{\mathrm{gap}}
\definecolor{ycol}{RGB}{30,90,200}
\definecolor{lcol}{RGB}{190,40,40}
\definecolor{ucol}{RGB}{40,130,60}

\title{\bfseries The Exhaustive Peres Transform:\\ Extraction and Universal Coding}
\author{Aur\'elien Garivier\thanks{ENS de Lyon, France. E-mail: \texttt{aurelien.garivier@ens-lyon.fr}. ORCID: 0000-0002-4906-9573.}}
\date{November 2026}

\begin{document}

\maketitle
\thispagestyle{plain}

\begin{abstract}
Run to exhaustion, Peres's iteration of von Neumann's procedure yields a bijection of $\bits^N$ that splits every string into fair bits and a residual. The residual is self-delimiting and determines the number of ones, so the fair bits are exactly uniform given the residual under every exchangeable law. Each residual one accounts for a power of two of the weight of the input, and the residual of every input of length $N$ has fewer than $\sqrt3\,N^{0.793}$ ones. This sparsity gives a short proof, with an explicit remainder, that the fair bits per symbol tend almost surely to the entropy for Bernoulli sources and every finite alphabet. The Bernoulli likelihood factorises along the recursion, and code B, which codes the residual at plug-in biases, has the length of the two-part code. Code G gamma-codes the gaps of the residual, with no estimate and no arithmetic coder: on every string its length is at most the logarithm of the type class plus $O(N^{0.793}\log N)$. What is more, a concatenation of $J$ pieces costs its piecewise empirical entropy plus a term that is sublinear when $J=o(N)$, without the cuts being sent. After a Burrows--Wheeler transform of the reversed string, both codes are universal for stationary ergodic sources: code B through a context tree of growing depth, with regret $O(2^k\log N)$ against Markov sources of order $k$, and code G with no tree, competing on every individual sequence with the empirical conditional entropy of every order. For laws invariant under permutations inside the pieces of a partition, the fair bits not mixed across the pieces are exactly uniform given everything else; after the BWT, the fair bits of every stationary source fall short of their length by $o(N)$ bits of entropy, with an exponential tail for Markov sources. Experiments measure the finite-length price of these guarantees.
\end{abstract}


\section{Introduction}

Von Neumann's procedure \cite{vonNeumann51} turns a coin of unknown bias into a fair one: read the bits in pairs, output the second bit of every unequal pair and discard the equal pairs. Its yield, $\theta(1-\theta)$ fair bits per input bit for a coin of bias $\theta$, is far from the entropy $h(\theta)$. Elias \cite{Elias72} attained the entropy by extracting from the index of the input in its type class, and Ryabko and Matchikina \cite{RyabkoMatchikina00} computed this index in quasi-linear time with fast multiplication. Peres \cite{Peres92} took another route: the sequence $u$ of pair XORs and the sequence $v$ of values of the equal pairs still contain randomness and can be fed back to the same procedure; truncated at depth $\nu$, the recursion extracts at a rate $r_\nu(\theta)$ that increases to $h(\theta)$. Pae generalised the recursion to loaded dice \cite{Pae15}, to binarisation trees \cite{Pae16} and to a general class of Peres-style recursions \cite{Pae18}, where the recursion run to exhaustion is called the Peres function and its asymptotic optimality is stated; Zhou and Bruck \cite{ZhouBruck12} gave a universal scheme to lift binary extractors to larger alphabets, and Prasitsupparote et al.\ \cite{PKS18} compared the extractors of Elias and Peres at finite lengths.

On the coding side, enumerative coding \cite{Cover73} codes the type of the input and then its index in the type class; the Krichevsky--Trofimov (KT) mixture \cite{KT81} attains a regret of $\frac12\log N+O(1)$, which is first-order optimal \cite{Rissanen84}, and Rissanen's two-part codes send an estimate of the parameter followed by the data coded with it. Context-tree weighting \cite{WST95} extends KT to tree sources. Without any source, Lempel and Ziv \cite{ZivLempel78} compress every individual sequence to its finite-state compressibility, and the empirical conditional entropy of order $k$ is the benchmark of the pointwise analyses of LZ78 \cite{KosarajuManzini99} and of the compressors \cite{Manzini01,KLV07} based on the Burrows--Wheeler transform (BWT) \cite{BW94}; the universality of BWT-based codes for stationary ergodic sources is studied in \cite{EVKV02}. Source codes and random number generators are related: the output of an asymptotically optimal source code is close to uniform in a normalised-divergence sense \cite{VKV98}, whereas the fair bits below are exactly uniform at every length. Finally, Ar{\i}kan's source polarization \cite{Arikan10} applies the same XOR of pairs, recursively and linearly.

Peres's recursion is usually assessed by the randomness it extracts. This paper studies the information it leaves behind. Run to exhaustion with \emph{everything} it does not extract kept, the recursion is a bijection $\T$ of $\bits^N$ onto pairs $(Y;R)$ of fair bits and residual bits, with $|Y|+|R|=N$. Two elementary facts about the residual carry everything. The residual strings form a complete prefix code that determines the weight of the input, so that each residual labels a block of $2^{|Y|}$ strings of the same type; this is what makes the fair bits exactly fair under exchangeable laws, where \emph{exactly fair} means that, conditionally on the residual, $Y$ is uniform on $\bits^{|Y|}$: its bits are independent, unbiased, and independent of the residual given their number. And each residual one accounts for a power of two of the weight of the input, an exact charge identity which forces the number of residual ones to be sublinear on every input; this deterministic constraint is a statement about every individual string, and it is the common source of the extraction and coding results below. The contributions are three.
\begin{enumerate}
\item \emph{A sparse reversible representation} (Section~\ref{sec:transform}). The residuals form a complete prefix code of the whole space $\bits^N$ that the decoder parses without knowing its length and that determines the weight (Proposition~\ref{prop:image}); each residual one carries the charge $2^{v}$, $v$ the number of $\vstep$ steps of its path, and the charges sum to the weight (Theorem~\ref{thm:charge}); hence the residual of every input of length $N$ has fewer than $\sqrt3\,N^{(\log3)/2}$ ones (Theorem~\ref{thm:sparse}; the exponent $0.7925$ improves to $0.7729$), and the Elias gamma codes of its gaps describe it in $O(N^{0.793}\log N)$ bits. The exhaustive recursion as an extractor and its asymptotic optimality are in \cite{Pae18}, uniformity conditional on the output length is Peres's Proposition~1 \cite{Peres92}, and the component bijections are Pae's structure lemma \cite[Lemma~4]{Pae18}; the prefix code of the whole space rather than of one type class, so that nothing is discarded, and the charge identity with its sparsity consequence are new.
\item \emph{Two consequences: exact extraction with a remainder, and a parameter-free code} (Sections~\ref{sec:extraction} and~\ref{sec:B}). The fair bits are exactly uniform given the residual for every exchangeable law, the extraction loss $H(X)-\E|Y|$ is at most $O(N^{0.793}\log N)$ with an exponential tail, and $|Y|/N$ tends to $h(\theta)$ almost surely for a Bernoulli source, with a remainder uniform in $\theta$ (Theorem~\ref{thm:rate}); binarisation extends this to every finite alphabet. Code G stores the fair bits and writes the gamma codes of the gaps of the residual, with no weight, no bias, no fitted parameter and no arithmetic coder; on every string its length is at most the logarithm of the type class plus a sublinear term (Theorem~\ref{thm:universal}), whose constant is large at practical lengths. A pricing identity for pairs of bits factorises any product of Bernoulli probabilities along the recursion; for a Bernoulli source it gives a causal factorisation with a factor $\frac12$ per fair bit and explicit biases for the residual bits (Theorem~\ref{thm:factor}), and code B, which codes the residual at plug-in biases, has the length of the two-part code \cite{Rissanen84}, at most $\log(N+1)+1$ bits above $-\log\Pr_\theta(s)$ for every $s$ and $\theta$ (Theorem~\ref{thm:B}). The rate is known \cite{Peres92,Pae18}; the proof by counting, its remainder, and the per-string factorisation with pair-dependent biases are new.
\item \emph{Adaptation to structure without transmitted boundaries} (Sections~\ref{sec:structure} and~\ref{sec:memory}). A string with $r$ runs costs $O(r(1+\log(N/r))^2)$ with code G, and the logarithmic square is unavoidable when the fair bits are stored as they are (Theorem~\ref{thm:tworuns}); a concatenation of $J$ pieces costs, without the cuts being sent, the sum of their empirical entropies plus a term sublinear when $J=o(N)$ (Theorem~\ref{thm:concat}; the rate is weaker than that of the sequential estimators for piecewise-stationary sources \cite{Merhav93,ShamirMerhav99,VWBG13}), and so does every residual coder whose cost per residual one is logarithmic (Theorem~\ref{thm:logcoders}). After the Burrows--Wheeler transform of the reversed string, the context intervals of every order are such partitions, so code G applied once to the BWT output competes on every individual sequence with the empirical conditional entropy of every order at once (Theorem~\ref{thm:bwtmarkov}), which gives a pointwise Markov regret and universality for every stationary ergodic source (Theorem~\ref{thm:bwtergodic}), on every finite alphabet. Universality through the BWT and the empirical-entropy benchmarks are those of \cite{EVKV02,KosarajuManzini99,Manzini01,KLV07}; the fixed second stage that attains them, and the argument behind its remainder, are new. The same fair bits are nearly fair: for every law invariant under permutations inside the pieces, the unmixed fair bits are exactly uniform given the rest, and after the BWT of any stationary source the fair bits fall short of their length by $o(N)$ bits of entropy while their expected number is $Nh_P+o(N)$ (Section~\ref{sec:bwtextract}). The BWT-tree code, which applies code B or KT to the leaves of a context tree, is universal as well when its depth grows (Corollary~\ref{cor:treeergodic}), with a smaller proved redundancy for a fixed Markov order, $O(2^k\log N)$ (Proposition~\ref{prop:oracle}) against $O(N^{\beta}\log N)$ for the tree-free code; the experiments confirm the difference at finite length.
\end{enumerate}
Section~\ref{sec:bwtextract} turns the certificate of Section~\ref{sec:structure} into an exact extraction statement for piecewise exchangeable laws and bounds the entropy deficiency of the fair bits of the BWT output for Markov and stationary sources. Section~\ref{sec:exp} measures three things: a gap of three to four orders of magnitude between the worst-case remainders and the observed losses, a strong adaptation to structured strings that no type-class code can see, and finite-length disadvantages against model-based coders on Markov sources, against partition tree weighting on piecewise sources and against PPM compressors on files. Section~\ref{sec:disc} compares with Elias's extractor, says which code to use when, and lists three open problems. Longer proofs, the residual coders A and C used in the experiments, further properties of the residual, the model-based BWT-tree code and two consequences of Section~\ref{sec:bwtextract}, a distribution-preserving replacement of the fair bits and a seeded extraction in total variation, are in the appendices; an extended version with worked examples, the information carried by each part of the codeword, further transforms and implementation details is available from the author. Table~\ref{tab:guide} lists, for each main statement, its input model, its nature and the type of guarantee it gives.

\begin{table}[t]
\caption{Guide to the main statements: input model, nature of the statement, and type of guarantee.}\label{tab:guide}
\centering\footnotesize\setlength{\tabcolsep}{4pt}
\resizebox{\columnwidth}{!}{%
\begin{tabular}{llll}
\toprule
Statement & Input model & Nature & Guarantee\\
\midrule
Residual prefix code (Prop.~\ref{prop:image}) & none & deterministic & every $N$; $F\le\log\binom Nw$\\
Charge, sparsity (Thms.~\ref{thm:charge}, \ref{thm:sparse}) & none & deterministic & every string; $z<\sqrt3\,N^{0.793}$\\
Exact uniformity (Thm.~\ref{thm:uniform}) & exchangeable & probabilistic & exact at every $N$\\
Extraction loss, rate (Prop.~\ref{prop:tail}, Thm.~\ref{thm:rate}, Prop.~\ref{prop:binar}) & exchangeable; i.i.d. & probabilistic & finite $N$, remainder $O(N^{0.793}\log N)$; a.s.\ rate\\
Code G, logarithmic coders (Thms.~\ref{thm:universal}, \ref{thm:logcoders}) & none & pointwise & $\le\log\binom Nw+O(N^{0.793}\log N)$\\
Pricing, factorisation, code B (Prop.~\ref{prop:pricing}, Thms.~\ref{thm:factor}, \ref{thm:B}) & any profile; Bernoulli & identity; pointwise & regret of B $\le\log(N+1)+1$ for every $s$, $\theta$\\
Runs, concatenations (Thms.~\ref{thm:tworuns}, \ref{thm:runs}, \ref{thm:concat}) & none & pointwise & explicit, every string\\
BWT without tree (Thm.~\ref{thm:bwtmarkov}) & none; Markov of order $k$ & pointwise; probabilistic & $\le E_k+\Delta_{N,k}$ for all $k$ at once; regret $\le\Delta_{N,k}$\\
Ergodic universality (Thm.~\ref{thm:bwtergodic}, Prop.~\ref{prop:qary}) & stationary ergodic & probabilistic & a.s.\ and $L^1$ rate, no speed\\
Unmixed extraction (Thm.~\ref{thm:piecewise}) & piecewise exchangeable & probabilistic & exact given residual, mixed slots and values\\
BWT extraction (Thm.~\ref{thm:bwtextract}, Prop.~\ref{prop:extracttail}) & stationary; Markov for the tail & probabilistic & $o(N)$ entropy deficiency; exponential tail\\
Tree code (Prop.~\ref{prop:oracle}, Appendix~\ref{app:tree}) & none; against a Markov law of order $k\le D$ & pointwise & regret $O(2^k\log N)$ for every string\\
\bottomrule
\end{tabular}}
\end{table}

\emph{Notation.} Strings are finite sequences of bits; $|s|$ is the length of $s$, $\wt(s)$ its number of ones (its weight), $s*t$ or $st$ the concatenation, $\oplus$ addition modulo~$2$, $a^k$ the string of $k$ symbols $a$, and $\epsilon$ the empty string. Logarithms are to base~$2$ unless written $\ln$. The Bernoulli source of bias $\theta\in[0,1]$ gives $\Pr_\theta(s)=\theta^{\wt(s)}(1-\theta)^{|s|-\wt(s)}$ to $s$, and $h(\theta)=-\theta\log\theta-(1-\theta)\log(1-\theta)$. A law on $\bits^N$ is \emph{exchangeable} if the probability of a string depends only on its weight. The regret of a code on $s$ with respect to a law $P$ is its length plus $\log P(s)$, its expected redundancy its expected length minus the entropy. The length $N$ is known to the decoder; a self-delimiting description of $N$ adds $O(\log(N+1))$ bits to every code below. The Elias gamma code \cite{Elias75} of an integer $a\ge1$ is $\lfloor\log a\rfloor$ zeros followed by the binary expansion of $a$; its length is $\gamma(a)=2\lfloor\log a\rfloor+1$. The KT probability of a string of length $m$ and weight $w$ is $\KT(m,w)=\Gamma(w+\frac12)\Gamma(m-w+\frac12)/(\pi\,\Gamma(m+1))$, with $\KT(0,0)=1$; its regret is at most $\frac12\log m+1$ for every string \cite{WST95}. Four benchmarks recur for a string $s$ of length $N$ and weight $w$: $\log\binom Nw$, the ideal length of the index of $s$ in its type class; $Nh(w/N)$, the plug-in length, at most $\log(N+1)$ bits larger; $-\log\Pr_\theta(s)$, the ideal length under the true law, at least $Nh(w/N)$; and $-\log\KT(N,w)$, the length of a universal code, at most $Nh(w/N)+\frac12\log N+1$, hence $\log N+O(1)$ above $\log\binom Nw$ (for $w=N/2$, $-\log\KT-\log\binom Nw=\log N+\log(\pi/2)+o(1)$). Header and rounding costs are counted only when passing from an ideal length to an actual code.

\section{A Reversible Transform With Sparse Residuals}\label{sec:transform}

\subsection{One Level}

Let $s\in\bits^m$, $m=2n+\rho$ with $\rho\in\{0,1\}$. With the pairs $(s_{2j-1},s_{2j})$, $1\le j\le n$, define $u\in\bits^n$ by $u_j=s_{2j-1}\oplus s_{2j}$; $y=(s_{2j})_{j:u_j=1}$, of length $k=\wt(u)$; $v=(s_{2j})_{j:u_j=0}$, of length $n-k$; and $r=(s_m)$ if $\rho=1$, $r=\epsilon$ otherwise. Write $\sigma(s)=(u,y,v,r)$. The string $y$ is von Neumann's output.

\begin{lemma}\label{lem:split}
$\sigma$ is a bijection from $\bits^m$ onto $\{(u,y,v,r):u\in\bits^n,\,y\in\bits^{\wt(u)},\,v\in\bits^{n-\wt(u)},\,r\in\bits^\rho\}$, $|u|+|y|+|v|+|r|=m$, and
\begin{equation}\label{eq:weight}
\wt(s)=\wt(u)+2\wt(v)+\wt(r).
\end{equation}
\end{lemma}

\begin{proof}
Given $(u,y,v,r)$, set $s_{2j}$ to the next unread bit of $y$ if $u_j=1$ and of $v$ if $u_j=0$, then $s_{2j-1}=s_{2j}\oplus u_j$, and $s_m=r$ if $\rho=1$. An unequal pair has weight $1$, an equal pair weight $2s_{2j}$.
\end{proof}

Equation~\eqref{eq:weight} does not involve $y$: re-orienting an unequal pair changes neither $u$, $v$, $r$ nor the weight.

\subsection{Iteration to Exhaustion}

\begin{definition}[Exhaustive Peres transform]\label{def:T}
For $s\in\bits^m$ with $m\le1$, $\T(s)=(\epsilon;s)$. For $m\ge2$, with $(u,y,v,r)=\sigma(s)$, $\T(u)=(Y_u;R_u)$ and $\T(v)=(Y_v;R_v)$,
\[
\T(s)=\big(Y_u*y*Y_v;\;R_u*R_v*r\big).
\]
$Y(s)$ is the string of \emph{fair bits} of $s$ and $R(s)$ its \emph{residual}; $F(s)=|Y(s)|$ is the number of fair bits, $z(s)=\wt(R(s))$ the number of residual ones, and $F(s)+|R(s)|=m$.
\end{definition}

\begin{figure}[t]\centering
\resizebox{0.8\columnwidth}{!}{%
\begin{tikzpicture}[
  every node/.style={font=\ttfamily\scriptsize},
  box/.style={draw,rounded corners,align=center,inner sep=2.5pt},
  leaf/.style={draw,rounded corners,align=center,inner sep=2pt,fill=lcol!10},
  lab/.style={font=\rmfamily\scriptsize,fill=white,inner sep=1pt}]
\node[box] (A) at (0,0) {$\epsilon$: 10\,01\,11\,01\,00\,11\\ {\color{ucol}u=110100} {\color{ycol}y=011} {\color{lcol}v=101}\\ $\theta=0.300$};
\node[box] (B) at (-2.3,-1.65) {u: 11\,01\,00\\ {\color{ucol}u=010} {\color{ycol}y=1} {\color{lcol}v=10}\\ $0.420$};
\node[box] (C) at (2.5,-1.65) {v: 10\,|\,{\color{lcol}1}\\ {\color{ucol}u=1} {\color{ycol}y=0} {\color{lcol}r=1}\\ $0.155$};
\node[box] (D) at (-3.5,-3.3) {uu: 01\,|\,{\color{lcol}0}\\ {\color{ucol}u=1} {\color{ycol}y=1} {\color{lcol}r=0}\\ $0.487$};
\node[box] (E) at (-1.1,-3.3) {uv: 10\\ {\color{ucol}u=1} {\color{ycol}y=0}\\ $0.344$};
\node[leaf] (Cu) at (2.5,-3.3) {vu: {\color{lcol}1}\\ $0.262$};
\node[leaf] (Du) at (-3.5,-4.75) {uuu: {\color{lcol}1}\\ $0.500$};
\node[leaf] (Eu) at (-1.1,-4.75) {uvu: {\color{lcol}1}\\ $0.451$};
\draw (A) -- node[lab]{u} (B);
\draw (A) -- node[lab]{v} (C);
\draw (B) -- node[lab]{u} (D);
\draw (B) -- node[lab]{v} (E);
\draw (C) -- node[lab]{u} (Cu);
\draw (D) -- node[lab]{u} (Du);
\draw (E) -- node[lab]{u} (Eu);
\end{tikzpicture}}
\caption{Exhaustive transform of $s=100111010011$. Each node shows its path, its string in pairs (a bar precedes the odd remainder), $u$, the fair bits $y$, the string $v$ passed to the $\vstep $-child and the remainder $r$; empty $\vstep $-children are omitted. The last line is the bias $\phi_\pi(0.3)$ of Section~\ref{sec:B}. The residual bits, in decoding order, are $1,0,1,1,1$ at paths $\ustep \ustep \ustep ,\ustep \ustep ,\ustep \vstep \ustep ,\vstep \ustep ,\vstep $; the fair bits are $Y=1100110$.}\label{fig:tree}
\end{figure}
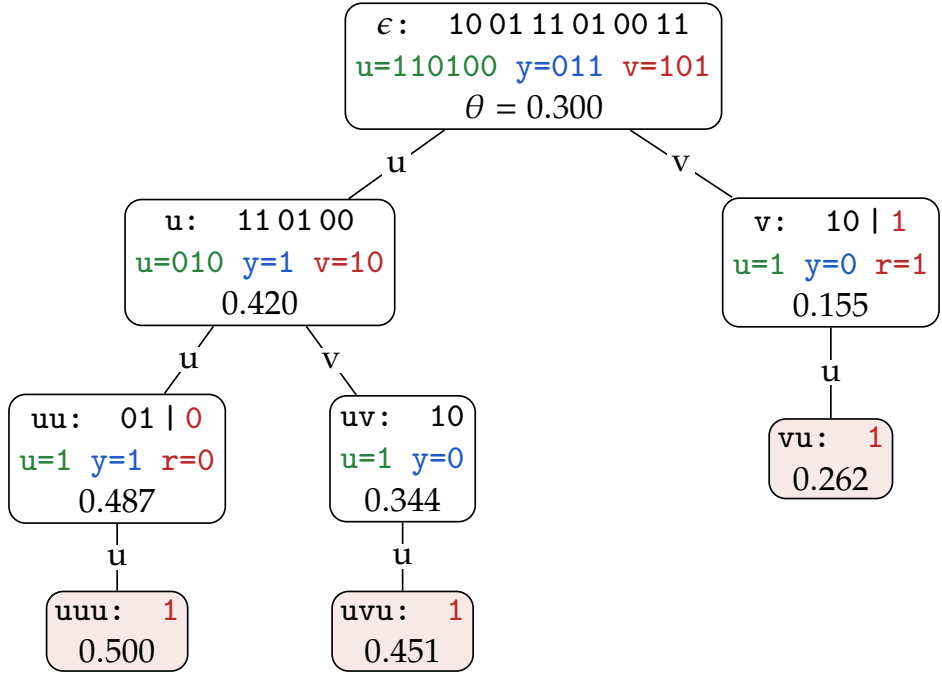

The calls of $\T$ form a binary tree: a node of length at least $2$ has a $\ustep $-child and a $\vstep $-child, nodes of length at most $1$ are leaves, and an odd remainder is attached to the node that produced it. A node is named by its \emph{path} $\pi\in\{\ustep ,\vstep \}^*$ from the root, and $v(\pi)$ denotes the number of letters $\vstep $ in $\pi$. Every node emits at most one residual bit, its odd remainder or, at a leaf of length one, its single bit, and this bit inherits the path of the node. A child has at most half the length of its parent, so a nonempty node has depth at most $\lfloor\log N\rfloor$. Figure~\ref{fig:tree} shows the example $s=100111010011$ of \cite{MIT6441}, for which $Y=1100110$ and $R=10111$.

\begin{remark}[The name of $Y$]\label{rem:name}
We call $Y$ the fair bits in every context, because this is their role: under an exchangeable law they are exactly uniform given the residual (Theorem~\ref{thm:uniform}). For an individual string, or under a law that is not exchangeable, the name refers to that role and not to a property: the fair bits of $(01)^m$ are $1^m$ (Section~\ref{sec:limits}), and there are stationary sources of entropy rate zero whose fair bits have a positive rate (Section~\ref{sec:notsay}).
\end{remark}

\subsection{Decoding and the Residual Prefix Code}

For each $m\ge0$, define $\Dec_m(Y;R)=(s;Y';R')$, which reads a prefix of each of the two streams $Y,R$ and returns $s\in\bits^m$ with the unread parts $Y',R'$. $\Dec_0$ reads nothing and returns $\epsilon$; $\Dec_1$ returns the first bit of $R$. For $m\ge2$, with $n=\lfloor m/2\rfloor$: $(u;Y_1;R_1)=\Dec_n(Y;R)$; with $k=\wt(u)$, $y$ is the first $k$ bits of $Y_1$ and $Y_2$ the rest; $(v;Y_3;R_2)=\Dec_{n-k}(Y_2;R_1)$; $r$ is the first $m-2n$ bits of $R_2$ and $R_3$ the rest; the output is $(\sigma^{-1}(u,y,v,r);Y_3;R_3)$.

\begin{proposition}\label{prop:bij}
For every $s\in\bits^N$, $\Dec_N(\T(s))=(s;\epsilon;\epsilon)$. Hence $\T$ is injective and is inverted by a decoder that reads both streams from left to right without knowing their lengths.
\end{proposition}

\begin{proof}
By induction on $m$, $\Dec_m(Y(s)*Y'';R(s)*R'')=(s;Y'';R'')$ for all $s\in\bits^m$ and all strings $Y'',R''$. The length of each piece is known when it is needed: $|u|=n$, and $|y|=\wt(u)$, $|v|=n-\wt(u)$ once $u$ is rebuilt; the encoder emits $Y_u,y,Y_v$ and $R_u,R_v,r$ in the order in which the decoder requests them.
\end{proof}

In Figure~\ref{fig:tree}, $\Dec_{12}$ calls $\Dec_6$ for the $u$-string of the root, which calls $\Dec_3$ for node $\ustep \ustep $, which calls $\Dec_1$: the first residual bit, $1$, is the $u$-string of $\ustep \ustep $. Then $k=1$, the first fair bit is $y$, the $v$-string is empty and the second residual bit is the odd remainder, so node $\ustep \ustep $ is $010$. Back at node $\ustep $, $k=1$, the next fair bit is its $y$, and $\Dec_2$ rebuilds node $\ustep \vstep =10$ from the third residual bit and the next fair bit. Back at the root, $k=3$, the next three fair bits are $y=011$, $\Dec_3$ rebuilds node $\vstep =101$ from the last two residual bits and the last fair bit, and both streams are exhausted exactly when the root returns.

The only decisions of the decoder are the weights $\wt(u)$ of the nodes, which by \eqref{eq:weight} are functions of residual bits already read. The \emph{residual parser} is the decoder restricted to the residual stream: it reads residual bits one at a time, computes from them the lengths and weights of all nodes, and stops exactly at the end of a residual, without reading any fair bit.

\begin{proposition}[The residual prefix code]\label{prop:image}
Let $\Rc_0=\{\epsilon\}$, $\Rc_1=\bits$, $w_m(R)=\wt(R)$ for $m\le1$, and for $m\ge2$, $n=\lfloor m/2\rfloor$, $k=w_n(R_u)$:
\begin{gather*}
\Rc_m=\{R_u*R_v*r:\ R_u\in\Rc_n,\ R_v\in\Rc_{n-k},\ r\in\bits^{m-2n}\},\\
w_m(R_u*R_v*r)=k+2\,w_{n-k}(R_v)+\wt(r).
\end{gather*}
Then $\Rc_m$ is a complete prefix-free set, $\sum_{R\in\Rc_m}2^{-|R|}=1$, and $\T$ is a bijection from $\bits^N$ onto $\{(Y;R):R\in\Rc_N,\ Y\in\bits^{N-|R|}\}$, with $\wt(s)=w_N(R(s))$. In particular, for each $R\in\Rc_N$, the \emph{fibre} $\{s:R(s)=R\}$ has exactly $2^{N-|R|}$ elements, all of weight $w_N(R)$, and
\begin{equation}\label{eq:typebound}
F(s)\le\log\binom N{\wt(s)}\le Nh(\wt(s)/N)\qquad\text{for every }s\in\bits^N.
\end{equation}
\end{proposition}

\begin{proof}
Induction on $m$. Since $\Rc_n$ and $\Rc_{n-k}$ are prefix-free, the decomposition $R_u*R_v*r$ is unique and $w_m$ is well defined. For $s\in\bits^m$, $R_u\in\Rc_n$ with $\wt(u)=w_n(R_u)$, so $|v|=n-k$, $R_v\in\Rc_{n-k}$, $R(s)\in\Rc_m$ and $\wt(s)=w_m(R(s))$ by \eqref{eq:weight}. Conversely, for $R\in\Rc_m$ and $Y\in\bits^{m-|R|}$, $\Dec_m$ consumes exactly $R_u$, $k$ fair bits, $R_v$, $r$, and the fair stream has exactly the $m-|R|$ bits it requests, so $(Y;R)$ is the transform of the output. If $R_u*R_v*r$ and $R'_u*R'_v*r'$ in $\Rc_m$ are comparable for the prefix order, then $R_u$ and $R'_u$ are comparable, hence equal by induction; then $k$ is the same, $R_v=R'_v$ likewise, and $r,r'$ have the same length: $\Rc_m$ is prefix-free. With Proposition~\ref{prop:bij} the map is a bijection, so $\sum_{R\in\Rc_N}2^{N-|R|}=2^N$, which is the completeness. The fibre of $R$ is the image of $\bits^{N-|R|}$ under $Y\mapsto\Dec_N(Y;R)$, so it has $2^{N-|R|}$ elements, all of weight $w_N(R)$; since they are among the $\binom Nw$ strings of weight $w=w_N(R)$, $F(s)\le\log\binom Nw$, and $\binom Nw2^{-Nh(w/N)}\le\sum_j\binom Nj(w/N)^j(1-w/N)^{N-j}=1$.
\end{proof}

The image of $\T$ is not a product set: the number of fair bits is dictated by the residual. Restricted to one weight, the Kraft equality reads $\sum_{R\in\Rc_N:w_N(R)=w}2^{N-|R|}=\binom Nw$, so the type class is split into blocks of $2^{F}$ strings sharing a residual. For $N=4$, $\Rc_4=\{0000,0001,001,01,10,11\}$, and $|\Rc_N|=21,100,202$ for $N=8,12,16$. Since $\Rc_N$ is prefix-free, the single string $R*Y$ of length $N$ determines $(Y;R)$, and $N$ is read off its length. Pae's structure lemma \cite[Lemma~4]{Pae18} maps each type class over the pair alphabet $\{00,01,10,11\}$ bijectively onto a product of type classes, one per component of the recursion, here $u$, $y$ and $v$; a type class of the binary weight alone is a disjoint union of such products, $\binom{2n}w=\sum_k\binom nk2^k\binom{n-k}{(w-k)/2}$. What distinguishes Proposition~\ref{prop:image} is the information retained: one residual code of all of $\bits^N$ at once, across the types, with its parser and its recursively computable weight $w_N$, so that the conditioning of Theorem~\ref{thm:uniform} below is on the complete residual, where Peres's Proposition~1 \cite{Peres92} conditions on the length of the output; and the charge of a residual one with its sparsity consequence, next, has no counterpart in either statement.

\subsection{The Charge of a Residual One}

\begin{theorem}[Charge identity]\label{thm:charge}
For every $s\in\bits^N$, with $\pi_i$ the path of the residual bit $R_i$,
\begin{equation}\label{eq:charge}
\wt(s)=\sum_{i:R_i=1}2^{v(\pi_i)}.
\end{equation}
In particular $z(s)\le\wt(s)$.
\end{theorem}

\begin{proof}
Let $C(s)$ be the right-hand side. For $|s|\le1$ the only possible residual bit is $s$ itself, at the empty path, so $C(s)=\wt(s)$. For $|s|\ge2$, the residual bits of the $\ustep $-child receive the prefix $\ustep $, which leaves their number of $\vstep $ steps unchanged, those of the $\vstep $-child receive the prefix $\vstep $, which doubles their charge, and the odd remainder has the empty path; hence $C(s)=C(u)+2C(v)+\wt(r)$, and \eqref{eq:weight} with induction gives $C(s)=\wt(s)$.
\end{proof}

The factor two records the duplication of an equal pair; it is not a likelihood or an expectation. On the string of Figure~\ref{fig:tree}, the residual ones at the paths $\ustep\ustep\ustep$, $\ustep\vstep\ustep$, $\vstep\ustep$ and $\vstep$ have charges $1,2,2,2$, and $1+2+2+2=7=\wt(s)$. The fair bits have a charge identity of their own, and the complement of a string flips only the residual bits of the all-$\vstep$ branch; these further properties, which the main line does not use, are in Appendix~\ref{app:residual}.

\subsection{Every Residual Is Sparse}

\begin{theorem}[Sparsity]\label{thm:sparse}
Let $\beta=\frac12\log3=0.79248\ldots$. For every $s\in\bits^N$ of weight $w\ge1$,
\[
z(s)^2<3w\,(3/2)^{\lfloor\log N\rfloor}\le3N^{2\beta},
\]
so $z(s)<\sqrt3\,N^\beta$, and $z(s)=0$ if $w=0$.
\end{theorem}

\begin{proof}
By Cauchy--Schwarz and \eqref{eq:charge},
\[
z^2=\Big(\sum_{i:R_i=1}2^{v(\pi_i)/2}\,2^{-v(\pi_i)/2}\Big)^2\le\Big(\sum_{i:R_i=1}2^{v(\pi_i)}\Big)\Big(\sum_{i:R_i=1}2^{-v(\pi_i)}\Big)=w\sum_{i:R_i=1}2^{-v(\pi_i)}.
\]
Distinct residual bits have distinct paths of depth at most $D=\lfloor\log N\rfloor$, so the last sum is at most the sum over all paths of depth at most $D$; at depth $d$, $\binom dj$ paths have $j$ letters $\vstep $, and $\sum_j\binom dj2^{-j}=(3/2)^d$. Hence $z^2\le w\sum_{d=0}^D(3/2)^d<3w(3/2)^D$, and $(3/2)^D\le N^{\log(3/2)}$, $w\le N$, $1+\log(3/2)=2\beta$.
\end{proof}

A residual can be as long as the input, $R(0^N)=0^N$, but its number of ones is always $o(N)$; if its length is comparable to $N$, its own density of ones therefore tends to zero (it need not be small for a short residual: for every $N$ the parser fed with ones stops on some $1^{m}\in\Rc_N$, with $m=105$ for $N=1024$). H\"older's inequality in place of Cauchy--Schwarz improves the exponent: for every $\lambda>0$, $z(s)\le(1+2^\lambda)^{1/(1+\lambda)}N^{\beta(\lambda)}$ with $\beta(\lambda)=\log(1+2^\lambda)/(1+\lambda)$, whose minimum is $\beta_*=1-\alpha=0.77290\ldots$, where $\alpha\in(0,\frac12)$ solves $h(\alpha)=1-\alpha$ (Appendix~\ref{app:holder}). We state the results below with $\beta$ and explicit constants; each holds with $\beta_*$ and other absolute constants. The exponent $\beta_*$ is probably not sharp (Section~\ref{sec:disc}). It enters every statement below only through remainders of the form $CN^\beta\log N$: every qualitative conclusion, the almost-sure rate, the pointwise universality of code G, the concatenation and BWT bounds and the stationary ergodic universality, needs only $\beta<1$, and a better exponent for $z$ would improve, by the same power of $N$, the remainders that come from the gap description and the extraction loss. The concatenation and BWT bounds involve a second estimate, that of the mixed fair bits (Lemma~\ref{lem:mixed}), proved by a separate argument on the charged weights of whole levels; it has the same exponent $\beta$ and the same H\"older refinement, but an improvement of the sparsity bound does not transfer to it automatically.

\subsection{Describing a Residual by Its Gaps}\label{sec:gapdesc}

\begin{definition}[Gap description]\label{def:gap}
The gap description of a residual $R$ writes, for each residual one, $\gamma$-coded, the number $g+1$, where $g$ is the number of zeros since the previous one or the start, and, if $R$ ends with a nonempty run of $t$ zeros, the number $t+1$, $\gamma$-coded. Its length is denoted $L_\gap(R)$.
\end{definition}

\begin{lemma}\label{lem:gap}
The gap descriptions of the strings of $\Rc_N$ form a prefix-free set, decoded by the residual parser alone. If $R\in\Rc_N$ has length $m$ and $z$ ones,
\begin{equation}\label{eq:gaplength}
L_\gap(R)\le(z+1)\Big[1+2\log\Big(1+\frac m{z+1}\Big)\Big]\le(z+1)\big[1+2\log(N+1)\big].
\end{equation}
\end{lemma}

\begin{proof}
Gamma words are self-delimiting. On decoding a word $g+1$, prepare $g$ zeros followed by an implicit one, and feed these bits one at a time to the residual parser. If the parser stops at the last of the zeros, the word described the final run and the implicit one is dropped. The parser cannot stop inside a run of zeros of a valid description, since a proper prefix of a residual would then be a residual. So the decoder stops exactly at the end of the description, and the descriptions are prefix-free. For the length, let $a_1,\dots,a_q$ be the arguments of the gamma words: if $R$ ends with a one, $q=z$ and $\sum_ja_j=m$; otherwise $q=z+1$ and $\sum_ja_j=m+1$. Adding in the first case a dummy argument equal to one, we have $z+1$ arguments summing to $m+1$, and concavity of $\log$ gives $L_\gap(R)\le\sum_j(1+2\log a_j)\le(z+1)[1+2\log((m+1)/(z+1))]$.
\end{proof}

The function $a\mapsto a[1+2\log(1+m/a)]$ is increasing for $a>0$, since the derivative of $a\ln(1+m/a)$ is $\ln(1+t)-t/(1+t)\ge0$ with $t=m/a$. With Theorem~\ref{thm:sparse} and $1+2\log(N+1)\le3\log(N+1)$ for $N\ge1$, Lemma~\ref{lem:gap} gives
\begin{equation}\label{eq:bN}
b_N:=\max_{R\in\Rc_N}L_\gap(R)\le\bar b_N:=3(\sqrt3+1)\,N^\beta\log(N+1)=o(N),
\end{equation}
and, since the descriptions are prefix-free, Kraft's inequality gives $|\Rc_N|\,2^{-b_N}\le\sum_R2^{-L_\gap(R)}\le1$:
\begin{equation}\label{eq:countR}
\log|\Rc_N|\le b_N.
\end{equation}
For $N=1024$, $\log|\Rc_N|$ is about $0.19N$, so \eqref{eq:countR} is an asymptotic statement. The residual parser tells the decoder where the gap description ends, and the parsed residual determines $F=N-|R|$, so the next $F$ bits can be copied as $Y$ and the transform inverted: appending the fair bits to the gap description of the residual is a prefix code for $s$, of length $L_\gap(R(s))+F(s)$. This is code G (Section~\ref{sec:G}); the same residual description that bounds the extraction loss in Section~\ref{sec:extraction} gives its pointwise coding guarantee.

\subsection{Cost}

One level of $\sigma$ on a string of length $m$ is an XOR of two half-length vectors and two compactions, $O(m)$ bit operations. The nodes at a given depth have total length at most $N$ and the depth is at most $\lfloor\log N\rfloor$, so $\T$ and its inverse cost $O(N\log N)$ bit operations and $O(N)$ memory, with no multiplication, and each level is data-parallel. In the file compressor of Section~\ref{sec:exp}, which adds a suffix sort, a context tree and a coder, the whole pipeline compresses text at $3$ to $9$\,MB/s and decompresses it at $3$ to $9$\,MB/s according to the mode, on one core (Table~\ref{tab:speed}), with a peak memory of about $20$ bytes per input byte on text and up to $140$ on incompressible input, where the dynamic programme that chooses the tree explores every node; these are pipeline measurements and do not isolate the cost of the transform.

\section{Exact Extraction and Code G}\label{sec:extraction}

\subsection{Exact Uniformity}

\begin{theorem}\label{thm:uniform}
Let $X\in\bits^N$ have an exchangeable law $P$, and let $P_R$ be the law of $R(X)$. Then $F(X)$ is a function of $R(X)$, and conditionally on $R(X)$, $Y(X)$ is uniform on $\bits^{F(X)}$. Moreover, for every $s$ with $P(s)>0$,
\begin{equation}\label{eq:balance}
-\log P(s)=F(s)-\log P_R(R(s)),
\end{equation}
and $H(X)=H(R)+\E F(X)$.
\end{theorem}

\begin{proof}
By Proposition~\ref{prop:image}, the fibre of $R$ consists of $2^{N-|R|}$ strings of the same weight, which have the same probability under $P$; hence $P_R(R(s))=2^{F(s)}P(s)$, which is \eqref{eq:balance}, and given $R$ the fair string is uniform. Taking expectations in \eqref{eq:balance} gives the entropy identity.
\end{proof}

Equation~\eqref{eq:balance} is an identity for each string, not only between entropies: the self-information of the input is the number of its fair bits plus the self-information of its residual.

\subsection{The Entropic Rate}

\begin{proposition}[Tail of the extraction deficit]\label{prop:tail}
For every exchangeable law $P$ on $\bits^N$ and every $t\ge0$,
\[
P\big\{-\log P(X)-F(X)>\log|\Rc_N|+t\big\}\le2^{-t},
\]
and $-\log P(X)-F(X)\ge0$ almost surely.
\end{proposition}

\begin{proof}
By \eqref{eq:balance}, $-\log P(X)-F(X)=-\log P_R(R(X))\ge0$, and the event of the statement is the set of residuals of probability less than $2^{-t}/|\Rc_N|$; there are at most $|\Rc_N|$ of them.
\end{proof}

\begin{theorem}[Entropic extraction]\label{thm:rate}
For every exchangeable law on $\bits^N$,
\begin{equation}\label{eq:loss}
0\le H(X)-\E F(X)=H(R(X))\le\log|\Rc_N|\le b_N=O(N^\beta\log N).
\end{equation}
If $X_1,X_2,\dots$ are i.i.d.\ Bernoulli$(\theta)$ and $F_N=F(X_1\cdots X_N)$, then $0\le Nh(\theta)-\E F_N\le b_N$ uniformly in $\theta\in[0,1]$, and $F_N/N\to h(\theta)$ almost surely and in $L^1$.
\end{theorem}

\begin{proof}
The identity in \eqref{eq:loss} is Theorem~\ref{thm:uniform}, and $H(R)\le\log|\Rc_N|$. Equivalently, the prefix code formed by the gap description of the residual followed by the fair bits has expected length at least $H(X)$, so $H(X)\le\E L_\gap(R(X))+\E F(X)\le b_N+\E F(X)$; this argument uses nothing but the deterministic bound \eqref{eq:bN}. For the almost-sure statement with $0<\theta<1$, apply Proposition~\ref{prop:tail} with $t=2\log N$: the probabilities $N^{-2}$ are summable, so by Borel--Cantelli, almost surely, for all large $N$, $0\le-\log\Pr_\theta(X_1\cdots X_N)-F_N\le\log|\Rc_N|+2\log N=o(N)$; by the strong law of large numbers, $-\frac1N\log\Pr_\theta(X_1\cdots X_N)=-\frac{W_N}N\log\theta-(1-\frac{W_N}N)\log(1-\theta)\to h(\theta)$, with $W_N$ the weight. Convergence in $L^1$ follows from $0\le F_N/N\le1$. For $\theta\in\{0,1\}$ the input is constant and $F_N=0$.
\end{proof}

The proof needs no study of the transformed biases and no passage from a fixed recursion depth to infinite depth. For a finite length there is also a purely combinatorial statement: in the class of the strings of weight $w$, the proportion of strings with $F<\log\binom Nw-\log|\Rc_N|-t$ is less than $2^{-t}$, because each of the at most $|\Rc_N|$ fibres meeting the class contains fewer than $\binom Nw2^{-t}/|\Rc_N|$ such strings.

The same tail bound says how the biases of the residual bits polarise, once these biases are defined (Corollary~\ref{cor:polar} in Section~\ref{sec:B}).

\subsection{What Extraction Does Not Say}\label{sec:notsay}

The rate is a statement about laws, not about every string: for $N$ a power of two, $0^{N/2}1^{N/2}$ has the empirical entropy $N$ and a single fair bit. Exchangeability, or at least uniformity on the fibres, is essential: for a general law, the bijection only gives $H(X)=H(R)+H(Y\mid R)$ with $H(Y\mid R)\le\E F$. Choose uniformly one of the four phases of the periodic sequence $00110011\cdots$: this stationary ergodic source has entropy rate zero, but on the phase that starts with $0011$ the prefix $(0011)^m$ has $m$ fair bits (Section~\ref{sec:limits}). In particular, after a Burrows--Wheeler transform the fair bits are not exactly fair; Section~\ref{sec:bwtextract} measures the defect, in entropy, and shows that it is $o(N)$ for every stationary source.

\subsection{Finite Alphabets by Binarisation}\label{sec:binar}

Let $\A$ be a finite alphabet of size $q\ge2$, and fix a binary tree with $q$ leaves labelled by the symbols. For $x\in\A^N$ and an internal vertex $\nu$, the \emph{binarised sequence} $x^{(\nu)}$ lists, in order, the branch (left $0$, right $1$) taken at $\nu$ by the symbols of $x$ whose leaf lies below $\nu$. Transform each of the $q-1$ binarised sequences with $\T$, the fair bits being the concatenation of their fair bits; this is the complete binarisation of \cite{ZhouBruck12} and the binarisation tree of \cite{Pae16}.

\begin{proposition}\label{prop:binar}
The residuals, read in an order where each vertex comes after its parent, determine the lengths and weights of all binarised sequences, hence the type of $x$; for every exchangeable law on $\A^N$ (probabilities depending only on the type) the fair bits are uniform given the residuals; and if $X_1,X_2,\dots$ are i.i.d.\ of law $p$ on $\A$, then $0\le NH(p)-\E F\le(q-1)\bar b_N$, with $\bar b_N$ as in \eqref{eq:bN}, and $F/N\to H(p)$ almost surely.
\end{proposition}

\begin{proof}
The root sequence has length $N$, and the weight of the sequence of a vertex, read from its residual, gives the lengths of the sequences of its children; the leaf counts follow. Given all residuals, every choice of the fair bits is valid and gives a string of the same type, so the fibres are products of cubes inside a type class, and the argument of Theorem~\ref{thm:uniform} applies. The concatenation of the gap descriptions of the residuals, in that order, is decodable, so it is a prefix-free description of the tuple of residuals, of length at most $\sum_\nu\bar b_{|x^{(\nu)}|}\le(q-1)\bar b_N$, the bound $\bar b_N$ of \eqref{eq:bN} being increasing in $N$. The tuple therefore takes at most $2^{(q-1)\bar b_N}$ values, and the proofs of Proposition~\ref{prop:tail} and Theorem~\ref{thm:rate} apply with $\log|\Rc_N|$ replaced by $(q-1)\bar b_N$; the strong law applies to $-\frac1N\log P(X_1\cdots X_N)$.
\end{proof}

The binarisation can also be replaced by a transform acting directly on the alphabet $\bits^b$, recursing on the XOR of each pair \cite{Pae15,Pae18}; we do not use it.

\subsection{Code G: a Universal Code Without Parameters}\label{sec:G}

\begin{definition}[Code G]\label{def:G}
For $s\in\bits^N$, $G_N(s)$ is the gap description of $R(s)$ (Definition~\ref{def:gap}) followed by the fair bits $Y(s)$, as they are. Its length is $L_G(s)=L_\gap(R(s))+F(s)$.
\end{definition}

The decoder of Lemma~\ref{lem:gap} parses the residual from the gap words, which gives $F=N-|R|$, reads $F$ fair bits and inverts the transform. Code G sends no weight, computes no bias, fits no parameter and uses no arithmetic coder; it counts runs of zeros and writes gamma codes (a gamma code restricted by the budget of Appendix~\ref{app:residual} is slightly shorter, which we do not use in the statements). For the string of Figure~\ref{fig:tree}, $R=10111$ has the gap arguments $1,2,1,1$, so the gap description is $1\,010\,1\,1$ and $G_{12}(s)=101011\,1100110$, of length $13$: code G does not shorten a short unstructured string, and $0^{2^e}1^{2^e}$ below shows what it does on a structured one.

\begin{theorem}[Pointwise universality]\label{thm:universal}
$G_N$ is a prefix code, and for every $s\in\bits^N$ of weight $w$, with $m=|R(s)|$ and $z=z(s)$,
\[
L_G(s)\le F(s)+(z+1)\Big[1+2\log\Big(1+\frac m{z+1}\Big)\Big]\le\log\binom Nw+3(\sqrt3+1)N^\beta\log(N+1).
\]
Consequently, for every $\theta\in[0,1]$, $0\le\E_\theta L_G(X)-Nh(\theta)\le3(\sqrt3+1)N^\beta\log(N+1)$.
\end{theorem}

\begin{proof}
Lemma~\ref{lem:gap}, \eqref{eq:bN} and \eqref{eq:typebound}. For the expectation, the lower bound is the entropy bound for a prefix code, and $\E\log\binom Nw\le N\E h(W/N)\le Nh(\theta)$ by concavity.
\end{proof}

\emph{How large the remainder is.} The bound is uniform over all strings but quantitatively loose. At $N=2^{20}$ its residual term $\bar b_N=3(\sqrt3+1)N^\beta\log(N+1)$ is $9.7\cdot10^6$ bits, more than $N=1.05\cdot10^6$, whereas the observed mean regrets of code G against the type class are $1.5\cdot10^3$ to $3.2\cdot10^3$ bits (Section~\ref{sec:expbern}); $\bar b_N<N$ only from $N=2^{41}$ on ($2^{37}$ with $\beta_*$), and the remainders of Theorems~\ref{thm:concat} and~\ref{thm:bwtmarkov} and of Proposition~\ref{prop:qary}, which contain $\bar b_N$ and grow with the number of pieces, the order and the alphabet, are larger still. The theorem therefore establishes the form of the bound, with constant one in front of $\log\binom Nw$, and asymptotic universality; it is not a useful worst-case redundancy at the lengths of the experiments, and Section~\ref{sec:expbern} says where the slack is.

What code G pays on a memoryless source is explicit at $\theta=\frac12$. Under the fair law, $P(R=r)=2^{-|r|}$ for every $r\in\Rc_N$, by Proposition~\ref{prop:image}: the residual is a fair stream stopped by its own parser, every requested residual bit is fair given the past, $\E z=\frac12\E|R|$, and storing $R$ raw before $Y$ is a prefix code of exactly $N$ bits, optimal in expectation. A residual can thus be sparse relative to $N$ while having density $\frac12$ within itself. For the complete geometric gaps of an infinite fair stream, gamma codes cost $1+2\sum_{j\ge1}2^{-(2^j-1)}=2.266$ bits per gap of mean length $2$, that is $1.13$ bits per residual bit; the stopped residual may end with an incomplete gap, so this is a reference value, and it explains the $13\%$ residual expansion of Table~\ref{tab:bernoulli} and the $0.5\%$ loss on random bytes in Section~\ref{sec:exp}.

\section{The Likelihood and Code B}\label{sec:B}

\subsection{Pricing a Pair}\label{sec:pricing}

For $a,b\in(0,1)$, put
\begin{equation}\label{eq:pairmaps}
U_{a,b}=a+b-2ab,\qquad q_{a,b}=\frac{(1-a)b}{U_{a,b}},\qquad V_{a,b}=\frac{ab}{1-U_{a,b}}.
\end{equation}
All three lie in $(0,1)$, and the four pairs of bits satisfy
\begin{equation}\label{eq:pairid}
\begin{aligned}
(1-a)b&=U_{a,b}\,q_{a,b},&a(1-b)&=U_{a,b}(1-q_{a,b}),\\
ab&=(1-U_{a,b})V_{a,b},&(1-a)(1-b)&=(1-U_{a,b})(1-V_{a,b}).
\end{aligned}
\end{equation}
In words: price the flag $u_j$, which is $1$ for an unequal pair, at $U_{a,b}$; if the pair is unequal, price its second bit, which is its fair bit, at $q_{a,b}$; if it is equal, price its common value at $V_{a,b}$. A \emph{profile} for $s\in\bits^m$ is a vector $p\in(0,1)^m$, and $P_p(s)=\prod_ip_i^{s_i}(1-p_i)^{1-s_i}$. For $\sigma(s)=(u,y,v,r)$, the profile of $u$ is $(U_{p_{2j-1},p_{2j}})_{j\le n}$, the profile of $v$ is $(V_{p_{2j-1},p_{2j}})_{j:u_j=0}$, and the fair bit of an unequal pair $j$ has the parameter $q_{p_{2j-1},p_{2j}}$; by \eqref{eq:pairid},
\[
P_p(s)=P_{p^u}(u)\prod_{j:u_j=1}q_j^{s_{2j}}(1-q_j)^{1-s_{2j}}\;P_{p^v}(v)\;P_{p_m}(r),
\]
where $p^u,p^v$ are these profiles, $r$ is priced at the parameter of the last position, and $P_{p_m}(\epsilon)=1$.

\begin{proposition}[Pricing along the recursion]\label{prop:pricing}
For every $s\in\bits^N$ and every profile $p\in(0,1)^N$, iterating the last identity at every node writes $P_p(s)$ as a product of factors in $(0,1)$, one per fair bit and one per residual bit: the fair bit $Y_j$ contributes $q_j^{Y_j}(1-q_j)^{1-Y_j}$ and the residual bit $R_i$ contributes $p'^{R_i}_i(1-p'_i)^{1-R_i}$, where $q_j$ and $p'_i$ are the parameters computed along the recursion. In decoding order, each parameter is a function of $p$ and of the bits already decoded.
\end{proposition}

\begin{proof}
Induction on the length, applying the one-level identity to $s$, then to $u$ with $p^u$ and to $v$ with $p^v$. The profile $p^u$ depends only on $p$; the profile $p^v$ depends on $u$, which the decoder rebuilds before it requests any bit of the $\vstep $-subtree, and the parameter of the remainder is $p_m$.
\end{proof}

The identity holds for every profile and every string; no probability is involved, and parameters chosen from the data cause no difficulty. Section~\ref{sec:concat} uses it with a profile that is constant on the pieces of a concatenation.

\subsection{Bernoulli Sources}

Define the bias maps
\begin{equation}\label{eq:maps}
\U(\theta)=2\theta(1-\theta),\qquad\V(\theta)=\frac{\theta^2}{\theta^2+(1-\theta)^2},
\end{equation}
so that $U_{t,t}=\U(t)$, $V_{t,t}=\V(t)$ and $q_{t,t}=\frac12$. For a path $\pi$, $\phi_\pi(\theta)$ composes the maps along $\pi$, first letter first: $\phi_\epsilon(\theta)=\theta$, $\phi_{\pi\ustep }=\U\circ\phi_\pi$, $\phi_{\pi\vstep }=\V\circ\phi_\pi$. In the natural log-odds $\lambda=\ln\frac\theta{1-\theta}$, $\V:\lambda\mapsto2\lambda$ and $\U:\lambda\mapsto-\ln\cosh\lambda$: a $\vstep $ step doubles the log-odds, a $\ustep $ step brings them towards $0$, quadratically when they are small and by about $\ln2$ per step when they are large.

\begin{theorem}[Causal factorisation]\label{thm:factor}
For every $\theta\in(0,1)$ and $s\in\bits^N$, with $\pi_i$ the path of the residual bit $R_i$,
\begin{equation}\label{eq:factor}
\Pr_\theta(s)=2^{-F(s)}\prod_{i=1}^{|R(s)|}\Pr_{\phi_{\pi_i}(\theta)}(R_i).
\end{equation}
Consequently, if $X\sim\Pr_\theta$, then conditionally on $R_1,\dots,R_{i-1}$, which determine whether an $i$-th residual bit is requested and its path $\pi_i$, $R_i$ is Bernoulli$(\phi_{\pi_i}(\theta))$; and conditionally on $R$, $Y$ is uniform.
\end{theorem}

\begin{proof}
With the constant profile $\theta$, every node at path $\pi$ has the constant profile $\phi_\pi(\theta)$, by induction, and every fair bit has the parameter $q_{t,t}=\frac12$; Proposition~\ref{prop:pricing} gives \eqref{eq:factor}. For the second statement, the decoder fed with fair bits and with residual bits drawn, at the moment of their request, from Bernoulli$(\phi_\pi(\theta))$ produces each $(Y;R)$ in the image of $\T$ with probability $2^{-F}\prod_i\Pr_{\phi_{\pi_i}(\theta)}(R_i)$, which by \eqref{eq:factor} is $\Pr_\theta$ of its output.
\end{proof}

The causal form is essential. Conditioning on the whole shape of the recursion, which depends on residual bits not yet requested, changes the biases: for $N=2$, an empty $\vstep $-child means that the pair is unequal, and the residual bit $u$ is then $1$ with probability one, not $\U(\theta)$. In Figure~\ref{fig:tree}, \eqref{eq:factor} reads $\theta^7(1-\theta)^5=2^{-7}\phi_{\ustep \ustep \ustep }(1-\phi_{\ustep \ustep })\phi_{\ustep \vstep \ustep }\phi_{\vstep \ustep }\phi_{\vstep }$, and both sides equal $3.68\cdot10^{-5}$ at $\theta=0.3$. Taking expectations at one level gives Peres's identity $h(\theta)=\theta(1-\theta)+\frac12h(\U\theta)+\frac12(1-\U\theta)h(\V\theta)$; globally, the chain rule along the residual bits in request order gives $H(R)=\E\sum_ih(\phi_{\pi_i}(\theta))$, which with Theorem~\ref{thm:uniform} is $Nh(\theta)-\E F$, the extraction loss.

The biases also say what a coder of the gaps of the residual faces. For $\theta\le\frac12$, since $\U$ maps $[0,1]$ into $[0,\frac12]$ and $\V$ preserves $[0,\frac12]$, every residual bias is at most $\frac12$: a one is never the more probable value, and what varies along the residual is the hazard of the next one, from about $\frac12$ on the all-$\ustep$ branch, whose first bit is the parity of the first $M=2^{\lfloor\log N\rfloor}$ input bits and is one with probability $\frac12(1-(1-2\theta)^M)$, close to $\frac12$ for every fixed $\theta\in(0,1)$ once $M$ is large, down to $\phi_{\vstep^d}(\theta)=\theta^{2^d}/(\theta^{2^d}+(1-\theta)^{2^d})$ on the all-$\vstep$ branch. A coder of the gaps that knows this hazard, as codes A and C of Appendix~\ref{app:logcoders} do, has something to gain over the gamma codes of code G.

The tail bound of Proposition~\ref{prop:tail} says how these biases polarise. Under $\Pr_\theta$, let $\rho_i=\phi_{\pi_i}(\theta)$ be the bias of the $i$-th residual bit in the causal factorisation of Theorem~\ref{thm:factor}, and for $0<\delta<\frac12$ let $A_{N,\delta}$ be the number of residual bits with $\delta\le\rho_i\le1-\delta$.

\begin{corollary}[Polarisation]\label{cor:polar}
For every $\theta\in(0,1)$, $0<\delta<\frac12$ and $t\ge0$, $\Pr_\theta\{A_{N,\delta}>(\log|\Rc_N|+t)/a_\delta\}\le2^{-t}$ with $a_\delta=-\log(1-\delta)$, and $\E_\theta A_{N,\delta}\le\log|\Rc_N|/h(\delta)$. Hence $A_{N,\delta}=O_\delta(N^\beta\log N)$ almost surely: for $\theta\ne\frac12$, where $|R|/N\to1-h(\theta)>0$, the fraction of residual bits whose bias is not within $\delta$ of $0$ or $1$ tends to zero almost surely; for $\theta=\frac12$ every residual bit has bias $\frac12$ and there are $o(N)$ of them.
\end{corollary}

\begin{proof}
By \eqref{eq:factor}, $-\log\Pr_\theta(X)-F(X)=-\sum_i\log\Pr_{\rho_i}(R_i)$, and each counted bit contributes at least $a_\delta$ whatever its value, so $a_\delta A_{N,\delta}\le-\log\Pr_\theta(X)-F(X)$; Proposition~\ref{prop:tail} gives the tail bound, and Borel--Cantelli with $t=2\log N$ the almost-sure statement. Taking expectations, $\E\sum_ih(\rho_i)=H(R)\le\log|\Rc_N|$ by the chain rule in request order, and each counted bit contributes at least $h(\delta)$.
\end{proof}

Ar{\i}kan's source polarization \cite{Arikan10} applies the same XOR of pairs, recursively and linearly, and shows that the conditional entropies of the transformed bits polarise to $0$ or $1$. The exhaustive transform is a nonlinear, parameter-free parallel: by Corollary~\ref{cor:polar}, for $\theta\ne\frac12$ all but a vanishing fraction of the residual bits of a Bernoulli string have biases within $\delta$ of $0$ or $1$, almost surely and for every $\delta$, with at most $O_\delta(N^\beta\log N)$ exceptions, and the fair bits are exactly fair; the polarisation here comes with an explicit remainder, through a deterministic sparsity bound.

\subsection{Code B}

The decoder knows $N$. Code B sends $w=\wt(s)$ in $\lceil\log(N+1)\rceil$ bits; encoder and decoder then use the plug-in bias $\hat p=w/N$ and carry $\phi_\pi(\hat p)$ down the recursion, so that the bias of each residual bit is known to both when it is produced or requested.

\begin{definition}[Code B]
Each residual bit is coded at the probability $\phi_{\pi_i}(\hat p)$ of its node, and the fair bits are stored as they are. The \emph{ideal length} of code B is $\ell_B(s)=\lceil\log(N+1)\rceil+F(s)-\sum_i\log\Pr_{\phi_{\pi_i}(\hat p)}(R_i)$.
\end{definition}

\begin{theorem}\label{thm:B}
For every $s\in\bits^N$ and $\theta\in(0,1)$,
\[
\ell_B(s)=\lceil\log(N+1)\rceil+Nh(\wt(s)/N)\le-\log\Pr_\theta(s)+\log(N+1)+1.
\]
\end{theorem}

\begin{proof}
By \eqref{eq:factor} at $\theta=\hat p$, $F$ plus the residual cost equals $-\log\Pr_{\hat p}(s)=Nh(\hat p)$, which is $\min_\theta-\log\Pr_\theta(s)$. For $w\in\{0,N\}$, \eqref{eq:factor} is an identity between rational functions continuous on $[0,1]$, so it holds at $\hat p\in\{0,1\}$: there is no fair bit and every residual bit has probability one.
\end{proof}

On the string of Figure~\ref{fig:tree}, $w=7$ and $\ell_B(s)=\lceil\log13\rceil+12\,h(7/12)=4+11.76$ bits, against $13$ bits for code G (Section~\ref{sec:G}) and the $12$ bits of the string itself: at this length the weight header dominates, and neither code shortens an unstructured string.

An actual code realises the ideal length within one bit. By Theorem~\ref{thm:factor}, $Q_{\hat p}(R)=\prod_i\Pr_{\phi_{\pi_i}(\hat p)}(R_i)$ is a probability law on the prefix code $\Rc_N$, the residual bits being drawn causally, so the weight, a Shannon codeword for $R$ under $Q_{\hat p}$ and the $F$ fair bits form a prefix code of length at most $\ell_B(s)+1$; an arithmetic realisation attains $\ell_B$ within a constant termination cost. Code B therefore has the length of the two-part code \cite{Rissanen84}: the transform does not shorten plug-in codes, it changes which bits need a coder. About $Nh(\theta)$ bits are stored as they are, and only the residual, at biases that are mostly close to $0$ or $1$, passes through the coder. The expected redundancy is $\log N+O(1)$; quantising $\hat p$ on a grid of about $\sqrt N$ points in the arcsine scale would bring it to $\frac12\log N+O(1)$ \cite{Rissanen84}.

\begin{remark}[Exact fair bits at the KT length]\label{rem:E}
For a weight $w$, let $Q_w(R)=2^{N-|R|}/\binom Nw$ for $R\in\Rc_N$ with $w_N(R)=w$. By the Kraft equality restricted to one weight, $Q_w$ is a probability law, and coding $R$ under $Q_w$ followed by its $N-|R|$ fair bits costs exactly $\log\binom Nw$ ideal bits given $w$. If $w$ is coded with the beta-binomial law $\mathrm{BB}(w)=\binom Nw B(w+\frac12,N-w+\frac12)/\pi$, the total ideal length is $-\log\mathrm{BB}(w)+\log\binom Nw=-\log\KT(N,w)$. Exactly fair bits and the KT length are therefore compatible. The law $Q_w$ factorises along the recursion (at a node of length $2n+1$ and weight $w$, the remainder is $b$ with probability $\binom{2n}{w-b}/\binom{2n+1}w$, and at a node of length $2n$ the number $k$ of unequal pairs has probability $\binom nk2^k\binom{n-k}{(w-k)/2}/\binom{2n}w$), which gives a recursive enumerative code \cite{Cover73} with the large binomial coefficients that code B avoids.
\end{remark}

\begin{remark}[The weight-free KT residual law]\label{rem:E2}
The weight need not be sent at all. The function
\[
Q_N(R)=2^{N-|R|}\,\KT(N,w_N(R)),\qquad R\in\Rc_N,
\]
is a probability law on $\Rc_N$, since the fibres partition $\bits^N$ and $\sum_s\KT(N,\wt(s))=1$, and coding $R$ under $Q_N$ followed by its fair bits costs exactly $-\log\KT(N,\wt(s))$ ideal bits; with Shannon lengths this is a prefix code of regret less than $\frac12\log N+2$ for every $s$ and $\theta$, with the original exactly fair bits. $Q_N$ is the mixture of the laws $Q_\theta$ of Theorem~\ref{thm:factor} under Jeffreys' prior; it has an exact causal form, since the probability of a residual prefix is a polynomial of degree at most $N$ in $\theta$, which a Gauss--Chebyshev rule with $\lceil(N+1)/2\rceil$ nodes integrates exactly, so that the next residual bit is priced with $O(N)$ posterior weights and no enumeration (the update is written out in the extended version); a fast implementation is an open problem (Section~\ref{sec:disc}).
\end{remark}

\subsection{What Code B Does Not See}\label{sec:Bblind}

The length of code B depends on the input only through its weight. This respects the symmetry of exchangeable laws, but it cannot distinguish arrangements within a type class: if $k$ is a power of two, $\T(0^k1^k)=(1;0^{2k-2}1)$ (Proposition~\ref{prop:dilation} below), a residual with a single one, yet code B prices its $2k-1$ residual bits at the biases $\phi_\pi(\frac12)=\frac12$ and spends $\lceil\log(2k+1)\rceil+2k$ bits, as for a string of fair coin flips. The transform has separated a structured string into one fair bit and a residual that is almost all zeros; the plug-in model, which computes every bias from the weight, cannot use it, whereas code G codes this string in $2e+2$ bits for $k=2^e$ (Section~\ref{sec:structure}). The two codes make opposite choices: code B computes every bias from the weight and is nearly optimal for memoryless sources, code G uses no estimate and loses there, but it pays for what the residual actually contains, which is what makes it adapt to structure (Section~\ref{sec:structure}). One mode bit selecting the shorter of the two gives both bounds within one bit.

\section{Structure Without Transmitted Boundaries}\label{sec:structure}

Code B prices a string through its type and cannot see the arrangement of a type class; code G pays for what the residual contains. This section bounds what it pays on strings with few runs and on concatenations whose cuts nobody transmits, with the pricing identity of Section~\ref{sec:pricing} as the main tool, and extends the bounds to every residual coder whose cost per residual one is logarithmic.

\subsection{Runs}\label{sec:runs}

\begin{proposition}[Dyadic dilation]\label{prop:dilation}
Let $x^{[a]}$ repeat each symbol of $x$ exactly $2^a$ times. Then $\T(x^{[a]})=\big(Y(x);\,0^{(2^a-1)|x|}R(x)\big)$. In particular, if $k$ is a power of two, $\T(0^k1^k)=(1;0^{2k-2}1)$.
\end{proposition}

\begin{proof}
For one duplication, $u=0^{|x|}$, $y=\epsilon$ and $v=x$, and $\T(0^n)=(\epsilon;0^n)$; hence $Y(x^{[1]})=Y(x)$ and $R(x^{[1]})=0^{|x|}R(x)$. Iterate over the lengths $|x|,2|x|,\dots,2^{a-1}|x|$, and use $\T(01)=(1;1)$.
\end{proof}

Alignment matters: off powers of two the fair bits of $0^k1^k$ are governed by the binary digits of $k$.

\begin{theorem}[Two runs]\label{thm:tworuns}
Write $k=\sum_{i=0}^db_i2^i$ with $b_d=1$. Then
\[
F(0^k1^k)=\sum_{i=0}^db_i(d-i+1),\qquad z(0^k1^k)=\sum_{i=0}^db_i.
\]
For $k=2^e-1$, $e\ge1$, $Y(0^k1^k)=1^{e(e+1)/2}$, $R(0^k1^k)=\big(\prod_{j=e,e-1,\dots,2}1\,0^{2^j-j-1}\big)1$ (a concatenation, empty for $e=1$), and $L_G(0^k1^k)=(3e^2+e)/2$.
\end{theorem}

The proof is in Appendix~\ref{app:tworuns}. Thus $0^{2^e}1^{2^e}$ has one fair bit and costs $2e+2$ bits in code G, whereas $0^{2^e-1}1^{2^e-1}$ has $e(e+1)/2$ fair bits: a code that stores the fair bits as they are pays $\Omega(\log^2N)$ on these strings, whatever it does with the residual. The family has a description of $O(\log N)$ bits; the logarithmic square is the price of keeping the fair bits raw, and it is attained:

\begin{theorem}[Few runs]\label{thm:runs}
Let $s\in\bits^N$ have $r$ runs and $D=\lceil\log(N/r)\rceil$. Then $z(s)\le r(D+1)$, $F(s)\le r\big[1+(1+\log\ee)D+D(D-1)/2\big]$, and $L_G(s)=O\big(r[1+\log(N/r)]^2\big)$ with an absolute constant.
\end{theorem}

\begin{proof}
Follow the $\vstep $-spine: $x_0=s$ and $x_{i+1}$ is the $v$-string of $x_i$. Each $x_i$ is a subsequence of $s$, so it has at most $r$ runs, and $|x_i|\le N2^{-i}$. At $x_i$, let $u_i$ be its $u$-string, $k_i=\wt(u_i)$ its number of unequal pairs and $\varepsilon_i$ the weight of its remainder; an unequal pair contains a run boundary, so $k_i\le r-1$. The recursion gives $z(x_i)=z(u_i)+z(x_{i+1})+\varepsilon_i$ and $F(x_i)=F(u_i)+k_i+F(x_{i+1})$. By Theorem~\ref{thm:charge}, $z(u_i)\le k_i$; telescoping down to $x_D$, of length at most $r$, gives $z(s)\le\sum_{i<D}(k_i+\varepsilon_i)+|x_D|\le r(D+1)$. For the fair bits, $n_i=|u_i|\le N2^{-i-1}$; if $n_i\ge r$, \eqref{eq:typebound} gives $F(u_i)\le\log\binom{n_i}{k_i}\le r\log(\ee n_i/r)\le r(\log\ee+D-i-1)$, since $t\mapsto t\log(\ee n/t)$ increases on $(0,n]$ and $N/r\le2^D$; if $n_i<r$, $k_i+F(u_i)\le2n_i<2r$. In both cases $k_i+F(u_i)\le r(1+\log\ee+D-i-1)$, and summing over $i<D$, with $F(x_D)\le r$, gives the bound on $F$. Finally, by the monotonicity noted after Lemma~\ref{lem:gap}, $z+1\le2r(D+1)$ and $|R|\le N$ give $L_\gap\le2r(D+1)\big[1+2\log\big(1+N/(2r(D+1))\big)\big]$, and $N/(2r(D+1))\le N/r\le2^D$ gives $\log\big(1+N/(2r(D+1))\big)\le D+1$, hence $L_\gap\le2r(D+1)(2D+3)$.
\end{proof}

The bound needs no knowledge of the runs, and it applies to concatenations ($r$ is at most the sum of the numbers of runs of the pieces) and to noisy versions: changing $e$ symbols creates at most $2e$ run boundaries, for every choice of the changed positions.

\subsection{Concatenations Without the Cuts}\label{sec:concat}

Let $s=s_1\cdots s_J$ with nonempty pieces of lengths $n_i$ and weights $w_i$, and let $\widehat H(s_1,\dots,s_J)=\sum_in_ih(w_i/n_i)$ be the piecewise empirical entropy. Code G is applied once to $s$; neither the transform nor the coder knows the cuts.

Attach to every bit of every node its \emph{ancestry}, the set of input positions it comes from: an input bit has its own position, and a bit of the $u$-string, or of the $v$-string, of a node has the union of the ancestries of its pair. Within one node the ancestries of its bits are disjoint and ordered, every position in the ancestry of a bit preceding every position in the ancestry of the next bit, though they need not be intervals. A fair bit is \emph{mixed} if the ancestry of its pair meets more than one piece, and $M(s;s_1,\dots,s_J)$ denotes the number of mixed fair bits.

\begin{lemma}[Pricing certificate]\label{lem:certificate}
For every partition, $F(s)\le\widehat H(s_1,\dots,s_J)+M(s;s_1,\dots,s_J)$.
\end{lemma}

\begin{proof}
Give every position of piece $i$ the parameter $p_i\in(0,1)$ and apply Proposition~\ref{prop:pricing}. At a node with path $\pi$, a bit whose ancestry lies in piece $i$ has the parameter $\phi_\pi(p_i)$, by induction, since $U_{t,t}=\U(t)$ and $V_{t,t}=\V(t)$; so an unmixed pair has two equal parameters and its fair bit the parameter $\frac12$. Keeping the factors $\frac12$ of the $F-M$ unmixed fair bits and bounding every other factor by one gives $\prod_ip_i^{w_i}(1-p_i)^{n_i-w_i}\le2^{-(F-M)}$. Choose $p_i=w_i/n_i$ for the nonconstant pieces and let $p_i$ tend to $0$ or $1$ for the constant ones; $F$ and $M$ do not depend on the parameters.
\end{proof}

\begin{lemma}[Mixed fair bits]\label{lem:mixed}
If $J\ge2$, $M(s;s_1,\dots,s_J)\le13\,N^\beta(J-1)^{1-\beta}$; and $M=0$ if $J=1$.
\end{lemma}

The proof (Appendix~\ref{app:mixed}) combines two facts: at one node, the ancestries of the pairs have disjoint spans, so at most $J-1$ of them straddle a cut; and the charge identity, stopped at a level of the tree, bounds $\sum_{|\pi|=d}2^{v(\pi)}\wt(x_\pi)$ by $N$, where $x_\pi$ is the string of node $\pi$.

\begin{theorem}[Concatenations]\label{thm:concat}
For every $s\in\bits^N$ and every partition of $s$ into $J$ consecutive nonempty pieces,
\[
L_G(s)\le\sum_{i=1}^Jn_ih(w_i/n_i)+13\,N^\beta(J-1)^{1-\beta}+3(\sqrt3+1)N^\beta\log(N+1).
\]
For a logarithmic coder, the last term is replaced by $H_{\rm hdr}(s)+O\big((N^\beta+K)\log N\big)$, the term in $(J-1)^{1-\beta}$ being unchanged.
\end{theorem}

\begin{proof}
$L_G=F+L_\gap(R)$, Lemma~\ref{lem:certificate}, Lemma~\ref{lem:mixed} and \eqref{eq:bN}.
\end{proof}

The inequality holds simultaneously for all partitions, so its right-hand side may be minimised afterwards; the excess is $o(N)$ whenever $J=o(N)$, and larger than $N$ at practical lengths (remark after Theorem~\ref{thm:universal}). If the cuts may be sent, a classical code does much better: the number of pieces, the cuts by their rank among the $\binom{N-1}{J-1}$ choices, and each weight and index in its type class cost $\widehat H+O(J\log(1+N/J))$. Nor do the cuts need to be sent to reach that order: for piecewise-stationary memoryless sources with a fixed number $J$ of pieces, Merhav \cite{Merhav93} shows that a redundancy of order $J\log N$ is unavoidable for almost every choice of the parameters and of the cuts, Shamir and Merhav \cite{ShamirMerhav99} attain it sequentially with low complexity, and the partition tree weighting (PTW) of Veness et al.\ \cite{VWBG13} mixes the dyadic partitions at a logarithmic cost per symbol, within $O(J\log^2N)$ with a KT base model, since a piece may need $O(\log N)$ dyadic intervals. These codes estimate the parameters of the pieces and mix over segmentations; Theorem~\ref{thm:concat} is weaker in rate, $N^\beta J^{1-\beta}$ against $J\log N$, but concerns one transform with no estimate and no cut sent. Code B, in contrast, pays $Nh(\wt(s)/N)$, the global empirical entropy, which can exceed $\widehat H$ by a linear amount. Nor is there an inequality against the separate lengths of code G: with $m=2^d$, $L_G((0011)^m)=m+2d+3$ and $L_G(0)=3$, whereas $L_G(0(0011)^m)\ge F(0(0011)^m)=2m+d$ by \eqref{eq:shift} below; one prepended bit costs a linear amount, which is no contradiction since the zero-order benchmark of $(0011)^m$ is $4m$.

\subsection{What a Raw Fair Part Cannot See}\label{sec:limits}

For $x=(01)^m$ every root pair is unequal, so $Y(x)=1^m$ and $R(x)=R(1^m)$: the fair bits are perfectly repetitive and any code that stores them as they are pays $N/2$ bits. A one-bit shift changes the number of fair bits linearly:
\begin{equation}\label{eq:shift}
F\big((0011)^m\big)=m,\qquad F\big(0(0011)^m\big)=2m-1+\lfloor\log(2m)\rfloor.
\end{equation}
For the first, every root pair is equal, $u=0^{2m}$ and $v=(01)^m$. For the second, the root pairs are $00$ and then $2m-1$ unequal pairs, with an odd remainder; the root emits $2m-1$ fair bits, its $\vstep $-child is a single zero, and $u=0\,1^{2m-1}$. A string $0\,1^{\ell-1}$ has one unequal first pair, constant $\vstep $-children, and a $u$-string with a single one in first position; each further halving emits one fair bit, $\lfloor\log(2m)\rfloor$ in all.

How much a residual-only model can recover is characterised exactly in Appendix~\ref{app:residual}: for a class $\mathcal K$ of inputs, the best code that stores the fair bits raw and codes the residual under any law pays $\log$ of the number of strings in the fibres met by $\mathcal K$, the Shtarkov normalisation \cite{Shtarkov87} of residual-only codes, so that even a singleton costs its $F(x)$ fair bits. Recursing also on the fair bits removes the obstruction of periodic strings, but no longer designates an exactly fair part; the extended version studies that transform, and we do not pursue it.

\subsection{Logarithmic Residual Coders}\label{sec:logcoders}

Sparsity is what makes code G universal, and the argument applies to every residual coder whose cost per residual one is logarithmic. Consider coders that send a header of $H_{\rm hdr}(s)$ bits, decodable on its own, store the fair bits as they are and code the residual bits in decoding order, after assigning each requested residual bit to one of $K$ \emph{classes} by a rule that the decoder can evaluate from $N$, from the header and from the bits already decoded; $K$ counts the classes actually used, and empty classes cost nothing. Two kinds are of interest. A \emph{gap coder} writes, within each class, the number of zeros before each one with an integer code, possibly adaptive, whose codeword for $g$ has length at most $c_0+c_1\log(g+1)$; the codeword of the next gap of a class is written where the decoder first requests a bit of that class after the previous gap is exhausted, and the final run of zeros of a class is written as a gap whose one is never requested. A \emph{sequential coder} codes each residual bit at the KT probability of the past of its class.

\begin{theorem}[Logarithmic coders]\label{thm:logcoders}
Let $s\in\bits^N$ have $z$ residual ones and $m=|R(s)|$. A gap coder with $K$ classes has length at most $H_{\rm hdr}(s)+F(s)+(z+K)\big[c_0+c_1\log\big(1+\frac{m+K}{z+K}\big)\big]$, and a sequential coder with $K$ classes has ideal length at most $H_{\rm hdr}(s)+F(s)+z\log(\ee m/z)+K(\frac12\log m+1)$, the middle term being zero if $z=0$. Hence both are within $\log\binom N{\wt(s)}+H_{\rm hdr}(s)+O\big((N^\beta+K)\log N\big)$ on every string, and a string with $r$ runs costs $O\big(r(1+\log(N/r))^2+K\log N\big)+H_{\rm hdr}(s)$. In particular, every such coder with a fixed number of classes and a header of $O(\log N)$ bits, and more generally with $\sup_{|s|=N}H_{\rm hdr}(s)+K_N\log(N+1)=o(N)$, is universal on individual sequences in the sense of Theorem~\ref{thm:universal}.
\end{theorem}

The proof is in Appendix~\ref{app:logcoders}. The control of $K$ is needed: assigning the $i$-th residual bit of $0^N$ to its own class costs $N$ bits with sequential KT and $3N$ with gamma codes, at zero empirical entropy, and an arbitrary header defeats any bound. Code G is the gap coder with one class, no header, $c_0=1$ and $c_1=2$. An adaptive Exp-Golomb code whose order is chosen from the past of the class and capped at $k_{\max}$ has $c_0=k_{\max}+1$ and $c_1=2$. An actual realisation of a sequential coder adds the rounding of the joint residual law, less than one bit with Shannon lengths. Two such coders appear in the experiments, both built on the plug-in biases of code B: code A, a sequential coder with at most $50$ classes, and code C, a gap coder with guarded Rice codes and an escape, with at most $18$ classes; a two-bit selector among code G, the same with Elias delta codes, the raw residual and code C costs at most $L_G(s)+2$ bits on every string and at most $N+2$. Their definitions, the role of their caps and the obstruction that an unprotected Rice code meets are in Appendix~\ref{app:logcoders}.

\section{The BWT Without a Context Tree}\label{sec:memory}

The concatenation bound applies to every partition of the input, although the encoder does not know its cuts. The Burrows--Wheeler transform of the reversed string supplies a useful family of such partitions: for each context order, the symbols that follow the same context occupy consecutive positions of its output. Applying code G once to that output therefore gives, with the same encoder, a bound in terms of the empirical conditional entropy of every order.

\subsection{The BWT of the Reversed String}

Let $x\in\bits^N$ and $t=x_Nx_{N-1}\cdots x_1\$$, the reversed string followed by a sentinel $\$$ smaller than $0$. Sort the $N+1$ suffixes of $t$; the BWT $L$ lists the symbol preceding each sorted suffix, cyclically. Let $p$ be the row of $\$$ in $L$ and $\ell\in\bits^N$ the string $L$ with $\$$ removed; $x\mapsto(\ell,p)$ is injective, inverted by the LF mapping, and $\ell$ is a permutation of $x$. Because $t$ is reversed, the suffix at row $i$ reads the bits preceding some position $j$ of $x$, most recent first, and $L_i=x_j$. Read contexts with zero padding ($x_i=0$ for $i\le0$); a suffix that reaches $\$$ sorts before every longer suffix with the same prefix, so it lies at the beginning of the intervals of its padded contexts. Hence the rows whose padded context starts with $c=c_1\cdots c_d$ ($c_1$ the most recent bit) form an interval $I(c)$, with $I(c)=I(c0)\cup I(c1)$ and $I(c0)$ first, where $cb$ extends $c$ by one older bit: the order-$k$ context subsequences are the intervals of depth $k$ of one binary tree, and every pruned subtree is a partition whose context lengths vary from branch to branch. The sentinel row carries no bit of $x$ and is ignored in the weights and likelihoods of the leaves below.

For $0\le k<N$ and $c\in\bits^k$, let $n_{c,b}$ be the number of positions $j>k$ such that $x_{j-1}x_{j-2}\cdots x_{j-k}=c$ and $x_j=b$, let $n_c=n_{c,0}+n_{c,1}$, and let
\[
E_k(x)=\sum_{c:\,n_c>0}n_ch(n_{c,1}/n_c),
\]
the empirical conditional entropy of order $k$, which excludes the first $k$ symbols; $E_0(x)=Nh(\wt(x)/N)$.

\begin{lemma}[Context intervals]\label{lem:bwtintervals}
For $0\le k<N$, $\ell$ is the concatenation of $J_k(x)=k+\#\{c:n_c>0\}\le\min(N,2^k+k)$ nonempty consecutive pieces whose zero-order empirical entropies sum to $E_k(x)$.
\end{lemma}

\begin{proof}
The lemma uses the unpadded intervals $I_{\rm raw}(c)$, the rows whose suffix begins with the $k$ binary symbols of $c$, which differ from the padded $I(c)$ above only by rows of the first $k$ positions. For $j>k$, the suffix preceding $x_j$ begins with the reversed context of $x_j$, and the suffixes with a given prefix of length $k$ are consecutive in sorted order. For $j\le k$, the suffix reaches $\$$ before $k$ binary symbols and lies in no $I_{\rm raw}(c)$; make each of these $k$ entries a piece of one symbol, of empirical entropy zero. Removing the sentinel row preserves the order. The piece of context $c$ has length $n_c$ and weight $n_{c,1}$.
\end{proof}

\subsection{Code G Needs No Tree}

Code G is applied to $\ell$ as a whole, with no tree, no weight and no split point:
\[
L_{\rm BG}(x)=\lceil\log(N+1)\rceil+L_G(\ell).
\]
This is a prefix code for each $N$: the index has a fixed length, $G_N$ is prefix-free and the BWT is invertible.

\begin{theorem}[Pointwise Markov comparison]\label{thm:bwtmarkov}
For every $x\in\bits^N$ and every $0\le k<N$,
\begin{gather*}
L_{\rm BG}(x)\le E_k(x)+\Delta_{N,k},\\
\Delta_{N,k}=13\,N^\beta(2^k+k-1)^{1-\beta}+3(\sqrt3+1)N^\beta\log(N+1)+\lceil\log(N+1)\rceil.
\end{gather*}
Consequently, for every Markov law $P$ of order $k$ with any initial distribution, $L_{\rm BG}(x)+\log P(x)\le\Delta_{N,k}$ whenever $P(x)>0$, and $0\le\E_PL_{\rm BG}(X)-H_P(X)\le\Delta_{N,k}$. The same holds for a logarithmic residual coder applied to $\ell$, with $3(\sqrt3+1)N^\beta\log(N+1)$ replaced by $H_{\rm hdr}(\ell)+O((N^\beta+K)\log N)$ in $\Delta_{N,k}$.
\end{theorem}

\begin{proof}
Theorem~\ref{thm:concat} applied to the partition of Lemma~\ref{lem:bwtintervals}, which has $J_k(x)-1\le2^k+k-1$ cuts, gives the first bound. If $\theta_c$ is the probability of a one after context $c$, then for each context $n_ch(n_{c,1}/n_c)\le-n_{c,1}\log\theta_c-n_{c,0}\log(1-\theta_c)$, the difference being $n_c$ times a binary divergence; summing gives $E_k(x)\le-\log P(x_{k+1}\cdots x_N\mid x_1\cdots x_k)\le-\log P(x)$. The expectation bound follows, and its lower bound is the entropy bound for a prefix code.
\end{proof}

For a stationary chain of order $k$ and entropy rate $h_P$, $H_P(X_1\cdots X_N)\le k+(N-k)h_P\le Nh_P+k$, so the excess over $Nh_P$ is at most $\Delta_{N,k}+k$, that is $O_k(N^{\beta-1}\log N)$ per symbol; at $N=2^{18}$ and $k=3$, $\Delta_{N,k}=3.3\cdot10^6$ bits against $NH\approx10^5$ in Table~\ref{tab:markov}, so the bound is empty at the lengths of the experiments (remark after Theorem~\ref{thm:universal}). The encoder is the same for every $k$: the order is chosen only in the analysis, and the bound holds for all orders at once. The rate is much slower than that of the tree codes, $O(2^k\log N/N)$ per symbol (Proposition~\ref{prop:oracle}), and Section~\ref{sec:exp} shows that the difference is real; what code G gains is that it needs no context tree, no counts and no split points. Since $\Delta_{N,k}/N\to0$ for each $k$, Theorem~\ref{thm:bwtmarkov} gives a statement on every infinite sequence:

\begin{corollary}[Individual sequences]\label{cor:individual}
For every infinite binary sequence $x$,
\[
\limsup_{N\to\infty}\frac{L_{\rm BG}(x_1\cdots x_N)}N\le\inf_k\limsup_{N\to\infty}\frac{E_k(x_1\cdots x_N)}N.
\]
\end{corollary}

The right-hand side is the benchmark of the empirical-entropy analyses of LZ78 \cite{KosarajuManzini99} and of BWT-based compressors \cite{Manzini01,KLV07}. For example, $x=(01)^m$ has $m$ fair bits before the BWT, but $\ell=0\,1^m0^{m-1}$ has at most three runs, so $L_{\rm BG}(x)=O(\log^2N)$ by Theorem~\ref{thm:runs}: sorting by context turns the periodic obstruction of Section~\ref{sec:limits} into a string with few runs.

\emph{What is new here, and at what rate.} Universality through the BWT is not new: for finite-memory sources, Effros et al.\ \cite{EVKV02} obtain, with a BWT followed by a sequential or piecewise estimator, a redundancy within a constant factor of the optimal rate, of order $\log N/N$ per symbol, and universality for stationary ergodic sources. Theorem~\ref{thm:bwtmarkov} gives $O_k(N^{\beta-1}\log N)=O(N^{-0.207}\log N)$ per symbol, polynomially slower, and Table~\ref{tab:markov} shows the difference at finite length. What Theorem~\ref{thm:bwtergodic} adds is the identity of the code that attains universality: a fixed second stage with no estimate, no counts and no arithmetic coder, the gamma codes of the gaps of one residual, with a pointwise bound for every string and every order at once. The form of that bound differs from the individual-sequence analyses of BWT compressors: \cite{KLV07} bound the output of the BWT followed by move-to-front and an order-$0$ coder by $\mu\,nH_k(s)+n\log\zeta(\mu)+g_k$ for every $\mu>1$, with $\zeta$ the Riemann zeta function and $g_k$ a constant, \cite{Manzini01} has multiplicative constants larger than one as well. LZ78 is \emph{coarsely optimal} in the terminology of \cite{KosarajuManzini99}: its length is at most $nH_k(s)+nf_k(n)$ with $f_k(n)\to0$ for every $k$ \cite{PWZ92}, with an $f_k$ of order $\log\log n/\log n$, and what \cite{KosarajuManzini99} disprove is $\lambda$-optimality, an excess of $o(nH_k(s))$ rather than $o(n)$. Theorem~\ref{thm:bwtmarkov} is a coarse-optimality statement of the same form, with $f_k(N)=O\big(2^{k(1-\beta)}N^{\beta-1}\log N\big)$, polynomially small; constant one with a vanishing additive term is therefore not new in itself, and the contribution is its attainment by a fixed residual gap coder, with the deterministic argument that ties the remainder to the exhaustive Peres transform, at the price of a constant so large that the bound is empty below $N\approx2^{41}$ (remark after Theorem~\ref{thm:universal}).

\begin{theorem}[Stationary ergodic universality]\label{thm:bwtergodic}
For every stationary ergodic binary source $P$ with entropy rate $h_P$,
\[
L_{\rm BG}(X_1\cdots X_N)/N\to h_P\quad\text{almost surely and in }L^1;
\]
in particular $\E_PL_{\rm BG}(X_1\cdots X_N)-H_P(X_1\cdots X_N)=o(N)$.
\end{theorem}

\begin{proof}
Fix $k$. By the ergodic theorem, $n_{c,b}/N$ converges almost surely to $P(X_kX_{k-1}\cdots X_1=c,X_{k+1}=b)$ for each of the finitely many $(c,b)$, and $(a,b)\mapsto(a+b)h(b/(a+b))$ is continuous on $[0,\infty)^2$; hence $E_k(X_1\cdots X_N)/N\to h_k=H_P(X_{k+1}\mid X_1\cdots X_k)$ almost surely, and Theorem~\ref{thm:bwtmarkov} gives $\limsup_NL_{\rm BG}/N\le h_k$. Intersecting over $k$, and since $h_k\downarrow h_P$, $\limsup_NL_{\rm BG}/N\le h_P$ almost surely. For the lower bound, Kraft's inequality gives
\[
P\{L_{\rm BG}(X_1\cdots X_N)<-\log P(X_1\cdots X_N)-\varepsilon N\}\le2^{-\varepsilon N},
\]
which is summable; by Borel--Cantelli and the Shannon--McMillan--Breiman theorem \cite{CoverThomas06}, almost surely $\liminf_NL_{\rm BG}/N\ge h_P-\varepsilon$, for every rational $\varepsilon>0$. Finally $L_{\rm BG}(x)\le N+\Delta_{N,0}$ for every $x$, and $\Delta_{N,0}/N$ is bounded, so dominated convergence gives the convergence in $L^1$.
\end{proof}

No mixing and no finite-memory assumption is needed. The theorem concerns lossless coding: the fair bits of the BWT output are not exactly fair (Section~\ref{sec:notsay}), but Section~\ref{sec:bwtextract} shows that their entropy falls short of their length by $o(N)$ bits only, for every stationary source.

\subsection{Finite Alphabets and Text}\label{sec:text}

Let $\A$ be a finite alphabet of size $q\ge2$. The BWT of the reversed string is defined in the same way, its context intervals are consecutive for every order, and the empirical conditional entropy of order $k$ is $E_k^{(q)}(x)=\sum_{c\in\A^k:\,n_c>0}\sum_{a\in\A}n_{c,a}\log(n_c/n_{c,a})$, with $n_{c,a}$ the number of positions $j>k$ with context $c$ (most recent symbol first) and $x_j=a$. The code $L^{(q)}_{\rm BG}$ sends the row of the sentinel and binarises $\ell$ (Section~\ref{sec:binar}): at every internal vertex of a binary tree whose leaves are the symbols, it collects in order the symbols of $\ell$ below that vertex and records their branches, and it codes the $q-1$ binary streams with code G, parents before children; the length of each stream is the number of symbols below its vertex, known from its parent, so no length and no context boundary is sent.

\begin{proposition}\label{prop:qary}
For every $x\in\A^N$ and $0\le k<N$,
\[
L^{(q)}_{\rm BG}(x)\le E_k^{(q)}(x)+(q-1)\big[13\,N^\beta(q^k+k-1)^{1-\beta}+3(\sqrt3+1)N^\beta\log(N+1)\big]+\lceil\log(N+1)\rceil,
\]
and for every stationary ergodic source $P$ on $\A$, $L^{(q)}_{\rm BG}(X_1\cdots X_N)/N\to h_P$ almost surely and in $L^1$.
\end{proposition}

\begin{proof}
The partition of $\ell$ into at most $q^k+k$ context pieces (Lemma~\ref{lem:bwtintervals}, whose proof does not use the alphabet) induces in each stream a partition into at most $q^k+k$ nonempty consecutive pieces. For a context $c$, the contributions $n_{c,\nu}h(n_{c,\nu1}/n_{c,\nu})$ of the internal vertices $\nu$, with $n_{c,\nu}$ the number of symbols of context $c$ below $\nu$ and $n_{c,\nu1}$ those below its right child, add up to $\sum_an_{c,a}\log(n_c/n_{c,a})$: expanding each contribution as $n_{c,\nu}\log n_{c,\nu}-n_{c,\nu0}\log n_{c,\nu0}-n_{c,\nu1}\log n_{c,\nu1}$, the terms of the internal vertices cancel. Theorem~\ref{thm:concat} applied to each stream, whose length is at most $N$, gives the bound. The proof of Theorem~\ref{thm:bwtergodic} applies verbatim, with $L^{(q)}_{\rm BG}(x)\le N\log q+o(N)$ for the domination.
\end{proof}

For $q=256$, $k=1$ and $N=2^{20}$ the remainder of Proposition~\ref{prop:qary} is $3\cdot10^9$ bits against $N\log q=8.4\cdot10^6$: like the binary bounds of Section~\ref{sec:G}, the finite-alphabet statement is asymptotic (see the remark after Theorem~\ref{thm:universal}), and the support reduction below does not change this. The bounds above concern the declared alphabet, and letters that do not occur still cost a stream each, although an empty stream costs nothing. They can be removed from the alphabet first: for a nonempty $x$ with support $S$ of size $d$, send $d$ with a gamma code and $S$ by its rank among the $\binom qd$ subsets, $H_S=\gamma(d)+\lceil\log\binom qd\rceil$ bits, then apply the code to the $d$-letter alphabet $S$, so that $L_{\rm sp}(x)\le N\widehat H(x)+(d-1)\bar b_N+H_S$ without the BWT, where $\widehat H(x)$ is the empirical entropy, and $L_{{\rm sp},\rm BG}(x)\le E_k^{(q)}(x)+\lceil\log(N+1)\rceil+H_S+(d-1)[13N^\beta(d^k+k-1)^{1-\beta}+\bar b_N]$ with it, for $d\ge2$ (the contexts that occur use only the letters of $S$, and a monotone relabelling preserves the order of the suffixes); for $d=1$ the sole letter is sent and repeated. Unused letters thus create no coded stream and their names are paid for once, in $H_S$; unknown names are not free, since the $q$ constant strings alone need $\log q$ bits on average.

For text, the compressor of Section~\ref{sec:exp} combines both approaches: the BWT is computed on bytes, the context tree is refined bit by bit of each context character, so that it stays binary with depth $8k$ for $k$ characters of context, and a leaf, a segment of bytes, is coded through its bit-plane binarisation: for every prefix $\alpha$ of a byte, the next bits of the symbols of the segment with prefix $\alpha$ form a binary sequence, of length known to the decoder once the shorter prefixes are decoded, which is transformed and coded with one of the codes above. With depth $0$ there is no tree, and the code is that of this section.

\subsection{The Model-Based Alternative: a Context Tree}

Code B prices a string by its weight. Applied to $\ell$ as a whole, its ideal body costs $E_0(x)$; to exploit the memory it must be applied to the context intervals, which the decoder must then learn. The BWT-tree code of Appendix~\ref{app:tree} sends the row of the sentinel, a pruned context tree chosen by dynamic programming with its split points, and codes each leaf interval with KT or with code B; against every Markov law of order $k\le D$ started from the zero context, its regret is $O(2^k\log N)$ for every string, with KT or with code-B leaves (Proposition~\ref{prop:oracle} gives the constants). Corollary~\ref{cor:treeergodic} there gives the pointwise form of this bound, against the empirical conditional entropy $E_k(x)$, and hence, when the depth grows with $N$, universality for stationary ergodic sources by the argument of Theorem~\ref{thm:bwtergodic}: both constructions are universal for stationary ergodic sources, and for a fixed Markov order the tree code has a smaller proved redundancy; what dropping the tree costs is that redundancy, and what it buys is a code with no tree to select or transmit. The split points cost $\log N$ each, which sequential conditional KT and context-tree weighting \cite{WST95} do not pay; the cost-driven partition of the BWT into context segments is compression boosting \cite{FGMS05}, which adapts to the order without being told it, so that adaptation to an unknown order is not what distinguishes code G: what does is the fixed second stage, with no tree. Coding BWT segments with KT is the piecewise coder of \cite{EVKV02}. Table~\ref{tab:markov} compares this code with code G without any tree.

\section{Extraction From Sources With Memory}\label{sec:bwtextract}

Theorem~\ref{thm:uniform} needs an exchangeable law, and Section~\ref{sec:notsay} shows that without it the fair bits of a stationary source can be far from fair. This section measures what the Burrows--Wheeler transform recovers. The result is a statement about entropy: for every stationary source, the fair bits of $\ell$, the BWT output of Section~\ref{sec:memory} transformed once as a whole, fall short of their length by $o(N)$ bits of entropy, with an explicit bound and an exponential tail for Markov sources, and their expected number is $Nh_P+o(N)$ (with almost-sure convergence for ergodic sources). Approximate fairness in this sense, a divergence from the uniform law sublinear in the length, is the notion in which the output of a universal source code is nearly uniform \cite{VKV98}; here the nearly fair part is designated in advance by one fixed transform. The route is that of Theorems~\ref{thm:concat} and~\ref{thm:bwtmarkov}, with three ingredients. The pricing certificate (Lemma~\ref{lem:certificate}) becomes an identity: for every law invariant under the permutations inside the pieces, the unmixed fair bits are exactly uniform given everything else. The strings of a Markov type class, once transformed by the BWT, form a subset of such a piecewise exchangeable product, whose density is given by a classical count of spanning trees. A stationary source is finally replaced by its Markov approximation of order $k$, at the price of the divergence between the two, which is at most $N(h_k-h_P)$.

\subsection{Piecewise Exchangeable Laws}

Fix a partition of $\{1,\dots,N\}$ into $J$ consecutive nonempty intervals, the pieces, and for $s\in\bits^N$ let $w_i(s)$ be its weight on piece $i$. A law $P$ on $\bits^N$ is \emph{piecewise exchangeable} if $P(s)$ depends on $s$ only through $(w_1(s),\dots,w_J(s))$, that is, if $P$ is invariant under every permutation of the positions that preserves each piece. Exchangeable laws are the case $J=1$; a product law whose profile is constant on each piece is piecewise exchangeable, and so is the uniform law on a product of type classes, one per piece.

Mixed fair bits are those of Section~\ref{sec:concat}. For $s\in\bits^N$ with $\T(s)=(Y;R)$, let $\mathcal S(s)\subseteq\{1,\dots,F(s)\}$ be the set of indices of the mixed fair bits in $Y$, the \emph{mixed slots}, so that $M(s)=|\mathcal S(s)|$, and let
\[
Z(s)=\big(R(s),\ \mathcal S(s),\ (Y_j)_{j\in\mathcal S(s)}\big)
\]
collect the residual, the mixed slots and their values. The fair bits outside $Z$ are exactly fair, because of the following property of the decoder.

\begin{lemma}[Flipping an unmixed fair bit]\label{lem:flip}
Let $s\in\bits^N$, $\T(s)=(Y;R)$, $j\notin\mathcal S(s)$, let $Y'$ be $Y$ with its $j$-th bit flipped and $s'=\Dec_N(Y';R)$. Then $\T(s')=(Y';R)$; the string $s'$ agrees with $s$ outside the span of the ancestry of the pair that emitted $Y_j$, the interval from the first to the last position of that ancestry, and has the same weight as $s$ on that span; and $Z(s')=Z(s)$.
\end{lemma}

\begin{proof}
$\T(s')=(Y';R)$ by Proposition~\ref{prop:image}, since $R\in\Rc_N$ and $|Y'|=N-|R|$. The decoder makes the same decisions for $(Y';R)$ as for $(Y;R)$, the lengths and weights of all nodes being functions of $R$ alone, so the tree of $s'$ has the shape of the tree of $s$. The two trees differ at the node $\pi$ that emitted $Y_j$, where the unequal pair $(x_{2t-1},x_{2t})$ of $x_\pi$ is reversed, and at the ancestors of $\pi$, each rebuilt from the modified child with its other components unchanged. Call $x'$ an \emph{$[a,b]$-variant} of $x$ if $x'$ agrees with $x$ outside the interval $[a,b]$ of positions and has the same weight as $x$ on $[a,b]$. The reversed pair makes $x'_\pi$ a $[2t-1,2t]$-variant of $x_\pi$. Let now $x'$ be an $[a,b]$-variant of the string $x$ of a node, and let $X$ be the string of its parent, $\sigma(X)=(u,y,v,r)$.

If $x=u$, the parent is rebuilt as $X'=\sigma^{-1}(x',y,v,r)$. The pairs of index $t<a$ read the same bits of $u$, consume the same prefixes of $y$ and $v$, and are unchanged. The pairs of index $t\in[a,b]$ contain as many ones of $u$ as before, so the pairs of index $t>b$ consume the same suffixes of $y$ and $v$ and are unchanged as well. On the pairs $a,\dots,b$, every unequal pair has weight one and the equal pairs receive the same segment of $v$ as before, so the weight is preserved: $X'$ is a $[2a-1,2b]$-variant of $X$.

If $x=v$, the parent is rebuilt as $X'=\sigma^{-1}(u,y,x',r)$. Every pair keeps its type and every unequal pair its bits; the $a$-th to $b$-th equal pairs, at the pair indices $t_a<\dots<t_b$, receive $x'_a,\dots,x'_b$ instead of $x_a,\dots,x_b$, which have the same sum: $X'$ is a $[2t_a-1,2t_b]$-variant of $X$.

In both cases the interval of the parent is the span, among the positions of $X$, of the ancestries in $X$ of the positions $a,\dots,b$ of $x$: the pair $t$ of $X$ for the $u$-bit $t$, the $t$-th equal pair of $X$ for the $v$-bit $t$. By induction up to the root, $s'$ is a variant of $s$ on the span of the ancestry of the reversed pair.

For $Z(s')=Z(s)$: the residual is $R$ by construction. At every node on the way up, the pairs outside the modified interval of pair indices are unchanged, so the fair bits they emit keep their slots, their values and their ancestries. The pairs inside the interval emit the same number of fair bits as before, in the same slots, and the ancestries of these pairs lie within the span of the ancestry of the reversed pair, before and after the change. Since $Y_j$ is unmixed, that span lies inside one piece, and all these fair bits are unmixed before and after. Hence the mixed slots and their values are the same, and $Y'$ differs from $Y$ at the slot $j$ only.
\end{proof}

\begin{theorem}[Exact fairness of the unmixed bits]\label{thm:piecewise}
Let $P$ be piecewise exchangeable. Conditionally on $Z$, the unmixed fair bits are independent and uniform: for every $s$ with $P(s)>0$,
\begin{equation}\label{eq:piecewise}
-\log P(s)=F(s)-M(s)-\log P_Z(Z(s)),
\end{equation}
where $P_Z$ is the law of $Z$. Hence $H(Y\mid Z)=\E[F-M]$ and $H(X)=H(Z)+\E[F-M]$.
\end{theorem}

\begin{proof}
Fix $z=(R,\mathcal S,m)$ with $P_Z(z)>0$, let $s$ be a string with $Z(s)=z$, and $F=N-|R|$. For every $y\in\bits^F$ that agrees with $m$ on $\mathcal S$, flip one by one the slots outside $\mathcal S$ where $y$ differs from $Y(s)$: Lemma~\ref{lem:flip} at each step gives a string $s_y$ with $\T(s_y)=(y;R)$, $Z(s_y)=z$ and $P(s_y)=P(s)$, since each flip preserves the weight of every piece. These $2^{F-M}$ strings are distinct, and they exhaust the fibre of $z$, since every string of the fibre has the residual $R$ and the values $m$ on $\mathcal S$. So $P_Z(z)=2^{F-M}P(s)$, which is \eqref{eq:piecewise}; the entropy identities follow by taking expectations, since $Z$ and $Y$ determine $X$.
\end{proof}

For $J=1$, $M=0$, $Z=R$ and the theorem is Theorem~\ref{thm:uniform}. Lemma~\ref{lem:certificate} is the inequality $P_Z\le1$ in \eqref{eq:piecewise}, for the product law whose profile is the empirical bias of each piece. The theorem says more: given the residual, the mixed slots and their values, the unmixed fair bits are exactly uniform, so the mixed fair bits are the only ones that a piecewise exchangeable law can bias, and the entropy identity prices all the rest, residual, slots and mixed values, by $H(Z)$. For a law that is not piecewise exchangeable, \eqref{eq:piecewise} fails, and the fair bits of $\ell$ after a BWT are a case in point, Section~\ref{sec:notsay}; the next two subsections quantify the failure.

\subsection{Markov Sources}

Let $P$ be a Markov law of order $k$ on $\bits^N$ with any initial distribution,
\[
P(x)=\mu(x_1\cdots x_k)\prod_{j>k}\theta(x_j\mid x_{j-1}\cdots x_{j-k}).
\]
With the counts $n_{c,b}$ of Section~\ref{sec:memory}, let
\[
\tau(x)=\big(x_1\cdots x_k,\ (n_{c,b})_{c\in\bits^k,\,b\in\bits}\big)
\]
be the \emph{order-$k$ type} of $x$ and $C_\tau=\{x:\tau(x)=\tau\}$ its class. The probability $P(x)$ is a function of $\tau(x)$, so conditionally on $\tau$, $X$ is uniform on $C_\tau$. The pieces of $\ell$ in Lemma~\ref{lem:bwtintervals} are determined by $\tau$: the piece of the context $c$ has length $n_c$, the pieces of the contexts follow the sorted order of the contexts, and the row of a position $j\le k$, whose suffix reaches $\$$ after $x_{j-1}\cdots x_1$, is the first row among those whose context begins with $x_{j-1}\cdots x_1$, whose number is a sum of counts $n_c$ and of rows of other positions $i\le k$, all read from $\tau$. Given $\tau$, let $A_\tau\subseteq\bits^N$ be the set of strings whose weight on each piece is the prescribed one, $x_j$ on the row of a position $j\le k$ and $n_{c,1}$ on the piece of $c$, so that $|A_\tau|=\prod_c\binom{n_c}{n_{c,1}}$, and let $\Pi_\tau$ be the uniform law on $A_\tau$, which is piecewise exchangeable for the partition of $\tau$. Let $p$ be the row of the sentinel; $x\mapsto(\ell,p)$ is injective, so
\[
V_{\tau,p}=\{\ell(x):x\in C_\tau,\ p(x)=p\}\subseteq A_\tau
\]
is in bijection with the strings of $C_\tau$ whose sentinel row is $p$, and conditionally on $(\tau,p)$, $\ell$ is uniform on $V_{\tau,p}$. The sets $V_{\tau,p}$ need not be disjoint: for $k=0$, the strings $101$ and $110$ have the same $\ell=110$ and different sentinel rows. Nevertheless $\sum_p|V_{\tau,p}|=|C_\tau|$, because the pairs $(\ell,p)$ are distinct; all counts below retain $p$. How much smaller than $A_\tau$ the class is has a classical answer.

\begin{lemma}[Markov type classes]\label{lem:markovtype}
Write the contexts most recent bit first, $c=c_1\cdots c_k$, and let $G_\tau^+$ be the directed multigraph on the visited contexts (including the initial and final contexts), with $n_{c,b}$ edges from $c$ to $b\,c_1\cdots c_{k-1}$ for each $(c,b)$, plus one distinguished edge from the final context $b^*=x_N\cdots x_{N-k+1}$ to the initial context $a^*=x_k\cdots x_1$. The final context is determined by the initial context and the in/out-degree imbalance; in the balanced case it equals the initial context. For $k=0$ there is one vertex, the empty context, and both symbols give loops. Then
\begin{equation}\label{eq:markovtype}
|C_\tau|=\rho_k(\tau)\prod_{c}\binom{n_c}{n_{c,1}},\qquad\rho_k(\tau)=\frac{t_{a^*}(G_\tau^+)}{\prod_{c\ne b^*}n_c},
\end{equation}
where $t_{a^*}(G_\tau^+)$ is the number of spanning trees of $G_\tau^+$ oriented towards $a^*$, parallel edges counted as distinct, and the product runs over the contexts $c\ne b^*$ with $n_c>0$. Moreover, with $d_\tau=\#\{c\ne b^*:n_c>0\}\le2^k-1$,
\[
0\le\log\frac1{\rho_k(\tau)}\le\sum_{c\ne b^*:n_c>0}\log n_c\le d_\tau\log\frac{N-k}{d_\tau}\le(2^k-1)\log N,
\]
where the middle expression is zero if $d_\tau=0$. In particular $\rho_0=1$, and $\rho_k(\tau)=2^{-k}$ when all the counts $n_{c,b}$ are equal and positive.
\end{lemma}

\begin{proof}
The contexts $c_{k+1},\dots,c_{N+1}$ of the positions $k+1,\dots,N+1$ of $x$ form a walk in $G_\tau^+$ from $a^*$ to $b^*$ that uses the edges from $c$ to $b\,c_1\cdots c_{k-1}$ exactly $n_{c,b}$ times; conversely such a walk from $a^*$ determines $x$. Adding the edge from $b^*$ to $a^*$ closes the walk into an Eulerian circuit of $G^+_\tau$. By the BEST theorem \cite{vAEdB51}, the Eulerian circuits of a connected Eulerian multigraph with distinguishable edges, counted with the distinguished closing edge fixed (equivalently, cutting the circuit just after it), number $t_{a^*}(G^+_\tau)\prod_c(d^+(c)-1)!$, with $d^+$ the out-degree; identifying the parallel edges divides by $\prod_{c,b}n_{c,b}!$. Since $d^+(c)=n_c+\mathbf 1\{c=b^*\}$, the quotient is \eqref{eq:markovtype}, which is Whittle's formula \cite{Whittle55} in a combinatorial form. The closed walk visits every vertex of $G^+_\tau$, so this active graph is strongly connected and $t_{a^*}\ge1$. There are at most $2^k-1$ denominator factors, whose sum is at most $N-k$; the lower bounds on $\rho_k$ follow from the arithmetic--geometric mean inequality. Unvisited contexts must be omitted: including isolated vertices would make the spanning-tree count zero. For $\rho_k\le1$, $x$ is determined by the sequences of symbols emitted at the successive visits of each context, one arrangement of length $n_c$ and weight $n_{c,1}$ per context, so $|C_\tau|\le|A_\tau|$. When all $n_{c,b}$ equal $n$, every context occurs $2n$ times, as many edges enter each vertex as leave it, so $a^*=b^*$ and the added edge is a loop, which belongs to no spanning tree. A spanning tree towards $a^*$ chooses one of the two out-edges of each other vertex, each of multiplicity $n$: $t_{a^*}=t\,n^{2^k-1}$, where $t$ is the number of spanning trees of the binary de Bruijn graph of order $k$ towards a vertex, that is $2^{2^k-k-1}$, the number of binary de Bruijn sequences of order $k+1$ \cite{vAEdB51}. Dividing by $(2n)^{2^k-1}$ gives $2^{-k}$.
\end{proof}

\begin{proposition}[Fair bits after the BWT of a Markov source]\label{prop:bwtextract}
Let $P$ be a law on $\bits^N$ under which $X$ is uniform on its class $C_\tau$ given its order-$k$ type $\tau$, for instance a Markov law of order $k$ with any initial distribution. Let $(Y;R)=\T(\ell)$, $F=|Y|$, and let $Z=Z(\ell)$ and $M=M(\ell)$ be defined with the partition of $\ell$ of order $k$. Then
\begin{equation}\label{eq:markovextract}
\begin{split}
\E F-H(Y\mid Z,\tau,p)
&\le\E M+\E\log\frac1{\rho_k(\tau)}+H(p\mid\tau)\\
&\le13\,N^\beta(2^k+k-1)^{1-\beta}+(2^k-1)\log N+\log(N+1).
\end{split}
\end{equation}
The sharper middle bound retains the actual mixed-bit count, the type-class density and the conditional entropy of the index. The left side is the mean divergence $\E\,\KL\big(P(Y\mid Z,\tau,p)\,\|\,U_F\big)$ of the conditional law of $Y$ from the uniform law $U_F$ on $\bits^F$, and the same bounds hold for $\E F-H(Y\mid R,p)$ and for $\E F-H(Y\mid F)$.
\end{proposition}

\begin{proof}
Fix $(\tau,p)$ of positive probability, $\ell\in V_{\tau,p}$, $z=Z(\ell)$ and $y=Y(\ell)$. Since $\ell$ is uniform on $V_{\tau,p}$ given $(\tau,p)$ and $\Pi_\tau$ is uniform on $A_\tau\supseteq V_{\tau,p}$,
\[
P(\ell\mid\tau,p)=\frac{\Pi_\tau(\ell)}{\Pi_\tau(V_{\tau,p})},\qquad P(z\mid\tau,p)=\frac{\Pi_\tau(\{Z=z\}\cap V_{\tau,p})}{\Pi_\tau(V_{\tau,p})}.
\]
Theorem~\ref{thm:piecewise} for $\Pi_\tau$ gives $\Pi_\tau(\ell)=2^{-(F-M)}\Pi_\tau(Z=z)$, $M$ being read from $z$. Hence
\begin{equation}\label{eq:valid}
P(y\mid z,\tau,p)=\frac{P(\ell\mid\tau,p)}{P(z\mid\tau,p)}=\frac{2^{-(F-M)}}{\Pi_\tau(V_{\tau,p}\mid Z=z)}:
\end{equation}
given $(z,\tau,p)$, the fair string is uniform on the \emph{valid} completions of $z$, those $y$ for which $\Dec_N(y;R)$ lies in $V_{\tau,p}$, a fraction $\Pi_\tau(V_{\tau,p}\mid Z=z)$ of the $2^{F-M}$ completions. Its divergence from $U_F$ is therefore $M+\log(1/\Pi_\tau(V_{\tau,p}\mid Z=z))$. Averaging over $z$ given $(\tau,p)$, with $P(z\mid\tau,p)=\Pi_\tau(V_{\tau,p}\mid Z=z)\,\Pi_\tau(Z=z)/\Pi_\tau(V_{\tau,p})$ and $\varphi(t)=t\log(1/t)$, which is concave,
\begin{multline*}
\sum_zP(z\mid\tau,p)\log\frac1{\Pi_\tau(V_{\tau,p}\mid Z=z)}=\frac{\sum_z\Pi_\tau(Z=z)\,\varphi\big(\Pi_\tau(V_{\tau,p}\mid Z=z)\big)}{\Pi_\tau(V_{\tau,p})}\\
\le\frac{\varphi(\Pi_\tau(V_{\tau,p}))}{\Pi_\tau(V_{\tau,p})}=\log\frac1{\Pi_\tau(V_{\tau,p})}
\end{multline*}
by Jensen's inequality. Finally $\Pi_\tau(V_{\tau,p})=|V_{\tau,p}|/|A_\tau|=\rho_k(\tau)\,|V_{\tau,p}|/|C_\tau|$ by Lemma~\ref{lem:markovtype}, and $P(p\mid\tau)=|V_{\tau,p}|/|C_\tau|$, so the mean of $\log(|C_\tau|/|V_{\tau,p}|)$ given $\tau$ is $H(p\mid\tau)\le\log(N+1)$. The expectation over $(\tau,p)$ gives the first inequality of \eqref{eq:markovextract}; the second is Lemma~\ref{lem:mixed} with the $J_k\le2^k+k$ pieces of Lemma~\ref{lem:bwtintervals} and the lower bound of Lemma~\ref{lem:markovtype}. Conditioning on less cannot decrease the conditional entropy, and $F$ is a function of $R$, which gives the last claim.
\end{proof}

The bound is about entropy, not about total variation. Given $W=(z,\tau,p)$, the fair string is uniform on a subset of $\bits^F$ of density $2^{-M}\Pi_\tau(V_{\tau,p}\mid Z=z)$, so its total variation distance from $U_F$ is one minus that density, and a sublinear entropy deficiency does not make the density tend to one: a flat source on a fraction $1/N$ of a cube has deficiency $\log N$ and total variation $1-1/N$. The fair bits of $\ell$ are thus a flat source of nearly full entropy, not necessarily a nearly uniform string: for the fair coin at $N=16$, with $k=0$ and so $M=0$, the mean total variation is already $0.90$, and it is $0.99$, $0.94$ and $0.51$ for the three chains of Table~\ref{tab:bwtextract} (\texttt{code/bwt\_extraction\_check.py}); the flatness gives a tail bound (Proposition~\ref{prop:extracttail}), and a seed turns it into a total variation statement (Appendix~\ref{app:flat}). Flatness holds under this full conditioning; the laws obtained by conditioning on $(R,p)$ or on $F$ only are mixtures of flat laws.

Table~\ref{tab:bwtextract} shows the exact values at $N=16$ for the stationary symmetric chain of order one that flips with probability $\varepsilon$, computed over all $2^{16}$ strings (\texttt{code/bwt\_extraction\_check.py}). The script checks Lemma~\ref{lem:flip} on random strings and partitions and \eqref{eq:markovtype} by exact rational determinants on all classes at $(N,k)=(12,0)$, $(12,1)$, $(12,2)$ and $(13,3)$. The independent \texttt{code/review3\_checks.py} checks every $N\le10$, $0\le k\le\min(3,N-1)$, the new tail bound and the general-law identity below, and all piecewise fibres through length $7$. Keeping $H(p\mid\tau)$ instead of replacing it by $\log17=4.09$ materially sharpens the bound, especially for small flip probabilities.

\begin{table}[t]\centering\footnotesize
\caption{Entropy deficiency of the fair bits of $\ell$ at $N=16$ for the symmetric chain of order one with flip probability $\varepsilon$, against the bound of Proposition~\ref{prop:bwtextract} with $k=1$, in bits.}\label{tab:bwtextract}
\setlength{\tabcolsep}{3pt}
\begin{tabular}{c|c|cc|cccc}
$\varepsilon$ & $\E F$ & $\E F-H(Y\mid R,p)$ & $\E F-H(Y\mid Z,\tau,p)$ & $\E M$ & $\E\log(1/\rho_1)$ & $H(p\mid\tau)$ & bound\\\hline
$0.5$ & $9.08$ & $3.61$ & $6.95$ & $4.22$ & $0.90$ & $2.78$ & $7.90$\\
$0.2$ & $7.06$ & $4.87$ & $6.27$ & $3.96$ & $1.79$ & $2.05$ & $7.80$\\
$0.05$ & $3.14$ & $2.92$ & $3.06$ & $2.10$ & $1.32$ & $0.42$ & $3.85$
\end{tabular}
\end{table}

The flat conditional laws of \eqref{eq:valid} give more than a mean. Let $W=(Z,\tau,p)$ and
\[
D(W)=F-\log\big|\operatorname{supp}P(Y\mid W)\big|=M+\log\frac1{\Pi_\tau(V_{\tau,p}\mid Z)},
\]
the deficiency of the conditional law of $Y$, in entropy and in min-entropy alike since the law is flat.

\begin{proposition}[Exponential tail]\label{prop:extracttail}
Under the assumptions of Proposition~\ref{prop:bwtextract}, for every $t\ge0$,
\[
P\Big\{D(W)>M+\log\frac1{\rho_k(\tau)}+\log(N+1)+t\Big\}\le2^{-t},\qquad\text{hence}\qquad P\{D(W)>c_{N,k}+t\}\le2^{-t}
\]
with $c_{N,k}=13N^\beta(2^k+k-1)^{1-\beta}+(2^k-1)\log N+\log(N+1)$.
\end{proposition}

\begin{proof}
Fix $\tau$, and write $a_{z,p}=\Pi_\tau(V_{\tau,p}\mid Z=z)$ and $q_z=\Pi_\tau(Z=z)$. Counting the distinct pairs $(\ell,p)$ gives $P(Z=z,p\mid\tau)=q_za_{z,p}/\rho_k(\tau)$, whether or not the sets $V_{\tau,p}$ overlap. On the event $a_{z,p}<\rho_k(\tau)2^{-t}/(N+1)$, this probability summed over $z$ and over the at most $N+1$ values of $p$ is at most $2^{-t}$, which is the first bound; the deterministic bounds on $M$ and $\rho_k$ give the second.
\end{proof}

Given $W$, the fair string is thus uniform on at least $2^{F-c_{N,k}-t}$ points except with probability $2^{-t}$, and a two-universal hash with a public seed turns this into a total variation statement, at the price of the seed, of an order bound for the output length and of the large constant $c_{N,k}$ (Proposition~\ref{prop:seeded}, Appendix~\ref{app:flat}). The same appendix records a steganographic reading of Theorem~\ref{thm:piecewise}: under a piecewise exchangeable law, replacing the unmixed fair bits by fresh uniform bits preserves the law of the input exactly.

\subsection{Stationary Sources}

Let $P$ be a stationary binary source, $h_k=H(X_{k+1}\mid X_1\cdots X_k)$, $h_P=\lim_kh_k$ its entropy rate, and $P_k$ its Markov approximation of order $k$, $P_k(x)=P(x_1\cdots x_k)\prod_{j>k}P(x_j\mid x_{j-1}\cdots x_{j-k})$, a Markov law of order $k$ with $P_k(x)>0$ whenever $P(x)>0$. Its divergence from $P$ on $\bits^N$ is explicit:
\begin{multline}\label{eq:markovapprox}
\KL(P\,\|\,P_k)=\E_P\big[-\log P_k(X)\big]-H(X_1\cdots X_N)\\
=H(X_1\cdots X_k)+(N-k)h_k-H(X_1\cdots X_N)\le(N-k)(h_k-h_P),
\end{multline}
since $H(X_1\cdots X_N)=H(X_1\cdots X_k)+\sum_{j>k}H(X_j\mid X_1\cdots X_{j-1})$ and every term of the sum is at least $h_P$. Ergodicity plays no role in what follows.

\begin{theorem}[Nearly fair bits from every stationary source]\label{thm:bwtextract}
For every stationary binary source $P$, every $N$ and every $0\le k<N$, with $Z$, $\tau$ and $p$ of order $k$ as in Proposition~\ref{prop:bwtextract},
\[
\E_PF-H_P(Y\mid Z,\tau,p)\le N(h_k-h_P)+13\,N^\beta(2^k+k-1)^{1-\beta}+(2^k-1)\log N+\log(N+1),
\]
and the same holds for $H_P(Y\mid R,p)$ and $H_P(Y\mid F)$. Consequently $\big(\E_PF-H_P(Y\mid R,p)\big)/N\to0$: the fair bits of $\ell$ carry, up to $o(N)$ bits, their full length in entropy, whatever the memory of the source, and the extractor does not depend on $k$.
\end{theorem}

\begin{proof}
Let $Q$ be the law with the same distribution of $\tau$ as $P$ and uniform on each class: $Q(x)=P(\tau(x))/|C_{\tau(x)}|$. Given $\tau$, both $Q$ and $P_k$ are uniform on $C_\tau$, so by the chain rule $\KL(P\|Q)=\E_\tau\KL(P(\cdot\mid\tau)\|P_k(\cdot\mid\tau))\le\KL(P\|P_k)\le N(h_k-h_P)$, by \eqref{eq:markovapprox}. Write $W=(Z,\tau,p)$; $X$ and $(Y,W)$ determine each other. For every $x$ with $P(x)>0$, with $y=Y(\ell)$, $z=Z(\ell)$ and $w=W(x)$,
\[
\log\big(2^FP(y\mid w)\big)=\log\frac{P(y\mid w)}{Q(y\mid w)}+\log\big(2^FQ(y\mid w)\big),
\]
and the proof of Proposition~\ref{prop:bwtextract}, which uses only the uniformity of $Q$ on the classes, gives $\log(2^FQ(y\mid w))=M+\log(1/\Pi_\tau(V_{\tau,p}\mid Z=z))$ exactly, by \eqref{eq:valid}. We take expectations under $P$. The first term has mean $A_1=\E_P\KL(P(Y\mid W)\|Q(Y\mid W))$. For the second, $Q(z\mid\tau,p)=\Pi_\tau(V_{\tau,p}\mid Z=z)\,\Pi_\tau(Z=z)/\Pi_\tau(V_{\tau,p})$ and $\Pi_\tau(V_{\tau,p})=\rho_k(\tau)Q(p\mid\tau)$ as in that proof, so
\[
\log\frac1{\Pi_\tau(V_{\tau,p}\mid Z=z)}=\log\frac{\Pi_\tau(Z=z)}{P(z\mid\tau,p)}+\log\frac{P(z\mid\tau,p)}{Q(z\mid\tau,p)}+\log\frac1{\rho_k(\tau)}+\log\frac1{Q(p\mid\tau)}.
\]
Under $P$, the first term has mean $-\E_{\tau,p}\KL(P(Z\mid\tau,p)\|\Pi_\tau(Z))\le0$, the second has mean $A_2=\E_{\tau,p}\KL(P(Z\mid\tau,p)\|Q(Z\mid\tau,p))$, and the last has mean $\E_\tau\big[H_P(p\mid\tau)+\KL(P(p\mid\tau)\|Q(p\mid\tau))\big]\le\log(N+1)+A_3$ with $A_3=\E_\tau\KL(P(p\mid\tau)\|Q(p\mid\tau))$. By the chain rule along $x\mapsto(\tau,p,Z,Y)$, and since $P$ and $Q$ give $\tau$ the same law, $A_1+A_2+A_3=\KL(P\|Q)$. Keeping the nonpositive term instead of discarding it gives the exact identity
\begin{equation}\label{eq:deficiencyidentity}
\begin{split}
\E_PF-H_P(Y\mid W)
={}&\KL(P\|Q)+\E_PM+\E_P\log\frac1{\rho_k(\tau)}+H_P(p\mid\tau)\\
 &-\E_{\tau,p}\KL\big(P(Z\mid\tau,p)\,\|\,\Pi_\tau(Z)\big).
\end{split}
\end{equation}
This identity holds for every law $P$ on $\bits^N$, with $Q$ its uniformization within types; stationarity is used only to bound $\KL(P\|Q)$. It identifies the slack in the mean bound as a nonnegative divergence. Dropping that term, bounding $H_P(p\mid\tau)$ by $\log(N+1)$, and using Lemmas~\ref{lem:mixed} and~\ref{lem:markovtype} proves the bound; coarser conditionings as in Proposition~\ref{prop:bwtextract}. For the limit, divide by $N$: the bound is $h_k-h_P+o(1)$ for every fixed $k$, and $h_k\downarrow h_P$.
\end{proof}

There is also a short coding proof of the qualitative conclusion under the coarser conditioning. For any input law, bijectivity gives
\begin{equation}\label{eq:codingdeficiency}
\begin{split}
\E L_{\rm BG}-H(X)
={}&\big[\E F-H(Y\mid R,p)\big]\\
 &+\big[\E L_\gap(R)+\lceil\log(N+1)\rceil-H(R,p)\big].
\end{split}
\end{equation}
Both brackets are nonnegative, the second because the index followed by the residual description is a prefix code for $(R,p)$. Thus the fair-bit deficiency is at most the redundancy of the full code. Theorem~\ref{thm:bwtmarkov} and the Markov approximation already imply that this deficiency is $o(N)$ for every stationary source. The additional content of the type-class argument is the stronger conditioning, removal of the residual-description term $b_N$ from the bound, the exact identity \eqref{eq:deficiencyidentity}, and the flat-source tail of Proposition~\ref{prop:extracttail}.

The theorem is relative to the length $N$. The number of fair bits itself follows the entropy rate, so that the mean deficiency is also small relative to $\E F$ when $h_P>0$:

\begin{corollary}[The entropic rate after the BWT]\label{cor:bwtrate}
For every stationary binary source $P$ and every $0\le k<N$,
\[
H_P(X_1\cdots X_N)-b_N-\log(N+1)\le\E_PF(\ell)\le (N-k)h_k+13\,N^\beta(2^k+k-1)^{1-\beta},
\]
so that $\E_PF(\ell)/N\to h_P$. If $P$ is also ergodic, $F(\ell)/N\to h_P$ almost surely and in $L^1$.
\end{corollary}

\begin{proof}
The pair $(\ell,p)$ determines $X$, so $H(X)=H(R,p)+H(Y\mid R,p)\le H(R)+\log(N+1)+\E F$, and $H(R)\le\log|\Rc_N|\le b_N$ by \eqref{eq:bN}. For the upper bound, Lemma~\ref{lem:certificate} with the partition of order $k$ gives $F(\ell)\le E_k(x)+M$ as in the proof of Theorem~\ref{thm:bwtmarkov}, and $E_k(x)\le-\log P_k(x_{k+1}\cdots x_N\mid x_1\cdots x_k)$ for the conditional probabilities of $P$, by the same argument; the mean of the right-hand side is $(N-k)h_k$, and $H(X_1\cdots X_N)\ge Nh_P$. Divide by $N$, let $N\to\infty$, then $k\to\infty$. In the ergodic case, $L_{\rm BG}=F+L_\gap(R)+\lceil\log(N+1)\rceil$ and $0\le L_\gap(R)\le b_N=o(N)$, so Theorem~\ref{thm:bwtergodic} gives the almost-sure and $L^1$ assertions directly.
\end{proof}

For a nonergodic source only the expected rate is asserted here: a mixture that chooses once between the all-zero process and a fair coin has sample rates zero or one, rather than the entropy rate of the mixture. The lower bound uses nothing about the source: for every law on $\bits^N$, $\E F(\ell)\ge H(X)-b_N-\log(N+1)$, and without the BWT $\E F(X)\ge H(X)-b_N$, which is the inequality behind Theorem~\ref{thm:rate}. Only the upper bound, and the fairness itself, need the memory to be finite or approximable.

\section{Finite-Length Behaviour}\label{sec:exp}

The experiments answer three questions. What does the residual see that the type cannot (Table~\ref{tab:structured})? How far are the finite-length extraction loss and coding regret from the bounds (Tables~\ref{tab:remainders} to~\ref{tab:markov})? And what does a file compressor built from the same pieces achieve against general-purpose tools (Tables~\ref{tab:files} to~\ref{tab:ablation}), a comparison meant to locate it, not to claim a new compressor: it has no match finder and no context mixing. Table~\ref{tab:settings} collects the settings.

\subsection{Implementations and Checks}

A Python implementation contains $\T$ and its inverse, codes B, G and A, the BWT and the tree codes; a C implementation is a file compressor. The Python code checks, for all strings of length at most $13$ ($16\,382$ strings), the round trip, that $F$ and the weight are functions of the residual, that the fair strings over each residual are exactly $\bits^{F}$, the Kraft equality per weight, the charge identity, the sparsity bounds, the prefix property and the round trip of code G, and the run bounds; it checks the two-run formula for $k\le1024$, the pricing certificate on all $43\,690$ pairs of a string of length at most $8$ and a partition of it, and the context partition of Lemma~\ref{lem:bwtintervals} for all strings of length at most $9$ and all orders. The C compressor is tested by unit tests, by round trips under the address and undefined-behaviour sanitisers and by malformed-input tests (a garbage header must be rejected, a truncated stream must not crash), and every compressed size reported below was decoded and compared with its input. The code, with the scripts that produce the tables (\texttt{code/version2\_experiments.py} for Tables~\ref{tab:bernoulli} to~\ref{tab:markov}, \texttt{tool/bench/version2\_files.py} for Table~\ref{tab:files}, which records the hashes of the inputs), is deposited with the paper \cite{code}.

\begin{table}[t]
\caption{Experimental settings, by table.}\label{tab:settings}
\centering\footnotesize
\begin{tabular}{lp{12.6cm}}
\toprule
Table & Setting\\
\midrule
\ref{tab:bernoulli} & Bernoulli strings, $N=2^{10},2^{12},\dots,2^{20}$, $\theta\in\{0.5,0.3,0.1,0.02\}$, five draws per cell from one pseudo-random stream (seed $0$); KT, code B and code A: ideal lengths; code G, code C and the selector: actual bit counts of the encoded stream; \texttt{code/version2\_experiments.py}.\\
\ref{tab:structured} & Strings of length $2^{16}$ whose pseudo-random parts use the fixed seeds of the script; lengths measured as in Table~\ref{tab:bernoulli}.\\
\ref{tab:markov} & $N=2^{12},\dots,2^{18}$, three draws per cell; transition probabilities of the order-$3$ and order-$6$ chains drawn once from Beta$(\frac12,\frac12)$; CTW and BWT-tree codes of depth $8$, ideal lengths; codes G and A on the BWT output measured as in Table~\ref{tab:bernoulli}.\\
\ref{tab:files} & C compressor (\texttt{tool/}): suffix array by SA-IS, BWT of the reversed input, context tree by dynamic programming with the cost of code B (of KT in mode \texttt{kt}), maximal depth $0$, $10$ or $32$ bits as in the header, bit-plane binarisation of the leaves; biases in $32$-bit fixed point, $16$-bit probabilities clipped to $[2^{-16},1-2^{-16}]$ and one binary range coder in modes \texttt{peres}, \texttt{adapt}, \texttt{local} and \texttt{kt}; integer codes only in modes \texttt{gap}, \texttt{gapb}, \texttt{gap2} and \texttt{gap3}; one core of a 12th-generation Intel Core i7-1265U, gcc~13.3 with \texttt{-O2}, one run; actual archive sizes, every archive decoded and compared with its input; \texttt{tool/bench/version2\_files.py}, which records the hashes of the inputs; gzip, bzip2 and xz at level $9$; 7-Zip 23.01 PPMd with \texttt{-m0=PPMd -mx=9 -mmem=256m -mo=8}, sizes including its container.\\
\ref{tab:ptw} & \texttt{code/ptw\_comparison.py}, seed $20261113$. Pieces: $N\in\{2^{12},2^{16}\}$, $J\in\{1,4,16,64\}$, biases alternating between $0.1$ and $0.9$ or between $0.3$ and $0.7$, pieces of equal lengths or cut at $J-1$ distinct uniform points, five draws per setting. BWT: orders $0,1,3,6$, $N=2^{12},2^{16},2^{18}$, three draws per setting, zero initial context, bias $0.3$ at order $0$, biases $(0.1,0.8)$ at order $1$, transition probabilities drawn once from Beta$(\frac12,\frac12)$ (seed $731$) at orders $3$ and $6$, distinct from those of Table~\ref{tab:markov}. Regret against the exact likelihood of each draw; PTW ideal lengths, G actual bit counts, $N$ known to both, the $\lceil\log(N+1)\rceil$-bit index charged to both on the BWT outputs.\\
\ref{tab:ablation} & The streams of mode \texttt{peres} of Table~\ref{tab:files}; ideal lengths of the same residual at exact and at clipped biases computed by \texttt{tool/bench/ablation.py}.\\
\bottomrule
\end{tabular}
\end{table}

\subsection{Structured Strings}

Table~\ref{tab:structured} codes strings of length $N=2^{16}$ (pseudo-random parts with fixed seeds): patternless strings, two halves of different compositions, $0^k1^k$ with $k=2^{15}$, strings of $8$ or $128$ runs, of $8$ or $128$ pieces of random compositions with random cuts, and of $8$ or $128$ periodic blocks of random periods at most $8$. The benchmark is a structural score, not a code: for pieces, the sum of $n_ih(w_i/n_i)+\log N$ over the pieces; for runs, $\log N$ per run; for periodic blocks, the period plus $\log N$ per block. KT and code B see only the weight and pay the global empirical entropy. Codes G and A (Appendix~\ref{app:logcoders}) see the structure that the transform puts in the residual: $32$ and $44$ bits for $0^k1^k$, against $65\,553$ for code B; $598$ bits for $8$ runs; within $2$ to $30\%$ of the benchmark on the halves and the pieces. Neither sees periodic blocks, whose fair bits are periodic and stored as they are: this row is a limitation (Section~\ref{sec:limits}), not a success. On the structured strings the plug-in biases of code C are wrong and the selector never chooses it; it is at most two bits above code G, and sometimes appreciably below it: by $15.6\%$ on $0^k1^k$ ($27$ against $32$ bits) and $5.5\%$ on eight runs ($565$ against $598$), where the delta codes win, and by $1$ to $3\%$ on the patternless strings, where code C wins.

\begin{table}[t]
\caption{Structured strings of length $2^{16}$: structural benchmark, lengths in bits of KT and code B (ideal), number $F$ of fair bits and $z$ of residual ones, and lengths of code G and of the selector of Appendix~\ref{app:logcoders} (actual) and of code A (ideal).}\label{tab:structured}
\centering\footnotesize\setlength{\tabcolsep}{4pt}
\begin{tabular}{l|r|rr|rr|rrr}
\toprule
string & benchmark & KT & B & $F$ & $z$ & G & A & selector\\
\midrule
patternless, 30\% ones & 57616 & 57609 & 57617 & 55344 & 1044 & 58115 & 57715 & 57680\\
patternless, 2\% ones & 9293 & 9285 & 9294 & 8830 & 193 & 9620 & 9393 & 9361\\
two halves, 5\% then 50\% ones & 41822 & 55601 & 55610 & 40309 & 969 & 46641 & 45736 & 46594\\
two halves, 10\% then 90\% ones & 31113 & 65544 & 65553 & 29851 & 581 & 31726 & 33942 & 31728\\
$0^k1^k$ & 32 & 65544 & 65553 & 1 & 1 & 32 & 44 & 27\\
8 runs & 128 & 54714 & 54722 & 304 & 40 & 598 & 915 & 565\\
128 runs & 2048 & 64988 & 64996 & 3503 & 165 & 4384 & 5231 & 4382\\
8 pieces, random compositions & 15564 & 62301 & 62309 & 15584 & 462 & 20238 & 19216 & 19922\\
128 pieces, random compositions & 42908 & 65340 & 65348 & 43831 & 1019 & 48394 & 49602 & 48396\\
8 periodic blocks & 169 & 58635 & 58644 & 33254 & 231 & 35500 & 35280 & 35351\\
128 periodic blocks & 2657 & 65087 & 65096 & 38403 & 557 & 41747 & 42359 & 41692\\

\bottomrule
\end{tabular}
\end{table}

\subsection{Extraction and Bernoulli Regret}\label{sec:expbern}

For $N=2^{10},2^{12},\dots,2^{20}$ and $\theta\in\{0.5,0.3,0.1,0.02\}$ we draw five strings per cell (one pseudo-random stream, seed $0$). Table~\ref{tab:bernoulli} gives the fraction of fair bits, the number of residual ones and the regrets. The fraction $F/N$ converges to $h(\theta)$ slowly: at $N=2^{20}$ the extraction loss is $1.7$, $1.6$, $1.7$ and $2.2\%$ of $h(\theta)$. The column $F_{\rm E}/N$ gives, for the same draws, the expected yield of Elias's extractor given the weight, $\sum_jj2^j/\binom Nw$ over the set bits $j$ of $\binom Nw$ \cite{Elias72}, which is $\log\binom Nw$ minus $0.7$ to $1.3$ bits in every cell: Elias loses $O(\log N)$ bits where the exhaustive transform loses a percentage at these lengths (Section~\ref{sec:disc}; \cite{PKS18} make the same comparison numerically). Table~\ref{tab:dispersion} gives the sample standard deviations over the five draws at $N=2^{16}$ and $2^{20}$: below $2$ bits for KT and code B, $15$ to $83$ bits for code G, and $1$ to $19$ bits for codes A and C and the selector, so that differences of a few bits between codes A and C are within the noise of five draws, whereas the growth of the regret of code G with $N$ is not. The number of residual ones is far below the bound of Theorem~\ref{thm:sparse}: $8\,808$ at $\theta=\frac12$ and $N=2^{20}$, against $\sqrt3\,N^\beta\approx1.02\cdot10^5$. Code B is about $\log N$ bits above $-\log\Pr_\theta(s)$ and $\frac12\log N$ above KT, consistently with Theorem~\ref{thm:B}. The regret of code G is polynomial, as Theorem~\ref{thm:universal} allows: $1\,465$ to $3\,184$ bits at $N=2^{20}$, $0.2$ to $1\%$ of $Nh(\theta)$. Code A, which uses the weight to sort the residual bits but learns their frequencies, stays within $35$ to $175$ bits; code C, with the same weight but guarded Rice codes instead of KT estimators, within $33$ to $156$, and the selector adds its two bits except on fair coin flips, where it stores the residual raw and costs $N+2$ bits exactly.

Table~\ref{tab:remainders} sets the proved remainder beside these measurements. The measured regret of code G against the type class, $L_G-\log\binom Nw$, is three to four orders of magnitude below $\bar b_N$, and it grows like a power of $N$, with rough slopes $0.57$ to $0.70$ between $N=2^{18}$ and $2^{20}$. Part of the slack is in Theorem~\ref{thm:sparse}: the largest measured $z$ is $12$ times below $\sqrt3\,N^\beta$ at $N=2^{20}$ and grows like $N^{0.70}$, close to the exponent of open problem~(i) of Section~\ref{sec:disc}, whereas Lemma~\ref{lem:gap} evaluated at the measured $z$ is within a factor of two of the measured cost of the gaps. The rest is a cancellation that the bound ignores: the regret decomposes as
\[
L_G-\log\binom Nw=L_\gap(R)-\Big(\log\binom Nw-F\Big),
\]
the cost of the residual minus the within-type extraction deficit, and the deficit pays back $68$ to $88\%$ of the cost of the gaps at $N=2^{20}$ ($12$ to $32\%$ of that cost remains as regret, the fraction growing as $\theta$ decreases); a substantial extraction deficit is thus not a substantial coding loss. The exact number of residuals, computed by the recursion of Proposition~\ref{prop:image} on the counts by weight up to $N=2^{12}$, gives $\log|\Rc_N|/N=0.188$ at $N=2^{10}$ and $0.133$ at $N=2^{12}$, decreasing like $N^{-0.25}$ over that range, consistent with the exponent and far below the constant. The finite-length behaviour of code G is thus an empirical observation that the theory bounds only loosely, and nothing excludes a much smaller remainder. Code B, by contrast, is within $21$ bits at $N=2^{20}$.

\begin{table}[t]
\caption{Proved against measured remainders, over the draws of Table~\ref{tab:bernoulli}: largest number of residual ones over the four biases, the bound $\sqrt3\,N^\beta$ of Theorem~\ref{thm:sparse}, the mean regret $L_G-\log\binom Nw$ of code G against the type class for $\theta=0.5,0.3,0.1,0.02$, and the bound $\bar b_N$ of \eqref{eq:bN}, in bits.}\label{tab:remainders}
\centering\footnotesize\setlength{\tabcolsep}{4pt}
\begin{tabular}{r|rr|rrrr|r}
\toprule
$N$ & $\max z$ & $\sqrt3\,N^\beta$ & \multicolumn{4}{c|}{$L_G-\log\binom Nw$} & $\bar b_N$\\
 & & & $\theta=0.5$ & $0.3$ & $0.1$ & $0.02$ & \\
\midrule
$2^{10}$ & 77 & 421 & 27 & 45 & 49 & 37 & $2.0\cdot10^{4}$\\
$2^{12}$ & 185 & 1263 & 48 & 89 & 123 & 81 & $7.2\cdot10^{4}$\\
$2^{14}$ & 500 & 3788 & 132 & 225 & 246 & 155 & $2.5\cdot10^{5}$\\
$2^{16}$ & 1304 & 11364 & 350 & 537 & 504 & 271 & $8.6\cdot10^{5}$\\
$2^{18}$ & 3397 & 34092 & 901 & 1258 & 1227 & 626 & $2.9\cdot10^{6}$\\
$2^{20}$ & 8848 & 102276 & 2388 & 3194 & 2691 & 1475 & $9.7\cdot10^{6}$\\
\bottomrule
\end{tabular}
\end{table}

\begin{table}[t]
\caption{Bernoulli sources, averages over five draws per cell: fraction of fair bits, expected fraction $F_{\rm E}/N$ of fair bits of Elias's extractor given the weight (Section~\ref{sec:disc}), number $z$ of residual ones, and regret in bits against $-\log\Pr_\theta(s)$ of KT and code B (ideal lengths), code G, code C and the selector of Appendix~\ref{app:logcoders} (actual bit counts) and code A (ideal length).}\label{tab:bernoulli}
\centering\footnotesize\setlength{\tabcolsep}{4pt}
\begin{tabular}{rr|rrr|rrrrrr}
\toprule
$\theta$ & $N$ & $F/N$ & $F_{\rm E}/N$ & $z$ & KT & B & G & A & C & selector\\
\midrule
0.50 & $2^{10}$ & 0.8615 & 0.9917 & 71 & 3.5 & 9.2 & 20 & 18 & 20 & 2\\
0.50 & $2^{12}$ & 0.9072 & 0.9981 & 180 & 6.1 & 12.8 & 42 & 20 & 15 & 2\\
0.50 & $2^{14}$ & 0.9401 & 0.9994 & 489 & 6.5 & 14.2 & 123 & 24 & 19 & 2\\
0.50 & $2^{16}$ & 0.9609 & 0.9998 & 1287 & 7.0 & 15.7 & 340 & 29 & 24 & 2\\
0.50 & $2^{18}$ & 0.9743 & 1.0000 & 3343 & 8.4 & 18.1 & 891 & 31 & 24 & 2\\
0.50 & $2^{20}$ & 0.9832 & 1.0000 & 8808 & 9.7 & 20.3 & 2377 & 35 & 33 & 2\\
0.30 & $2^{10}$ & 0.7656 & 0.8746 & 55 & 4.2 & 9.9 & 38 & 53 & 51 & 40\\
0.30 & $2^{12}$ & 0.8074 & 0.8853 & 150 & 5.7 & 12.4 & 82 & 73 & 54 & 56\\
0.30 & $2^{14}$ & 0.8309 & 0.8834 & 397 & 6.7 & 14.3 & 217 & 92 & 67 & 69\\
0.30 & $2^{16}$ & 0.8472 & 0.8809 & 1049 & 8.0 & 16.7 & 528 & 115 & 81 & 83\\
0.30 & $2^{18}$ & 0.8597 & 0.8819 & 2815 & 7.7 & 17.4 & 1247 & 132 & 94 & 96\\
0.30 & $2^{20}$ & 0.8670 & 0.8815 & 7466 & 10.0 & 20.6 & 3184 & 160 & 119 & 121\\
0.10 & $2^{10}$ & 0.4080 & 0.4724 & 27 & 4.7 & 10.4 & 44 & 66 & 53 & 45\\
0.10 & $2^{12}$ & 0.4264 & 0.4697 & 80 & 5.6 & 12.3 & 116 & 84 & 60 & 62\\
0.10 & $2^{14}$ & 0.4409 & 0.4695 & 209 & 7.2 & 14.9 & 240 & 103 & 74 & 76\\
0.10 & $2^{16}$ & 0.4508 & 0.4701 & 567 & 8.0 & 16.7 & 496 & 121 & 87 & 89\\
0.10 & $2^{18}$ & 0.4551 & 0.4677 & 1504 & 8.6 & 18.3 & 1218 & 144 & 116 & 118\\
0.10 & $2^{20}$ & 0.4612 & 0.4694 & 4045 & 9.1 & 19.7 & 2681 & 175 & 156 & 158\\
0.02 & $2^{10}$ & 0.0980 & 0.1165 & 8 & 3.9 & 9.5 & 33 & 61 & 55 & 33\\
0.02 & $2^{12}$ & 0.1252 & 0.1400 & 25 & 5.4 & 12.1 & 75 & 80 & 65 & 64\\
0.02 & $2^{14}$ & 0.1270 & 0.1373 & 68 & 6.8 & 14.4 & 149 & 98 & 76 & 78\\
0.02 & $2^{16}$ & 0.1347 & 0.1415 & 185 & 8.0 & 16.6 & 264 & 114 & 87 & 89\\
0.02 & $2^{18}$ & 0.1369 & 0.1415 & 512 & 9.1 & 18.8 & 618 & 134 & 96 & 98\\
0.02 & $2^{20}$ & 0.1383 & 0.1413 & 1395 & 9.1 & 19.8 & 1465 & 156 & 121 & 123\\
\bottomrule
\end{tabular}
\end{table}

\begin{table}[t]
\caption{Dispersion of Table~\ref{tab:bernoulli}: sample standard deviations, over the five draws, of the regrets of KT, code B, code G, code A, code C and the selector, at $N=2^{16}$ and $2^{20}$.}\label{tab:dispersion}
\centering\footnotesize\setlength{\tabcolsep}{4pt}
\begin{tabular}{rr|rrrrrr}
\toprule
$\theta$ & $N$ & KT & B & G & A & C & selector\\
\midrule
0.50 & $2^{16}$ & 1.9 & 1.9 & 18.8 & 0.8 & 9.6 & 0.0\\
0.50 & $2^{20}$ & 0.7 & 0.7 & 41.1 & 1.7 & 2.5 & 0.0\\
0.30 & $2^{16}$ & 0.4 & 0.4 & 21.3 & 3.1 & 4.8 & 4.8\\
0.30 & $2^{20}$ & 0.1 & 0.1 & 82.6 & 6.4 & 5.5 & 5.5\\
0.10 & $2^{16}$ & 0.3 & 0.3 & 43.1 & 6.1 & 11.7 & 11.7\\
0.10 & $2^{20}$ & 1.5 & 1.5 & 37.4 & 5.9 & 12.3 & 12.3\\
0.02 & $2^{16}$ & 0.4 & 0.4 & 15.4 & 6.2 & 8.8 & 8.8\\
0.02 & $2^{20}$ & 1.6 & 1.6 & 58.4 & 8.5 & 18.6 & 18.6\\
\bottomrule
\end{tabular}
\end{table}

\subsection{Binary Markov Sources}

Sources, with one fixed transition matrix per chain across the draws: i.i.d.\ ($\theta=0.3$); an order-$1$ chain with $\Pr(1\mid0)=0.1$, $\Pr(1\mid1)=0.8$; order-$3$ and order-$6$ chains with transition probabilities drawn once from Beta$(\frac12,\frac12)$. Table~\ref{tab:markov} gives regrets against the true likelihood (ideal lengths for CTW, conditional KT and the tree codes, three draws per cell). On these sources and lengths, sequential conditional KT with the true order is the best of the codes tested, and CTW of depth $8$ is within a factor $1.4$ of it without knowing the order. The BWT-tree code with KT leaves costs $3.2$ to $4.1$ times conditional KT at $N=2^{18}$; for order $3$ the bound of Proposition~\ref{prop:oracle} is $236$ bits, $133$ of them for the split points, against $205$ observed. Codes G and A applied to the BWT output, with no tree, are universal but their mean regret grows much faster with $N$ than that of the tree codes: at $N=2^{18}$, the mean regret of code G is $0.6$, $2.8$, $16$ and $15\%$ of $NH$ for orders $0$, $1$, $3$ and $6$. On binary Markov sources the code without context tree is not competitive; its interest is that it needs no context model and has a guarantee on every individual sequence. The sample standard deviations over the three draws at $N=2^{18}$ are $0$ to $7$ bits for CTW, $0$ to $9$ for conditional KT, $0$ to $14$ and $0$ to $16$ for the tree codes, and $37$ to $99$ and $2$ to $245$ bits for codes G and A on the BWT output, over the four sources; the differences discussed here are far outside these ranges.

\begin{table}[t]
\caption{Binary sources with memory: average regret in bits against the true likelihood, three draws per cell; $NH$ is $N$ times the entropy rate of the stationary chain. CTW of depth $8$; cKT$(k)$: sequential conditional KT of the true order $k$; tree-KT, tree-B: BWT-tree code of depth $8$ with KT or code-B leaves; BG, BA: codes G and A applied to the BWT output, no tree.}\label{tab:markov}
\centering\footnotesize\setlength{\tabcolsep}{4pt}
\begin{tabular}{l|r|r|rr|rr|rr}
\toprule
source & $N$ & $NH$ & CTW & cKT$(k)$ & tree-KT & tree-B & BG & BA\\
\midrule
iid ($k=0$) & $2^{12}$ & 3610 & 6 & 6 & 20 & 26 & 102 & 89\\
iid ($k=0$) & $2^{14}$ & 14439 & 7 & 6 & 22 & 29 & 217 & 110\\
iid ($k=0$) & $2^{16}$ & 57756 & 9 & 8 & 26 & 35 & 556 & 132\\
iid ($k=0$) & $2^{18}$ & 231025 & 10 & 9 & 29 & 39 & 1299 & 156\\
M1 ($k=1$) & $2^{12}$ & 2266 & 14 & 11 & 40 & 52 & 256 & 361\\
M1 ($k=1$) & $2^{14}$ & 9065 & 15 & 12 & 45 & 59 & 584 & 887\\
M1 ($k=1$) & $2^{16}$ & 36261 & 18 & 15 & 52 & 68 & 1515 & 2554\\
M1 ($k=1$) & $2^{18}$ & 145046 & 20 & 17 & 58 & 75 & 4081 & 7909\\
M3 ($k=3$) & $2^{12}$ & 1546 & 35 & 28 & 101 & 126 & 664 & 624\\
M3 ($k=3$) & $2^{14}$ & 6183 & 48 & 37 & 136 & 171 & 1803 & 1564\\
M3 ($k=3$) & $2^{16}$ & 24731 & 58 & 45 & 181 & 226 & 5385 & 4507\\
M3 ($k=3$) & $2^{18}$ & 98925 & 66 & 52 & 205 & 262 & 15781 & 13172\\
M6 ($k=6$) & $2^{12}$ & 1570 & 173 & 135 & 538 & 648 & 851 & 846\\
M6 ($k=6$) & $2^{14}$ & 6280 & 261 & 191 & 768 & 934 & 1957 & 1910\\
M6 ($k=6$) & $2^{16}$ & 25121 & 335 & 249 & 1015 & 1252 & 5444 & 5463\\
M6 ($k=6$) & $2^{18}$ & 100485 & 408 & 305 & 1256 & 1579 & 15528 & 15662\\

\bottomrule
\end{tabular}
\end{table}

\subsection{Comparison With Partition Tree Weighting}\label{sec:ptw}

The natural partition-adaptive benchmark for Theorem~\ref{thm:concat} is PTW with a KT base model \cite{VWBG13}. For a known horizon $N=2^d$ it is the mixture $Q_d(s)=\frac12\KT(|s|,\wt(s))+\frac12Q_{d-1}(s_{\rm left})Q_{d-1}(s_{\rm right})$, $Q_0(b)=\frac12$, which \texttt{code/ptw\_comparison.py} evaluates in the log domain (checked against the direct mixture, and for its total mass, through $N=8$). Table~\ref{tab:ptw} compares its ideal length with the actual length of code G, both knowing $N$, on two inputs: piecewise Bernoulli strings of $J$ pieces of alternating biases $0.1$ and $0.9$, cut at $J-1$ uniform random points, against the true likelihood with the cuts known; and the BWT outputs $\ell$ of binary Markov sources, the same string for both coders, with the index of the sentinel charged to both (settings in Table~\ref{tab:settings}).

\begin{table}[t]
\centering\footnotesize
\caption{Selected PTW comparisons: mean regret in bits, and sample standard deviation of the paired difference G minus PTW. Piecewise rows use unequal segment lengths and alternating biases $(0.1,0.9)$; BWT rows use the same transformed string for both coders.}\label{tab:ptw}
\begin{tabular}{lrr|rrr}
\toprule
Input & $N$ & $J$ or order & G & PTW-KT & paired SD\\
\midrule
pieces & $2^{12}$ & $J=4$ & 235 & 93 & 35\\
pieces & $2^{12}$ & $J=16$ & 417 & 273 & 38\\
pieces & $2^{12}$ & $J=64$ & 839 & 695 & 45\\
pieces & $2^{16}$ & $J=4$ & 781 & 188 & 46\\
pieces & $2^{16}$ & $J=16$ & 1226 & 637 & 21\\
pieces & $2^{16}$ & $J=64$ & 2419 & 1821 & 103\\
\midrule
BWT & $2^{18}$ & $k=0$ & 1267 & 29 & 48\\
BWT & $2^{18}$ & $k=1$ & 4136 & 116 & 125\\
BWT & $2^{18}$ & $k=3$ & 11842 & 331 & 68\\
BWT & $2^{18}$ & $k=6$ & 17304 & 1866 & 78\\
\bottomrule
\end{tabular}
\end{table}

PTW has the smaller mean regret in every setting of the grid. In the displayed rows, the mean regret of code G is $1.2$ to $4.2$ times larger on the pieces, where equal-length pieces favour the dyadic partitions of PTW even more ($33$, $120$ and $415$ bits against $789$, $1134$ and $2264$ for $J=4,16,64$ at $N=2^{16}$), and $9$ to $44$ times larger on the BWT outputs, where the comparison on identical strings places the gap of Table~\ref{tab:markov} in the second stage, not in the suffix sort. These measurements locate code G as Table~\ref{tab:markov} does; they show no compression advantage over PTW. The full results, with equal-length pieces and with the biases $0.3$ and $0.7$, are in \texttt{code/results/v3\_ptw.csv}.

\subsection{A File Compressor}

Our C compressor computes the suffix array by SA-IS \cite{NZC11}, the BWT of the reversed input, a context tree by dynamic programming with contexts refined bit by bit up to a given depth, and the bit-plane binarisation of the leaves (Section~\ref{sec:text}); every binary sequence is then transformed and coded in one of the following modes. Mode \texttt{peres} is the clipped implementation of code B: the biases, in $32$-bit fixed point, are handed to a range coder as $16$-bit probabilities clipped to $[2^{-16},1-2^{-16}]$, which changes the code (Table~\ref{tab:ablation}). Mode \texttt{gap} is code G, and mode \texttt{gapb} is code G with the gamma codes restricted by the budget of Appendix~\ref{app:residual}; the archive of mode \texttt{gap} is code G up to its container, which costs $22$ bytes on the four books at depth $0$ ($779\,188$ bytes against $779\,166$ for the sum of the lengths $L_G$ of the $171$ bit-plane sequences), so that Table~\ref{tab:files} measures code G itself. Mode \texttt{gap2} is a logarithmic gap coder with $128$ classes, by the numbers of $\vstep$ and $\ustep$ steps of the path of each residual bit (capped at $15$ and $7$), continued over all the sequences of the file; each class is coded with an Exp-Golomb code whose order is the mean of $\lfloor\log(g+1)\rfloor$ over the past gaps of the class, minus one, the sums being halved every $64$ gaps and the order capped at $40$, so that $c_0=41$ and $c_1=2$ in Theorem~\ref{thm:logcoders}; the caps $15$ and $7$ cover the depths at which residual ones occur on the files, the halving every $64$ gaps is a forgetting horizon that lets the order follow the drift of the gaps along a sequence, and the cap $40$ is the logarithm of the largest possible gap. Mode \texttt{gap3} is mode \texttt{gap2} with, in each class, a choice between two code trees for each gap at the adaptive order: the Exp-Golomb word, or a guarded Rice word (unary quotient up to $7$, then an escape to a gamma code); the tree whose words would have cost less on the class so far is used, so no choice bit is sent. Mode \texttt{adapt} is code A with its classes shared over the file. Mode \texttt{local} is code A in which the bias of each $\vstep$-child, and of each odd remainder, is re-estimated from the decoded $\ustep$-sibling, respectively the decoded pairs of its node, when a chi-square test (a two-part MDL test with a penalty of $\frac12\log(n+1)$ and a shrinkage towards the inherited bias) rejects the inherited bias. Mode \texttt{kt} codes each binary sequence with adaptive KT, without the transform. Except in modes \texttt{gap}, \texttt{gapb}, \texttt{gap2} and \texttt{gap3}, which use no arithmetic coder, one binary range coder carries everything, and the KT probabilities of modes \texttt{kt}, \texttt{adapt} and \texttt{local} are also clipped to the $16$-bit grid. The tree is chosen with the cost of code B (of KT in mode \texttt{kt}); the guarantees of Sections~\ref{sec:G} and~\ref{sec:memory} concern the ideal codes, not this heuristic choice nor the finite precision of the probabilities. Inputs: \emph{Alice in Wonderland}, the concatenation of four Project Gutenberg books, \emph{Alice} among them ($2\,614\,437$ bytes), the file samba of the Silesia corpus \cite{Silesia} (source code), the executable \texttt{/bin/bash} of Ubuntu 24.04, $1$\,MiB of pseudo-random bytes, and a $2$\,MiB binary Markov source of order $3$ packed into bytes, of entropy rate $3.019$ bits per byte.

\begin{table}[t]
\caption{File compressor, bits per byte (actual sizes, round trips verified, every column produced by \texttt{tool/bench/version2\_files.py}, which records the hashes of the inputs), at the maximal context depth in bits given in parentheses; depth $0$ is no tree at all.}\label{tab:files}
\centering\footnotesize\setlength{\tabcolsep}{1.5pt}
\begin{tabular}{l|rrrrrr|rrrrr|rrrr}
\toprule
& \multicolumn{6}{c|}{one binary range coder} & \multicolumn{5}{c|}{integer codes only} & \multicolumn{4}{c}{standard compressors}\\
\cmidrule(lr){2-7}\cmidrule(lr){8-12}\cmidrule(lr){13-16}
file & \texttt{peres} & \texttt{peres} & \texttt{kt} & \texttt{kt} & \texttt{adapt} & \texttt{local} & \texttt{gap} & \texttt{gap} & \texttt{gapb} & \texttt{gap2} & \texttt{gap3} & gzip & bzip2 & xz & PPMd\\
 & ($0$) & ($32$) & ($0$) & ($32$) & ($0$) & ($10$) & ($0$) & ($10$) & ($10$) & ($10$) & ($10$) & \texttt{-9} & \texttt{-9} & \texttt{-9} & o8\\
\midrule
\emph{Alice} & 3.808 & 2.807 & 4.676 & 2.602 & 2.770 & 2.531 & 2.733 & 2.768 & 2.703 & 2.621 & 2.611 & 2.822 & 2.262 & 2.521 & 2.039\\
four books & 3.636 & 2.383 & 4.584 & 2.296 & 2.418 & 2.222 & 2.384 & 2.341 & 2.333 & 2.262 & 2.258 & 3.045 & 2.298 & 2.421 & 1.975\\
samba & 4.785 & 2.148 & 6.093 & 2.110 & 2.066 & 1.879 & 2.012 & 1.986 & 1.980 & 1.896 & 1.890 & 2.002 & 1.685 & 1.394 & 1.476\\
bash & 5.276 & 3.951 & 6.202 & 3.729 & 4.081 & 3.811 & 4.039 & 3.998 & 3.892 & 3.762 & 3.736 & 3.826 & 3.594 & 3.229 & 2.899\\
random & 8.004 & 8.004 & 8.002 & 8.002 & 8.004 & 8.004 & 8.043 & 8.043 & 8.042 & 8.042 & 8.000 & 8.001 & 8.040 & 8.001 & 8.211\\
Markov(3) & 4.104 & 3.042 & 4.516 & 3.030 & 3.231 & 3.043 & 3.142 & 3.105 & 3.096 & 3.109 & 3.105 & 3.973 & 3.705 & 3.644 & 3.301\\
\bottomrule
\end{tabular}
\end{table}

Table~\ref{tab:files} shows the trade-off of the previous sections on real data. Without a context tree, mode \texttt{kt} codes each bit-plane sequence of the whole BWT output through its counts and pays essentially the order-$0$ cost, $4.584$ bits per byte on the four books, which is also, within $0.001$, the ideal length of exact code B, whose length depends only on the weights (Table~\ref{tab:ablation}, computed by \texttt{tool/bench/ablation.py} on the stream of mode \texttt{peres}). The clipped implementation costs $3.636$ there: the ideal length of the same residual at the exact biases clipped to $2^{-16}$ is $3.641$, so clipping accounts for the bulk of the difference, the fixed-point biases and the coder for the rest. Clipping caps the price of the residual events that the plug-in model declares nearly impossible and that structured data produce, a crude form of what codes G and A do systematically: with no tree at all, code G codes the four books at $2.384$, and the clipped code B needs a context tree of $32$ bits to reach $2.383$, where exact code B is at $2.417$. On the texts and on samba, the logarithmic coders with a short tree are the most efficient of our modes: mode \texttt{gap2}, without arithmetic coder, codes the four books at $2.262$, below bzip2 and xz, and mode \texttt{local} at $2.222$. Mode \texttt{gap3} improves on mode \texttt{gap2} by $0.1$ to $0.7\%$ on every file. On \emph{Alice}, mode \texttt{gap2} is above mode \texttt{kt} with its deep tree, on the executable mode \texttt{kt} is the best of our modes, and on the binary Markov file modes \texttt{kt} and \texttt{peres} are slightly ahead. On source code and the executable our compressor stays behind bzip2 and xz, which exploit long exact repetitions, though ahead of gzip; on random bytes the gamma codes lose $0.5\%$, because they spend more than one bit per nearly fair residual bit, a loss that mode \texttt{gap3} removes. A PPM compressor \cite{CW84} is ahead on the texts, the source code and the executable: 7-Zip's PPMd (order $8$, $256$\,MB of memory, last column) codes the four books at $1.975$, \emph{Alice} at $2.039$, samba at $1.476$ and bash at $2.899$ bits per byte, $11$ to $22\%$ below the best of our modes, which have no context mixing and no match finder; on the binary Markov file the binary-context modes are ahead of it ($3.030$ against $3.301$), and on random bytes it expands by $2.6\%$. The comparison locates the compressor, as said at the start of this section. Table~\ref{tab:speed} gives the speeds recorded with the sizes by the same script (decimal megabytes per second, one core of a 12th-generation Intel Core i7-1265U, gcc~13.3 with \texttt{-O2}, one run; with frequency scaling, repeated runs vary by up to a factor of two): the modes of the tool compress the two large files at $3$ to $9$\,MB/s and decompress them at $3$ to $9$\,MB/s, against $17$ and $34$\,MB/s for bzip2 on the four books; these are single-run pipeline measurements, not timed by component, and they support orders of magnitude, not a ranking.

\begin{table}[t]
\caption{Compression (c) and decompression (d) speeds in MB/s (one core of a 12th-generation Intel Core i7-1265U, gcc~13.3 with \texttt{-O2}, one run; \texttt{tool/bench/v2\_files.csv}); context depth in bits in parentheses.}\label{tab:speed}
\centering\footnotesize\setlength{\tabcolsep}{3.5pt}
\begin{tabular}{l|rr|rr|rr|rr|rr|rr|rr|rr|rr}
\toprule
file & \multicolumn{2}{c|}{\texttt{peres} (32)} & \multicolumn{2}{c|}{\texttt{kt} (32)} & \multicolumn{2}{c|}{\texttt{gap} (10)} & \multicolumn{2}{c|}{\texttt{gap3} (10)} & \multicolumn{2}{c|}{\texttt{local} (10)} & \multicolumn{2}{c|}{gzip} & \multicolumn{2}{c|}{bzip2} & \multicolumn{2}{c|}{xz} & \multicolumn{2}{c}{PPMd}\\
 & c & d & c & d & c & d & c & d & c & d & c & d & c & d & c & d & c & d\\
\midrule
four books & 5.3 & 6.7 & 6.2 & 8.7 & 5.6 & 5.3 & 4.3 & 5.9 & 4.4 & 4.1 & 15.5 & 174 & 17.3 & 34 & 2.7 & 87 & 8.4 & 6.4\\
samba & 3.6 & 6.7 & 4.2 & 9.1 & 3.5 & 4.1 & 2.9 & 4.0 & 2.8 & 2.8 & 21 & 193 & 17.0 & 52 & 3.6 & 112 & 12.5 & 11.1\\
\bottomrule
\end{tabular}
\end{table}

\begin{table}[t]
\caption{Exact and clipped code B on the BWT output, bits per byte, at depths $0$ and $32$: mode \texttt{peres} as shipped (actual size); ideal length of the same residual at the exact biases clipped to $2^{-16}$; ideal length of exact code B; modes \texttt{adapt} and \texttt{kt} (actual sizes).}\label{tab:ablation}
\centering\footnotesize\setlength{\tabcolsep}{4pt}
\begin{tabular}{lr|r|r|r|rr}
\toprule
& & actual & \multicolumn{2}{c|}{ideal length} & \multicolumn{2}{c}{actual}\\
\cmidrule(lr){3-3}\cmidrule(lr){4-5}\cmidrule(lr){6-7}
file & depth & \texttt{peres} & clipped & exact B & \texttt{adapt} & \texttt{kt}\\
\midrule
\emph{Alice} & 0 & 3.808 & 3.821 & 4.680 & 2.770 & 4.676\\
 & 32 & 2.807 & 2.814 & 2.837 & 2.739 & 2.602\\
four books & 0 & 3.636 & 3.641 & 4.584 & 2.418 & 4.584\\
 & 32 & 2.383 & 2.388 & 2.417 & 2.339 & 2.296\\
\bottomrule
\end{tabular}
\end{table}

\section{Discussion}\label{sec:disc}

\emph{Peres versus Elias.} Elias's extractor \cite{Elias72}, with the type and the index of the interval of the binary expansion of the class size as its residual, is also a bijection whose residual determines the type, and Theorem~\ref{thm:uniform} holds for it verbatim. By the Kraft equality per weight, both procedures partition each type class into dyadic blocks. Elias retains the type and the identifier of the dyadic block, one block per nonzero binary digit of $\binom Nw$, and outputs a uniform index within the block; this decomposition into block sizes maximises the expected number of output bits on a type class, at least $\log\binom Nw-2$. Peres can have a smaller yield at finite length: at $N=2^{20}$ the exhaustive transform still loses $1.6$ to $2.2\%$ of the entropy (column $F/N$ of Table~\ref{tab:bernoulli}), where the loss of Elias's procedure is $O(\log N)$ bits (column $F_{\rm E}/N$). In exchange its blocks come from a local recursion: $O(N\log N)$ bit operations with no multiplication instead of the large integers of the index, explicit biases for the residual bits, which make code B possible, and a sparse residual, which makes code G and its structural bounds possible. The standard uniform coding of a type index prices every arrangement of a type class alike; rankings or block partitions that favour simple arrangements are another possibility.

\emph{Which code when.} The codes of this paper are not in competition with one another. Code B, or the KT residual law of Remark~\ref{rem:E2}, is the code for data that are exchangeable or close to it: its Bernoulli regret is at most $\log(N+1)+1$ ideal bits, and the KT residual law needs no weight. Code G, and more generally a logarithmic residual coder, is the code when nothing is known about the structure, when no estimate is to be made and no arithmetic coder is wanted, or when the structure lies in runs and in pieces, as it does in the Burrows--Wheeler output of text: its remainder is polynomial, $O(N^{0.793}\log N)$, but it is paid once for every structure at the same time, and one mode bit selecting the shorter of the two codes gives both guarantees (Section~\ref{sec:logcoders}). For sources with memory of low order and at finite lengths, the tree code of Section~\ref{sec:memory}, universal as well (Corollary~\ref{cor:treeergodic}), and sequential coding with context remain far more efficient, at the price of a context model, as Table~\ref{tab:markov} shows; the code without tree is an option when no such model is available: its guarantee on every individual sequence is obtained without selecting or transmitting a context tree, and its full pipeline has not been shown cheaper than model-based universal coders.

\emph{Open problems.} (i) \emph{The growth of the residual weight.} An exact dynamic programme over lengths and weights, since $z(s)=z(u)+z(v)+\wt(r)$ with $u$ and $v$ free given their lengths and weights, gives the largest number of residual ones at a given length; at the lengths $2^e-1$ and $2^e$, $9\le e\le14$, it lies between $0.87$ and $0.95$ times $N^{\log((1+\sqrt5)/2)}=N^{0.6942}$, and the ratio rises slowly with $N$: the exponent fitted between $2^{10}$ and $2^{14}$ is $0.71$ to $0.72$, and $0.69$ to $0.73$ between consecutive doublings; the programme (\texttt{code/maxz.c}) agrees with exhaustive search up to $N=14$. The programme is $Z(m,w)=\max_{k,b}\{Z(n,k)+Z(n-k,(w-k-b)/2)+b\}$ with $n=\lfloor m/2\rfloor$ and $b\in\{0,m-2n\}$, $u$ and $v$ being free given their lengths and weights. The exponent $0.6942$ is explained, heuristically, for the mean under the fair law: $a_N=\E_{1/2}|R|$ satisfies exactly $a_{2n+\rho}=a_n+\E\,a_{B}+\rho$ with $B\sim\mathrm{Bin}(n,\frac12)$, since $u$ is uniform and $v$ uniform given its length, and $a_N\approx a_{N/2}+a_{N/4}$ gives $2^{-\alpha}+4^{-\alpha}=1$, $\alpha=\log\varphi$. We conjecture that the maximum grows like $N^{\alpha}$ up to a slowly varying factor, with $\alpha$ between $0.69$ and $0.73$, well below $\beta_*=0.7729$; this would improve the remainders of the gap description and of the extraction loss; the bound on mixed fair bits (Lemma~\ref{lem:mixed}) uses the charged weights of whole levels rather than the residual ones, and a matching improvement there is a related question. The fair-length charge of Proposition~\ref{prop:faircharge} does not change the exponent of the relaxation: on the paths that the weight charge allows, those with at most $0.227\,d$ letters $\vstep$ at depth $d$, the fair charges sum to at most $N\log N$. (ii) \emph{The boundary penalty.} Theorem~\ref{thm:tworuns} shows that $\Omega(\log^2N)$ is necessary for two pieces when the fair bits are stored as they are, and Theorem~\ref{thm:concat} charges $13N^\beta(J-1)^{1-\beta}$; the gap between the two is the open question, distinct from that of the residual weight, and a sharper penalty would also bring the rate of the BWT code without tree, $O(N^{\beta-1}\log N)$ per symbol, closer to the $O(2^k\log N/N)$ of the tree codes. (iii) \emph{A simple residual coder near the KT length.} The KT residual marginal of Remark~\ref{rem:E2} shows that storing the fair bits raw does not force a polynomial regret on Bernoulli sources, and its quadrature form computes it causally with $O(N)$ weights; can it be computed in near-linear time, and can a simple local residual coder, in particular one without arithmetic coder, reach a polylogarithmic regret without sending the weight?

\appendix

\section{The H\"older Exponent}\label{app:holder}

\begin{proposition}\label{prop:holder}
For every $\lambda>0$ and $s\in\bits^N$, $z(s)\le C_\lambda N^{\beta(\lambda)}$ with $\beta(\lambda)=\log(1+2^\lambda)/(1+\lambda)$ and $C_\lambda=(1+2^\lambda)^{1/(1+\lambda)}$. The minimum of $\beta$ is $\beta_*=1-\alpha=0.7729078\ldots$, where $\alpha\in(0,\frac12)$ solves $h(\alpha)=1-\alpha$, attained at $\lambda_*=1.7670182\ldots$.
\end{proposition}

\begin{proof}
Write $1=(2^{v(\pi_i)})^{\lambda/(1+\lambda)}(2^{-\lambda v(\pi_i)})^{1/(1+\lambda)}$ at each residual one and apply H\"older's inequality with exponents $(1+\lambda)/\lambda$ and $1+\lambda$:
\[
z^{1+\lambda}\le\Big(\sum_{R_i=1}2^{v(\pi_i)}\Big)^\lambda\sum_{R_i=1}2^{-\lambda v(\pi_i)}\le N^\lambda\sum_{d=0}^{\lfloor\log N\rfloor}(1+2^{-\lambda})^d,
\]
by \eqref{eq:charge} and the binomial theorem over all paths. With $\mu=1+2^{-\lambda}>1$, $\sum_{d\le\log N}\mu^d<\mu^{\log N+1}/(\mu-1)=(1+2^\lambda)N^{\log\mu}$, and $\lambda+\log\mu=\log(1+2^\lambda)$; take the $(1+\lambda)$-th root. The derivative of $\beta$ has the sign of $(1+\lambda)2^\lambda/(1+2^\lambda)-\log(1+2^\lambda)$, which with $\alpha=(1+2^\lambda)^{-1}$ is $1-\alpha-h(\alpha)$; $h(\alpha)+\alpha$ increases from $0$ to $\frac32$ on $(0,\frac12)$, so the derivative changes sign once, from negative to positive, at $h(\alpha)=1-\alpha$, where $\beta=1-\alpha$.
\end{proof}

The same refinement applies to Lemma~\ref{lem:mixed}. Factoring
\[
m_\pi=\big(m_\pi2^{v(\pi)}\big)^{\lambda/(1+\lambda)}\big(m_\pi2^{-\lambda v(\pi)}\big)^{1/(1+\lambda)}
\]
in the proof of Appendix~\ref{app:mixed} gives $M\le C'_\lambda N^{\beta(\lambda)}(J-1)^{1-\beta(\lambda)}$ with an absolute constant $C'_\lambda$, so every remainder of Sections~\ref{sec:G} and~\ref{sec:memory} holds with $\beta_*$ in place of $\beta$.

\section{Proof of Theorem~\ref{thm:tworuns}}\label{app:tworuns}

Let $f(k)=F(0^k1^k)$ and $g(k)=z(0^k1^k)$, with $f(0)=g(0)=0$. If $k=2m$, the boundary falls between pairs, all pairs are equal, $u=0^{2m}$ and $v=0^m1^m$; the zero child contributes nothing, so $f(2m)=f(m)$ and $g(2m)=g(m)$. If $k=2m+1$, exactly one pair, $01$, is unequal: its fair bit is $1$, the $u$-string is $0^m1\,0^m$ and the $v$-string is $0^m1^m$. Consider the successive $\ustep $-children of the singleton string $0^m1\,0^m$, of length $k$: at depth $j$, the length is $\ell_j=\lfloor k/2^j\rfloor$ and the one is at position $p_j=\lfloor\ell_j/2\rfloor+1$; indeed pairing sends position $p_j$ to $\lceil p_j/2\rceil=\lfloor\ell_{j+1}/2\rfloor+1$. While $\ell_j\ge2$, $p_j\le2\lfloor\ell_j/2\rfloor$, so the one belongs to a pair, which is unequal and emits one fair bit, while every other pair is $00$ and every $\vstep $-child and remainder is zero. After $\lfloor\log k\rfloor$ halvings the one becomes a residual bit, the first of the residual of that subtree. Hence
\[
f(2m+1)=f(m)+1+\lfloor\log(2m+1)\rfloor,\qquad g(2m+1)=g(m)+1.
\]
With $k=\sum_{i=0}^db_i2^i$, $b_d=1$, the successive arguments $\lfloor k/2^i\rfloor$ have last binary digit $b_i$ and binary length $d-i+1$; an even argument contributes nothing and an odd one $d-i+1$ to $f$ and $1$ to $g$, which proves the formula.

For $k=2^e-1$ all $e$ binary digits are ones, so $f(k)=\sum_{i=0}^{e-1}(e-i)=e(e+1)/2$. In the singleton subtree the lengths are $2^j-1$ and the one is at the even position $2^{j-1}$ until length one, so every fair bit is $1$, and so is the root fair bit; induction gives $Y=1^{e(e+1)/2}$. At the top, the singleton subtree has length $2^e-1$, emits $e-1$ fair bits, and leaves the residual $1\,0^{2^e-e-1}$, followed by the residual of the problem with $2^{e-1}-1$; with $R(01)=1$ this gives the residual of the statement. Its gap description costs one bit for the first one and, for $2\le j\le e$, $\gamma(2^j-j)=2j-1$ bits, since $2^{j-1}\le2^j-j<2^j$; in all $1+\sum_{j=2}^e(2j-1)=e^2$, and adding $F=e(e+1)/2$ gives $(3e^2+e)/2$.\hfill$\blacksquare$

\section{Proof of Lemma~\ref{lem:mixed}}\label{app:mixed}

Let $\kappa=J-1\ge1$ be the number of cuts, $m_\pi$ the number of mixed fair bits emitted at node $\pi$, and $M_d=\sum_{|\pi|=d}m_\pi$. At one node, the ancestries of the pairs are ordered and disjoint, so their spans, from first to last position, are disjoint intervals. A mixed pair has a cut inside its span, and a cut lies in at most one span, so at most $\kappa$ pairs of a node are mixed. A node at depth $d$ also has length at most $N2^{-d}$, so
\[
m_\pi\le c_d:=\min(\kappa,N2^{-d}).
\]
Every mixed fair bit comes from an unequal pair, so $m_\pi\le\wt(u_\pi)\le\wt(x_\pi)$, where $x_\pi$ is the string of the node and $u_\pi$ its $u$-string. The one-level identity \eqref{eq:weight} at node $\pi$ reads $2^{v(\pi)}\wt(x_\pi)=2^{v(\pi\ustep )}\wt(x_{\pi\ustep })+2^{v(\pi\vstep )}\wt(x_{\pi\vstep })+2^{v(\pi)}\wt(r_\pi)$, so the charged weight of a level does not increase from one level to the next: $\sum_{|\pi|=d}2^{v(\pi)}\wt(x_\pi)\le\wt(s)\le N$. By Cauchy--Schwarz,
\[
M_d^2\le\Big(\sum_{|\pi|=d}m_\pi2^{v(\pi)}\Big)\Big(\sum_{|\pi|=d}m_\pi2^{-v(\pi)}\Big)\le N\,c_d\sum_{\pi\in\{\ustep ,\vstep \}^d}2^{-v(\pi)}=N\,c_d\,(3/2)^d.
\]
Let $d_0=\lfloor\log(N/\kappa)\rfloor$, $\eta=\sqrt{3/2}$ and $\zeta=\sqrt{3/4}$. For $d\le d_0$, $M_d\le\sqrt{N\kappa}\,\eta^d$; for $d>d_0$, $M_d\le N\zeta^d$. Summing, with $\eta^{d_0}\le(N/\kappa)^{\log\eta}$ and $\zeta^{d_0+1}\le(N/\kappa)^{\log\zeta}$,
\[
M\le\frac\eta{\eta-1}\sqrt{N\kappa}\,(N/\kappa)^{\log\eta}+\frac1{1-\zeta}N(N/\kappa)^{\log\zeta}=\Big(\frac\eta{\eta-1}+\frac1{1-\zeta}\Big)N^\beta\kappa^{1-\beta},
\]
since $\log\eta=\beta-\frac12$ and $\log\zeta=\beta-1$; the constant is less than $5.46+7.47<13$.\hfill$\blacksquare$

\section{Logarithmic Residual Coders: Proof and the Codes A and C}\label{app:logcoders}

\emph{Proof of Theorem~\ref{thm:logcoders}.} A gap coder writes one word per residual one and at most one final word per class, and the arguments $g+1$ of the words of a class sum to at most its number of bits plus one. Completing with dummy words of argument one to exactly $z+K$ words, which only increases the bound, and using the concavity of $\log$ gives the first bound. For a sequential coder, a class with $m_c$ bits of which $z_c$ are ones costs $-\log\KT(m_c,z_c)\le m_ch(z_c/m_c)+\frac12\log m_c+1$ \cite{WST95}; moreover $mh(z/m)\le z\log(\ee m/z)$, because $-(1-p)\ln(1-p)\le p$, and $\sum_cz_c\log(m_c/z_c)\le z\log(m/z)$ by the log-sum inequality. Finally $z<\sqrt3\,N^\beta$ and $F\le\log\binom N{\wt(s)}$ by \eqref{eq:typebound}; for $r$ runs, insert the bounds of Theorem~\ref{thm:runs} below, using that $a\mapsto a\log(1+M/a)$ is increasing.
\hfill$\blacksquare$

\begin{definition}[Code A]\label{def:A}
Code A sends the weight in $\lceil\log(N+1)\rceil$ bits, computes the plug-in biases $\phi_\pi(\hat p)$ as code B, assigns each residual bit to the class $(k,b)$ where $b=1$ if its bias $\phi$ is at least $\frac12$, $b=0$ otherwise, and $k=\min\big(24,\max(0,\mathrm{round}\log(\ln2/\min(\phi,1-\phi)))\big)$, with $k=24$ when $\phi\in\{0,1\}$, and codes each class with a sequential KT estimator; the fair bits are stored as they are.
\end{definition}

The classes group biases within a factor of about two of their minority probability; code A uses the biases of code B to sort the residual bits and learns, in each class, the frequency that the bits actually have. The cap at $24$ only bounds the number of classes: a bias below $2^{-24}$ is below the resolution of the coders of Section~\ref{sec:exp}, and on the Bernoulli strings of that section the classes above $k=7$ contain no minority symbol at all. The cap nevertheless matters a little, since every class used pays its KT estimator about half a logarithm: on the draws of Table~\ref{tab:bernoulli} at $N=2^{16}$ and $2^{20}$ and $\theta\ne\frac12$, a cap of $16$ shortens code A by $28$ to $41$ bits and a cap of $32$ lengthens it by as much (\texttt{code/cap\_sensitivity.py}); at $\theta=\frac12$ all residual bits share one class and the cap is immaterial. By Theorem~\ref{thm:logcoders}, with at most $50$ classes and a header of $\lceil\log(N+1)\rceil$ bits, it is universal on every string, and on Bernoulli strings it is within $160$ bits of code B at $N=2^{20}$ (Table~\ref{tab:bernoulli}). By contrast, an unprotected Rice code with a fixed parameter $k$ writes a gap $g$ in $\lfloor g/2^k\rfloor+1+k$ bits, linearly in $g$, so coding the classes of code A with such codes, the parameter of a class growing with its index $k$ up to a cap, is not covered by Theorem~\ref{thm:logcoders}, and is indeed not universal. Take $s=0^{N-1}1$ with $N=2^d+1$ and the cap $24$ of Definition~\ref{def:A}, with $d\ge25$: the residual is $0^{N-1}1$, the single one is the odd remainder of the root, and its plug-in bias $1/N$ puts it in the capped class; so do the biases of all but a number independent of $N$ of the $N-1$ zeros that precede it, and its Rice codeword has about $N/2^{24}$ bits, against $Nh(1/N)=\log N+O(1)$. An escape repairs this.

\begin{definition}[Code C, a protected variant of code A]\label{def:C}
Code C sends the weight in $\lceil\log(N+1)\rceil$ bits and computes the plug-in biases as code A. A residual bit of bias $\phi$ is replaced by its \emph{error bit}, its XOR with $\mathbf1\{\phi>\frac12\}$, and assigned to the class $(k,\mathbf1\{\phi>\frac12\})$, where $k$ is the Rice order that minimises the expected Rice length of a geometric gap of hazard $\varepsilon=\min(\phi,1-\phi)$, $k=\min\big(8,\max(0,\lceil\log(\ln\frac{\sqrt5-1}2/\ln(1-\varepsilon))\rceil)\big)$ ($k=8$ if $\varepsilon=0$). The zero gaps of each class are written, as in Theorem~\ref{thm:logcoders}, with the \emph{guarded Rice code} of order $k$: for $g=q2^k+r$, the word $0^q1$ followed by $r$ in $k$ bits if $q<8$, and $0^8$, the gamma code of $q-7$ and $r$ in $k$ bits otherwise. The fair bits are stored as they are.
\end{definition}

The guarded codeword of $g$ has at most $17+2\log(g+1)$ bits, so code C is a logarithmic coder with at most $18$ classes and a header of $\lceil\log(N+1)\rceil$ bits; the error bits have at most $z+1+\lfloor\log N\rfloor$ ones, since a bias exceeds $\frac12$ only on the all-$\vstep$ branch, and Theorem~\ref{thm:logcoders} gives $L_C(s)\le\log\binom N{\wt(s)}+O(N^\beta\log N)$ on every string, with the concatenation and BWT bounds. The cap at $8$ trades the number of classes against the long gaps: beyond it the escape writes a gap $g$ in $17+2\log(g+1)$ bits instead of about $\log g$, a loss that Theorem~\ref{thm:logcoders} bounds. On the same draws a cap of $6$ costs up to $71$ bits at $N=2^{20}$, where long gaps escape to the gamma code, and a cap of $10$ changes the length by at most $17$ bits. Short gaps keep the short Rice words that the cap obstruction above forbids to a fixed parameter, and the escape keeps every long gap logarithmic. Two more bits select the shortest of code G, of the same with Elias delta codes in place of gamma codes, of the residual stored raw, and of code C: this selector costs at most $L_G(s)+2$ on every string, which keeps all the bounds of this section, and at most $N+2$ bits, which removes the expansion on fair data. On Bernoulli strings the selector is within $2$ to $89$ bits of $-\log\Pr_\theta(s)$ at $N=2^{16}$ where code G is $264$ to $528$ bits above (Table~\ref{tab:bernoulli}), and code C alone is at that length shorter than the KT classes of code A for $\theta\ne\frac12$, by $27$ to $34$ bits on average over the five draws with paired standard deviations of $3$ to $13$ (at $N=2^{20}$ the paired differences are $19$ to $40$ bits, with standard deviations up to $23$), without an arithmetic coder; on Bernoulli strings the delta codes lose to the gamma codes, whose short words suit the short gaps, and on structured strings they win on several examples, among them both two-run families ($24$ against $31$ residual bits for $0^{2^{15}}1^{2^{15}}$, $188$ against $225$ for $0^{2^{15}-1}1^{2^{15}-1}$).

\section{Further Properties of the Residual}\label{app:residual}

\emph{The fair-length charge.} \begin{proposition}[Fair-length charge]\label{prop:faircharge}
Let $A(\epsilon)=0$, $A(\pi\ustep)=A(\pi)+1$ and $A(\pi\vstep)=2A(\pi)$, that is $A(\pi)=\sum_{j:\pi_j=\ustep}2^{v(\pi_{j+1}\cdots\pi_{|\pi|})}$. For every $s\in\bits^N$,
\begin{equation}\label{eq:faircharge}
F(s)=\sum_{i:R_i=1}A(\pi_i),\qquad N=\sum_{i=1}^{|R|}\big(1+R_iA(\pi_i)\big).
\end{equation}
Consequently, after reading $j$ residual bits, the residual parser knows
\[
T=N-j-\sum_{i\le j:R_i=1}A(\pi_i),
\]
the exact number of additional zeros if all subsequent residual bits are zero; in particular the next gap has at most $T$ zeros, with equality only for the terminal run of zeros.
\end{proposition}

\begin{proof}
$F(s)$ is the sum of $\wt(u)$ over the internal nodes, and by \eqref{eq:charge} applied to the $u$-string of a node, a residual one contributes $2^{v(\cdot)}$ of its path below that $\ustep$ step to $\wt(u)$; summing over the $\ustep$ steps of its path gives $A(\pi_i)$. The second identity adds $|R|=N-F$. For the consequence, the continuation by zeros ends, since $\Rc_N$ is complete and the parser stops exactly when a residual is complete, and \eqref{eq:faircharge} applied to the completed residual gives the length of the continuation.
\end{proof}

Since the decoder knows, by Proposition~\ref{prop:faircharge}, that the next gap argument $g+1$ of a gap description is at most $T+1$, the gamma code of code G may be restricted to $[1,T+1]$ (prune the words above $T+1$ and contract every node left with one child), which never lengthens a word, needs no header and keeps every bound of the paper. The gain is small: $0.03$ to $0.8\%$ of the residual cost on Bernoulli strings of $2^{16}$ bits, $17$ instead of $33$ bits for $R=0^{65536}$, and on the file compressor (mode \texttt{gapb}, Table~\ref{tab:files}) nothing on long sequences but, with a context tree of $10$ bits, where thousands of short sequences each pay a final gap, $2.4\%$ of the archive of \emph{Alice}, $0.3\%$ of that of the four books and $2.7\%$ of that of the executable.

\emph{Charges and biases; complements.} The two charges are also the leading coefficient and exponent of the bias of a residual bit for a rare symbol: since $\U(p)\sim2p$ and $\V(p)\sim p^2$ as $p\to0$, for every fixed path $\pi$,
\[
\phi_\pi(\theta)\sim2^{A(\pi)}\,\theta^{2^{v(\pi)}}\qquad(\theta\to0),
\]
and for $0<\theta\le\frac12$ the log-odds $\log\frac{1-\phi_\pi(\theta)}{\phi_\pi(\theta)}$ is at least $2^{v(\pi)}\log\frac{1-\theta}\theta-A(\pi)$ (a $\ustep$ step lowers the log-odds by less than one, a $\vstep$ step doubles them). The order of the steps therefore matters: at $\theta=0.1$, $\phi_{\ustep^6\vstep^4}\simeq0.49999$ and $\phi_{\vstep^4\ustep^6}\simeq3.5\cdot10^{-14}$, with charges $A=96$ and $A=6$. The residual prefers the symbol $0$, not the minority symbol of the input. Complementing $s$ leaves the root $u$-string unchanged and complements $v$ and $r$; recursively, it complements the strings of the nodes at the paths $\vstep^j$ and leaves every other node unchanged, so it leaves the tree unchanged and flips exactly the residual bits whose path contains no $\ustep $, at most one per depth. With \eqref{eq:charge} applied to $s$ and to its complement, $z(s)\le\min(\wt(s),N-\wt(s))+1+\lfloor\log N\rfloor$; for instance $z(1^N)$ is the number of ones in the binary expansion of $N$.

\emph{The price of a raw fair part.} How much a residual-only model can recover is characterised exactly. For a nonempty class $\mathcal K\subseteq\bits^N$ of inputs, let $\sat(\mathcal K)=\{x:R(x)=R(a)\text{ for some }a\in\mathcal K\}$ be its completion into fibres and $S_{\mathcal K}=|\sat(\mathcal K)|=\sum_{R\in R(\mathcal K)}2^{N-|R|}$.

\begin{theorem}[The price of a raw fair part]\label{thm:saturation}
Among the ideal codes $L_Q(x)=F(x)-\log Q(R(x))$, with $Q$ a probability law on residuals, $\inf_Q\max_{x\in\mathcal K}L_Q(x)=\log S_{\mathcal K}$, and a prefix code attains $\log S_{\mathcal K}+1$ on $\mathcal K$ (and $\log S_{\mathcal K}+2$ while covering all of $\bits^N$).
\end{theorem}

\begin{proof}
$L_Q$ is constant on a fibre. If $\max_{x\in\mathcal K}L_Q\le L$, then $Q(R)\ge2^{F(R)-L}$ for every $R\in R(\mathcal K)$, and summing gives $1\ge2^{-L}S_{\mathcal K}$. Conversely $Q_{\mathcal K}(R)=2^{F(R)}/S_{\mathcal K}$ on $R(\mathcal K)$ gives $L_{Q_{\mathcal K}}=\log S_{\mathcal K}$ on $\mathcal K$; Shannon lengths $\lceil-\log Q_{\mathcal K}(R)\rceil$ satisfy Kraft's inequality, and one more bit selects between this code and a code for all inputs.
\end{proof}

This is the Shtarkov normalisation \cite{Shtarkov87} restricted to residual-only codes: the issue is the size of the completion into fibres, not of $\mathcal K$ itself; even a singleton has $S_{\mathcal K}=2^{F(x)}$. Recursing also on the fair bits removes the obstruction of periodic strings at the price of the exactly fair part; see the extended version.

\section{The BWT-Tree Code}\label{app:tree}

The model-based code of Section~\ref{sec:memory}, used as a baseline in Table~\ref{tab:markov} and in the file compressor, is the following.

\begin{definition}[BWT-tree code]\label{def:bwt}
Fix a maximal depth $D$ and a pruned context tree. Emit $p$ in $\lceil\log(N+1)\rceil$ bits, then traverse the tree: at a node with interval $[lo,hi)$ of rows, a flag says whether it is a leaf; an internal node emits its split point in $\lceil\log(hi-lo+1)\rceil$ bits and recurses; a leaf codes its bits of $\ell$ with a code for strings of known length (code B or KT; a leaf with no bit of $\ell$ costs nothing). Nodes with at most one bit of $\ell$, or at depth $D$, are leaves without a flag. The tree minimises the total length by a bottom-up dynamic programme over the nonempty intervals.
\end{definition}

\begin{proposition}\label{prop:oracle}
Let $0\le k\le D$, $a_N=\lceil\log(N+1)\rceil$ and $a_N'=\lceil\log(N+2)\rceil$. For every $x\in\bits^N$ and every Markov law $P$ of order $k$ started from the zero context, the ideal length $\ell(x)$ of the BWT-tree code with KT leaves satisfies
\[
\ell(x)+\log P(x)\le a_N+(2^k-1)\,a_N'+2^{k-1}\log\big(1+N2^{-k}\big)+3\cdot2^k,
\]
and with code-B leaves the same holds with $2^k\log(1+N2^{-k})$ in place of $2^{k-1}\log(1+N2^{-k})$.
\end{proposition}

\begin{proof}
The dynamic programme is never worse than the complete tree of depth $k$, in which nodes with at most one bit are made leaves. That tree costs $a_N$ for $p$, at most $2^{k+1}-1$ flags, and at most $2^k-1$ split points, each in at most $a_N'$ bits. A leaf $c$ with $m_c$ bits costs at most $-\log P(\ell|_{I(c)})+\frac12\log(m_c+1)+1$ with KT, where $P(\ell|_{I(c)})$ is the Bernoulli likelihood of its bits at the transition probability of its context; a leaf of depth less than $k$ has at most one bit and costs at most one bit. With zero padding these likelihoods multiply to $P(x)$, and $\sum_c\frac12\log(m_c+1)\le2^{k-1}\log(1+N2^{-k})$ by concavity. For code-B leaves the regret of a leaf is at most $\log(m_c+1)+1$ (Theorem~\ref{thm:B}).
\end{proof}

\begin{corollary}[The tree code on individual sequences and ergodic sources]\label{cor:treeergodic}
For every $x\in\bits^N$ and $0\le k\le\min(D,N-1)$, the ideal length of the BWT-tree code with KT leaves satisfies
\[
\ell(x)\le E_k(x)+a_N+(2^k-1)\,a_N'+2^{k-1}\log\big(1+N2^{-k}\big)+3\cdot2^k+2k\log(N+2),
\]
and with code-B leaves the same holds with $2^k\log(1+N2^{-k})$ in place of $2^{k-1}\log(1+N2^{-k})$. Consequently, if the depth $D_N$ tends to infinity with $N$, then for every stationary ergodic binary source $P$ with entropy rate $h_P$, $\ell(X_1\cdots X_N)/N\to h_P$ almost surely and in $L^1$.
\end{corollary}

\begin{proof}
In the proof of Proposition~\ref{prop:oracle}, price each leaf by its own zero-order empirical entropy instead of a Markov likelihood: a KT leaf with $m_c$ bits and $w_c$ ones costs at most $m_ch(w_c/m_c)+\frac12\log(m_c+1)+1$ \cite{WST95}, and a code-B leaf $m_ch(w_c/m_c)+\log(m_c+1)+1$. The padded intervals of the complete tree of depth $k$ are the pieces of Lemma~\ref{lem:bwtintervals} with the bits of the first $k$ positions added to some of them, and adding one bit to a piece of $m$ bits and $w$ ones raises $mh(w/m)$ by at most $2\log(m+2)$, since $(m+1)h(\frac{w+b}{m+1})\le\log\binom{m+1}{w+b}+\log(m+2)\le\log\binom mw+\log(m+1)+\log(m+2)$; so the leaf entropies sum to at most $E_k(x)+2k\log(N+2)$, and the flags and split points are counted as before. For the second statement, fix $k$; for $N$ large, $k\le D_N$, and the bound gives $\limsup_N\ell(X_1\cdots X_N)/N\le\limsup_NE_k/N=h_k$ almost surely by the ergodic theorem, as in the proof of Theorem~\ref{thm:bwtergodic}; the lower bound by Kraft's inequality and the Shannon--McMillan--Breiman theorem, and the domination by $\ell(x)\le N+O(\log N)$ (the dynamic programme is never worse than the single leaf), are as in that proof.
\end{proof}

The split points cost $\log N$ each, which sequential conditional KT and context-tree weighting do not pay. They can be computed from counts: with the padded intervals of all $N+1$ rows, if $o(c)$ and $a(c)$ are the numbers of ones and of zeros of $L$ in $I(c)$, the sentinel counting for neither, then $|I(1c)|=o(c)$ and $|I(0c)|=a(c)+\mathbf1\{c=0^{|c|}\}$, since sending the position $j\le N$ to $j+1$ prepends $x_j$ to its padded history and reaches every row but that of position $1$, whose history is all zeros. This is the occurrence count of backward search in the FM-index \cite{FM00,FM05}. On a suffix-closed tree (whenever $bc$ is a node, so is $c$) the counts of the leaves therefore determine every interval, but a code that uses them is a different format from Definition~\ref{def:bwt}: the tree and the leaf counts must be sent before the bodies, in fixed widths determined by $N$, and for KT leaves these counts are information that the leaf code would otherwise learn; we do not pursue it, and Proposition~\ref{prop:oracle} keeps its split term. 

\section{Two Consequences of the Flat Conditional Laws}\label{app:flat}

\emph{Distribution-preserving replacement.} Theorem~\ref{thm:piecewise} says that, under a piecewise exchangeable law, the unmixed fair bits are uniform given $Z$; resampling them therefore changes nothing.

\begin{proposition}[Distribution-preserving replacement]\label{prop:replacement}
Let $P$ be piecewise exchangeable, $(Y;R)=\T(X)$, and let $Y'$ agree with $Y$ on the mixed slots $\mathcal S$ and be uniform on $\bits^{F-M}$ on the other slots, independently of $X$ given $Z$. Then $X'=\Dec_N(Y';R)$ has the law $P$ of $X$, and $Z(X')=Z(X)$.
\end{proposition}

\begin{proof}
By Theorem~\ref{thm:piecewise}, given $Z=z$ the law of $Y$ is uniform on the completions of the mixed values, which is the conditional law of $Y'$ given $Z=z$; so $(Y';R)$ has the law of $(Y;R)$ and $X'$ that of $X$. The string $X'$ is reached from $X$ by flipping unmixed fair bits one at a time, and each flip preserves $Z$ by Lemma~\ref{lem:flip}.
\end{proof}

For a partition known to both parties, the unmixed slots can thus carry a payload masked by a fresh uniform string, without changing the law of the cover: $\E(F-M)$ bits per block, at least $Nh(\theta)-b_N$ for a Bernoulli cover and $J=1$. Perfect steganography at the entropy rate is the subject of \cite{RyabkoRyabko11}; the transform gives an explicit realisation, for piecewise exchangeable covers as well. The slots are computed in $O(N\log N)$ operations and $O(N)$ memory by keeping the smallest and the largest piece label of each ancestry. After a BWT the decoded string must moreover be a valid BWT output for the row $p$, a constraint that the proposition does not cover.

\emph{Seeded extraction in total variation.} Proposition~\ref{prop:extracttail} bounds the tail of the deficiency $D(W)$ of the flat conditional law of $Y$ given $W=(Z,\tau,p)$; the leftover hash lemma turns it into a statement in total variation.

\begin{proposition}[Seeded extraction]\label{prop:seeded}
Under the assumptions of Proposition~\ref{prop:bwtextract}, let $t\ge0$, $0<\varepsilon<1$, $c_{N,k}$ as in Proposition~\ref{prop:extracttail},
\[
K=\max\{0,\lfloor F-c_{N,k}-t-2\log(1/\varepsilon)\rfloor\},
\]
and let $(h_S)$ be a two-universal family of hash functions from $\bits^F$ to $\bits^K$, one family for each pair of lengths $(F,K)$, with a seed $S$ uniform over the family and independent of $X$. The joint law of $(h_S(Y),S,W)$ is within total variation $2^{-t}+\varepsilon/2$ of the law with the same marginal of $(S,W)$ under which, given $(S,W)=(s,w)$, the output is uniform on $\bits^{K(w)}$.
\end{proposition}

\begin{proof}
Outside the exceptional event of Proposition~\ref{prop:extracttail}, of probability at most $2^{-t}$, the law of $Y$ given $W$ is flat on $A\ge2^{F-c_{N,k}-t}$ points. For a flat law on $A$ points and a two-universal hash into $\bits^K$, the collision probability of the output, averaged over the seed, is at most $2^{-K}+1/A$, and the Cauchy--Schwarz inequality bounds the mean total variation of the output from the uniform law by $\frac12\sqrt{2^K/A}$, which is at most $\varepsilon/2$ by the choice of $K$ when $K>0$; the empty output is at distance zero. Averaging over $W$ and adding the exceptional probability gives the claim, with the seed disclosed.
\end{proof}

For a fixed order $k$, $c_{N,k}=o(N)$, and $t$ and $\log(1/\varepsilon)$ of order $\log N$ lose $o(N)$ bits while the distance tends to zero: by Corollary~\ref{cor:bwtrate}, for a stationary Markov source of order $k$, $\E K/N\to h_P$, and if the source is also ergodic, $K/N\to h_P$ almost surely and in $L^1$. The post-processing needs a seed and an order bound for its output length, which the transform itself does not, and the constant is the one of the paper: whenever the budget $c_{N,k}+t+2\log(1/\varepsilon)$ exceeds the number of fair bits, the guarantee returns $K=0$. At $N=2^{20}$, $c_{N,0}\approx20$, $c_{N,1}\approx0.8454N$, and $c_{N,k}>N$ for $k\ge2$; the extraction budget additionally includes $t+2\log(1/\varepsilon)$. Beyond the memoryless case this is an asymptotic extraction statement, not a competitive generator. Exact and efficient extraction from Markov sources, of known or unknown order, is known \cite{ZhouBruckMarkov12,SeroussiWeinberger15}; the statement here concerns the fixed output of one transform.

\section*{Acknowledgment}
Aur\'elien Garivier acknowledges the support of Chaire SeqALO (ANR-20-CHIA-0020-01) and PEPR-IA project FOUNDRY (ANR-23-PEIA-0003).


\begin{thebibliography}{99}

\bibitem{vonNeumann51} J.~von Neumann, ``Various techniques used in connection with random digits,'' in \emph{Monte Carlo Method}, Nat.\ Bur.\ Standards Appl.\ Math.\ Ser., vol.~12, pp.~36--38, 1951.

\bibitem{Elias72} P.~Elias, ``The efficient construction of an unbiased random sequence,'' \emph{Ann.\ Math.\ Statist.}, vol.~43, no.~3, pp.~865--870, 1972.

\bibitem{Elias75} P.~Elias, ``Universal codeword sets and representations of the integers,'' \emph{IEEE Trans.\ Inf.\ Theory}, vol.~21, no.~2, pp.~194--203, 1975.

\bibitem{RyabkoMatchikina00} B.~Ryabko and E.~Matchikina, ``Fast and efficient construction of an unbiased random sequence,'' \emph{IEEE Trans.\ Inf.\ Theory}, vol.~46, no.~3, pp.~1090--1093, 2000.

\bibitem{Peres92} Y.~Peres, ``Iterating von Neumann's procedure for extracting random bits,'' \emph{Ann.\ Statist.}, vol.~20, no.~1, pp.~590--597, 1992.

\bibitem{Pae15} S.~Pae, ``A generalization of Peres's algorithm for generating random bits from loaded dice,'' \emph{IEEE Trans.\ Inf.\ Theory}, vol.~61, no.~2, pp.~751--757, 2015.

\bibitem{Pae16} S.~Pae, ``Binarization trees and random number generation,'' \emph{IEEE Trans.\ Inf.\ Theory}, vol.~66, no.~4, pp.~2581--2587, 2020.

\bibitem{Pae18} S.~Pae, ``Peres-style recursive algorithms,'' arXiv:1805.04610v2, 2018.

\bibitem{ZhouBruck12} H.~Zhou and J.~Bruck, ``A universal scheme for transforming binary algorithms to generate random bits from loaded dice,'' arXiv:1209.0726, 2012.

\bibitem{PKS18} A.~Prasitsupparote, N.~Konno, and J.~Shikata, ``Numerical and non-asymptotic analysis of Elias's and Peres's extractors with finite input sequences,'' \emph{Entropy}, vol.~20, no.~10, art.~729, 2018.

\bibitem{Cover73} T.~M.~Cover, ``Enumerative source encoding,'' \emph{IEEE Trans.\ Inf.\ Theory}, vol.~19, no.~1, pp.~73--77, 1973.

\bibitem{KT81} R.~E.~Krichevsky and V.~K.~Trofimov, ``The performance of universal encoding,'' \emph{IEEE Trans.\ Inf.\ Theory}, vol.~27, no.~2, pp.~199--207, 1981.

\bibitem{Rissanen84} J.~Rissanen, ``Universal coding, information, prediction, and estimation,'' \emph{IEEE Trans.\ Inf.\ Theory}, vol.~30, no.~4, pp.~629--636, 1984.

\bibitem{Shtarkov87} Yu.~M.~Shtarkov, ``Universal sequential coding of single messages,'' \emph{Probl.\ Inf.\ Transm.}, vol.~23, no.~3, pp.~175--186, 1987.

\bibitem{WST95} F.~M.~J.~Willems, Y.~M.~Shtarkov, and T.~J.~Tjalkens, ``The context-tree weighting method: Basic properties,'' \emph{IEEE Trans.\ Inf.\ Theory}, vol.~41, no.~3, pp.~653--664, 1995.

\bibitem{ZivLempel78} J.~Ziv and A.~Lempel, ``Compression of individual sequences via variable-rate coding,'' \emph{IEEE Trans.\ Inf.\ Theory}, vol.~24, no.~5, pp.~530--536, 1978.

\bibitem{KosarajuManzini99} S.~R.~Kosaraju and G.~Manzini, ``Compression of low entropy strings with Lempel--Ziv algorithms,'' \emph{SIAM J.\ Comput.}, vol.~29, no.~3, pp.~893--911, 1999.

\bibitem{BW94} M.~Burrows and D.~J.~Wheeler, ``A block-sorting lossless data compression algorithm,'' Digital Equipment Corporation, Systems Research Center, Palo Alto, CA, Res.\ Rep.\ 124, 1994. [Online]. Available: \url{https://www.cs.jhu.edu/~langmea/resources/burrows_wheeler.pdf}

\bibitem{Manzini01} G.~Manzini, ``An analysis of the Burrows--Wheeler transform,'' \emph{J.\ ACM}, vol.~48, no.~3, pp.~407--430, 2001.

\bibitem{KLV07} H.~Kaplan, S.~Landau, and E.~Verbin, ``A simpler analysis of Burrows--Wheeler-based compression,'' \emph{Theoret.\ Comput.\ Sci.}, vol.~387, no.~3, pp.~220--235, 2007.

\bibitem{EVKV02} M.~Effros, K.~Visweswariah, S.~R.~Kulkarni, and S.~Verd\'u, ``Universal lossless source coding with the Burrows--Wheeler transform,'' \emph{IEEE Trans.\ Inf.\ Theory}, vol.~48, no.~5, pp.~1061--1081, 2002.

\bibitem{Merhav93} N.~Merhav, ``On the minimum description length principle for sources with piecewise constant parameters,'' \emph{IEEE Trans.\ Inf.\ Theory}, vol.~39, no.~6, pp.~1962--1967, 1993.

\bibitem{ShamirMerhav99} G.~I.~Shamir and N.~Merhav, ``Low-complexity sequential lossless coding for piecewise-stationary memoryless sources,'' \emph{IEEE Trans.\ Inf.\ Theory}, vol.~45, no.~5, pp.~1498--1519, 1999.

\bibitem{VWBG13} J.~Veness, M.~White, M.~Bowling, and A.~Gy\"orgy, ``Partition tree weighting,'' in \emph{Proc.\ Data Compression Conf.\ (DCC)}, 2013, pp.~321--330.

\bibitem{FM00} P.~Ferragina and G.~Manzini, ``Opportunistic data structures with applications,'' in \emph{Proc.\ 41st IEEE Symp.\ Found.\ Comput.\ Sci.\ (FOCS)}, 2000, pp.~390--398.

\bibitem{FM05} P.~Ferragina and G.~Manzini, ``Indexing compressed text,'' \emph{J.\ ACM}, vol.~52, no.~4, pp.~552--581, 2005.

\bibitem{FGMS05} P.~Ferragina, R.~Giancarlo, G.~Manzini, and M.~Sciortino, ``Boosting textual compression in optimal linear time,'' \emph{J.\ ACM}, vol.~52, no.~4, pp.~688--713, 2005.

\bibitem{VKV98} K.~Visweswariah, S.~R.~Kulkarni, and S.~Verd\'u, ``Source codes as random number generators,'' \emph{IEEE Trans.\ Inf.\ Theory}, vol.~44, no.~2, pp.~462--471, 1998.

\bibitem{Arikan10} E.~Ar{\i}kan, ``Source polarization,'' in \emph{Proc.\ IEEE Int.\ Symp.\ Inf.\ Theory (ISIT)}, 2010, pp.~899--903.

\bibitem{PWZ92} E.~Plotnik, M.~J.~Weinberger, and J.~Ziv, ``Upper bounds on the probability of sequences emitted by finite-state sources and on the redundancy of the Lempel--Ziv algorithm,'' \emph{IEEE Trans.\ Inf.\ Theory}, vol.~38, no.~1, pp.~66--72, 1992.

\bibitem{CoverThomas06} T.~M.~Cover and J.~A.~Thomas, \emph{Elements of Information Theory}, 2nd ed.\ Hoboken, NJ: Wiley, 2006.

\bibitem{MIT6441} Y.~Polyanskiy and Y.~Wu, \emph{Information Theory: From Coding to Learning}. Cambridge, U.K.: Cambridge Univ.\ Press, 2025, Sec.~9.4. [Online]. Available: \url{https://people.lids.mit.edu/yp/homepage/data/itbook-export.pdf}

\bibitem{code} A.~Garivier, ``ExhaustivePeres: companion code of \emph{The exhaustive Peres transform: extraction and universal coding},'' version~2.1, software, HAL Id hal-05802360, 2026. [Online]. Available: \url{https://hal.science/hal-05802360}

\bibitem{NZC11} G.~Nong, S.~Zhang, and W.~H.~Chan, ``Two efficient algorithms for linear time suffix array construction,'' \emph{IEEE Trans.\ Comput.}, vol.~60, no.~10, pp.~1471--1484, 2011.

\bibitem{CW84} J.~G.~Cleary and I.~H.~Witten, ``Data compression using adaptive coding and partial string matching,'' \emph{IEEE Trans.\ Commun.}, vol.~32, no.~4, pp.~396--402, 1984.

\bibitem{vAEdB51} T.~van Aardenne-Ehrenfest and N.~G.~de Bruijn, ``Circuits and trees in oriented linear graphs,'' \emph{Simon Stevin}, vol.~28, pp.~203--217, 1951.

\bibitem{Whittle55} P.~Whittle, ``Some distribution and moment formulae for the Markov chain,'' \emph{J.\ Roy.\ Statist.\ Soc.\ Ser.\ B}, vol.~17, no.~2, pp.~235--242, 1955.

\bibitem{ZhouBruckMarkov12} H.~Zhou and J.~Bruck, ``Efficient generation of random bits from finite state Markov chains,'' \emph{IEEE Trans.\ Inf.\ Theory}, vol.~58, no.~4, pp.~2490--2506, 2012.

\bibitem{SeroussiWeinberger15} G.~Seroussi and M.~J.~Weinberger, ``Optimal algorithms for universal random number generation from finite memory sources,'' \emph{IEEE Trans.\ Inf.\ Theory}, vol.~61, no.~3, pp.~1277--1297, 2015.

\bibitem{RyabkoRyabko11} B.~Ryabko and D.~Ryabko, ``Constructing perfect steganographic systems,'' \emph{Inform.\ and Comput.}, vol.~209, no.~9, pp.~1223--1230, 2011.

\bibitem{Silesia} S.~Deorowicz, ``Silesia compression corpus.'' [Online]. Available: \url{https://sun.aei.polsl.pl/~sdeor/index.php?page=silesia}

\end{thebibliography}
\end{document}